\documentstyle{article}
\def\Cop{{\rm l\!\!\!C}}
\def\Zop{Z\!\!\!Z}
\def\Rop{I\!\!R }
\def\Nop{I\!\!N}

\def\bbbz {{\sf Z\!\!Z}}

\renewcommand{\today}{May 1996}

\begin{document}
\thispagestyle{empty}
\begin{flushright}
  hep-th/9605020\\DAMTP 96-46
\end{flushright}
\vskip 2em
\begin{center}\LARGE
On Zhu's Associative Algebra as a \protect\\  Tool in the Representation Theory of \protect\\ Vertex Operator Algebras
\end{center}\vskip 1.5em
\begin{center}\large
 Klaus Lucke
\end{center}
\begin{center}\it
Department of Applied Mathematics and Theoretical
Physics, \\
University of Cambridge, 
Cambridge CB3 9EW, U.K.
\end{center}
\vskip 1em
\begin{center}
  \today
\end{center}
\vskip 1em
\begin{abstract}
We describe an approach to classify (meromorphic) representations of a
given vertex operator algebra by calculating Zhu's algebra
explicitly. We demonstrate this for FKS lattice theories and subtheories corresponding to the
$Z\!\!\!Z_2$ reflection twist and the $Z\!\!\!Z_3$ twist. Our work is mainly offering a novel uniqueness tool, but, as shown in the   $Z\!\!\!Z_3$ case, it can also be used to extract enough information to construct new representations.
We prove the existence and some properties of a new non-unitary representation of the $Z\!\!\!Z_3$-invariant subtheory of the (two dimensional) Heisenberg algebra. 
\end{abstract}

\section{Introduction} \label{real intro}
A conformal field theory consists of (two copies of) a vertex operator algebra and the set of its representations, which are combined subject to certain duality and modular properties \cite{ms}. To classify conformal field theories, and hence  critical phenomena in two dimensions, one has to obtain all vertex operator algebras and calculate all their representations. Then, for each vertex operator algebra, all possible ways to assemble the representations to obtain a conformal field theory have to be classified (cf \cite{itzykson} for the Virasoro vertex operator algebra).

Here we discuss a new approach to classify representations for a given vertex operator algebra. Complete knowledge of the representations of subalgebras can be used to classify all possible orbifolds associated with the given vertex operator algebra and  subalgebra (after checking further locality conditions). As orbifolding is one of the ways to construct new vertex operator algebras  from old ones, this is a further motivation for developing tools in representation theory.

Zhu noted \cite{zhu} that the irreducible  representations of a vertex operator algebra $V$ are in 1-1 correspondence with the irreducible representations of an associative algebra $A(V)$, which is given in terms of $V$ as a quotient $A(V):=\frac{V}{O(V)}$. More generally the correspondence is between the subclass of ``reconstructible'' representations of $V$ and representations of $A(V)$  (see section \ref{intro} for a summary of basic definitions). 

Our approach requires the explicit  form of the vertex operators, i.e. an implicit knowledge of $V$ is not sufficient. So far only few  vertex operator algebras have been fully constructed. Among these are those associated with Euclidean integral even lattices and their subtheories, the representation theory of which is not yet understood. 

Here we {\it explicitly} construct $A(V)$ for some of these vertex operator algebras. In the construction we can make explicit use of the ground states of known representations, so in fact our approach, in its simplest version,  is most useful as a uniqueness tool. However a more detailed study of the structure of $A(V)$
can also provide existence tools, as we demonstrate in section \ref{z3}. In the case of the Virasoro and affine Kac-Moody vertex operator algebras $A(V)$ had previously been constructed by Frenkel, Zhu and Wang 
\cite{fz,w}. However their approach seems to be quite hard to apply in a more general situation (compare section \ref{non-tensor}).

The dimension of $A(V)$ is closely related to the number of inequivalent irreducible representations $M^{(i)}$ weighted with the square of the dimension of the space of ground states $M_0^{(i)}$. In section \ref{conjecture}, where we discuss the general structure of $A(V)$,  we see that for rational $V$
\begin{equation} \label{blabla}
  A(V) = \oplus_i  End M^{(i)}_0 .
\end{equation}

To calculate $A(V)$ explicitly for a given vertex operator algebra $V$, we generally proceed as follows. We write down a number of easy vectors in 
$O(V)$, spanning some $\tilde O(V) \subset O(V)$, and check if a basis of  $\tilde A(V):= \frac{V}{\tilde O(V)}$ has an independent action on the ground states of representations $\oplus_i  M^{(i)}_0$, which would ensure that indeed $A(V) = \tilde A(V)$, as the zero mode of any vector in $O(V)$ has a trivial action on ground states \cite{zhu}. If the action were not independent, we would have to look for more representations {\it or} more vectors in $O(V)$. In fact, if we introduce inductive
arguments (in weight and/or number of oscillators in theories associated with lattices), we only need to calculate few terms in vectors in $\tilde O(V)$ to construct $\tilde A(V)$. 

In the case that there are no new representations we rarely have to calculate the multiplication rule in $A(V)$. If needed, it would follow automatically from the multiplication of zero modes on the ground states and the fact that the action of a basis of $A(V)$ on these ground states is independent. In section \ref{sec2} this is the case  for $V$ being the Heisenberg algebra or the full FKS lattice theory
with the lattice $\Lambda = \sqrt{2} \Zop$. In section \ref{pure oscillators}, \ref{sec4} and \ref{sec6} we construct $A(V)$ for the $\Zop_2$ projected Heisenberg algebra, for its generalizations to higher dimension and for the $\Zop_2$ projected FKS lattice theory with $\Lambda = \sqrt{2} \Zop$. As a result we rederive the ``uniqueness of the twisted representation'' (cf. \cite{montague}). 

If, however, we seem to have missed representations, seen as $\tilde A (V)$ containing non-zero elements  acting trivially on the ground states of all known representations, we have to proceed as follows. First we map out $O(V)$ by calculating $a \Box b $ systematically for all $a,b$ up to a certain weight (see section \ref{intro} for the notation). [Such brute force exercise has to be done on a computer.] This should leave us with a fairly strong conjecture about $A(V)$ as a space. Next we can calculate all $\star$ products for the conjectured $A(V)$. This can again be a fairly time consuming exercise on the computer, as we have to bring the products $a \star b$ into normal form, meaning we have to move to our selected representant in the coset $[ a \star b]$. Having a conjecture for $(A(V),\star)$ as an associative algebra, we can easily determine its representation theory. This leaves us with (conjectured) knowledge about the ground states of ``missed'' representations. 
In section \ref{z3} we demonstrate this  for the $\Zop_3$ projection   of the two dimensional Heisenberg algebra. Here the conjecture is even suggesting a simple way to prove, that we indeed have a new representation. The proof is given in section \ref{pf of repnconj}.
 We obtain a surprising new indecomposable reducible (hence non-unitary) representation of this unitary 
vertex operator algebra. We find that this representation, together with the  known (reducible and irreducible) untwisted and twisted representations \cite{z3montague}, completes the set of reconstructible representations.

Our approach is very close in spirit to work by Montague \cite{montague} in that it is most useful as  a uniqueness tool and e.g. no use is made of  information about the classification of non-meromorphic representations of the full theory when we discuss subtheories. In fact it is possible to exploit partial results of this work in Montague's approach, by using whatever we know about $O(V)$ for his selection of ground states. There is however a crucial difference between this work and \cite{montague}. Here we do not restrict to unitary representations. In fact it is not quite clear whether the conditions in \cite{montague} do rule out non-unitary representations.

For the convenience of those readers that are only interested in parts of the article, we note that sections \ref{real intro}-\ref{pure oscillators}  are essential to understand the remainder. Sections \ref{sec4}, \ref{sec6}, \ref{z3} can be read in any order afterwards.

\section{Basic Definitions} \label{intro}
One way to axiomatically describe what is known as the chiral algebra in two dimensional conformal field theory \cite{bpz,ms} is to use the notion of {\it vertex operator algebra} \cite{b,fhl,flm}. It consists of a graded vector space $V=\oplus_{n \in \Nop} V_n$ of states and vertex operators (field operators) $V(\psi,z)$ for every state $\psi$. The vertex operators are formal Laurent series in $z$ with linear operators on $V$ as coefficients (modes). The correspondence between states and fields is linear and bijective. The vertex operator corresponding to the vacuum $\vert 0 \!>\in V_0$ is the unit operator. The field $V(\psi_L,z)$ corresponding to the energy momentum tensor has modes $L_n$ satisfying the Virasoro algebra with fixed  central charge $c \in \Cop$.  Furthermore the correspondence of states and fields is meant to respect conformal transformations, which can be assured by assuming 
\begin{eqnarray}
L_0 \vert_{V_n} &=& n \ id_{V_n}  \label{L0 axiom} \\
V(L_{-1} \psi,z) &=& \frac{d}{dz} V(\psi,z) \quad .
\end{eqnarray}
To assure locality and duality of the correlation functions, one of a variety of equivalent axioms (see \cite{fhl}) has to hold. We assume (e.g.) for $a,b,c \in V$
\begin{eqnarray*}
\lefteqn{Res_{z\!-\!w} \Bigl(V(V(a,z-w) b,w) c \ \iota_{w,z\!-\!w}((z-w)^m z^n))\Bigr)=}\\
&&Res_{z} \Bigl(V(a,z) V(b,w) c \ \iota_{z,w}((z\!-\!w)^m z^n))\Bigr)\\
&&-Res_{z} \Bigl(V(b,w) V(a,z) c \ \iota_{w,z}((z\!-\!w)^m z^n))\Bigr)
\end{eqnarray*}
with $\iota_{x,y}$ denoting that the formal expansion is done as in the domain $|x| > |y|$ and the result is regarded as a formal Laurent series in $x$ and $y$.

We are using the notion of  a {\it representation} $M=\oplus_{n \in \Nop} M_n$ (with $M_0 \neq \lbrace  \rbrace$) of $V$, if there are also field operators $V_M(\psi,z)$ ($\psi \in V$),  that act on $M$ with just the ``same'' set of axioms\footnote{We often write $V(\psi,z)$ or $U(\psi,z)$ for $V_M(\psi,z)$.}. Only  (\ref{L0 axiom}) will not be assumed. It is replaced by the weaker version
$$ \psi_n \ M_m \subset M_{m+n} $$
with $\psi_n$ from $V_M(\psi,z) = \sum_{n \in \Zop} \psi_n z^{-n-h}$ for $\psi \in V_h$, so that we allow for arbitrary weights in the space of ground states $M_0$. Also the locality-duality property (for $a,b \in V$ and $c \in M$) has to be interpreted as
\begin{eqnarray*}
\lefteqn{Res_{z\!-\!w} \Bigl(V_M(V(a,z-w) b,w) c \ \iota_{w,z\!-\!w}((z-w)^m z^n))\Bigr)=}\\
&&Res_{z} \Bigl(V_M(a,z) V_M(b,w) c \ \iota_{z,w}((z\!-\!w)^m z^n))\Bigr)\\
&&-Res_{z} \Bigl(V_M(b,w) V_M(a,z) c \ \iota_{w,z}((z\!-\!w)^m z^n))\Bigr) \ .
\end{eqnarray*}

Note that this axiomatization  does not assume any positive definite inner product structure as in other axiomatic approaches \cite{g,dgm95}.

To define the subclass of {\it reconstructible representations} we need some auxiliary concepts: Given a representation $M$, we can define 
{\it correlation functions} 
$$S: M_0' \otimes V \otimes \dots \otimes V \otimes M \rightarrow \Cop \, [[ z_1,\dots,z_n]]$$
by using the natural bilinear pairing of $M_0'$ with $M_0$, extended to a pairing of $M_0'$ with $M$ by $(m',m):=(m',m_0)$ (for 
$m' \in M_0'$, $m=\sum_k m_k$, $m_k \in M_k$), as
\begin{equation}
S(m';\psi_1,\dots,\psi_n;m):= (m', V_M(\psi_1,z_1) \dots   V_M(\psi_n,z_n) m) \ .
\end{equation}
It can be shown \cite{fhl} that the formal series converges (in the domain of radially ordered $z_i$) to a rational function and the
expected locality and duality properties hold. Now we can define a subrepresentation $Rad M$:  
\begin{eqnarray} 
Rad M &:= \Bigl\lbrace m \in M : & \!\! S(m';\psi_1,\dots,\psi_n;m)=0 \ \mbox{for all} \ n \in \Nop, \nonumber \\
&& m' \in M_0', \psi_1,\dots,\psi_n \in V \Bigr\rbrace \quad .\label{def rad}
\end{eqnarray}
We can define another subrepresentation $\bar M $ as the subspace of $M$ generated by the action of vertex operators on the space of ground states $M_0$. We finally define $\tilde M:= \bar M/ Rad \bar M$, the {\it reconstructible representation associated with $M$}. [$M$ is called {\it reconstructible}, iff $M$ is generated from $M_0$ (by the action of vertex operators) and $Rad M ={0}$, so that the weak topology, induced by the correlation functions, seperates points in $M$.]

Following Zhu, we define (confer \cite{zhu,fz,w,montague}) for $a,b \in V$\footnote{Strictly speaking we always restrict to homogeneous elements if the weight appears in a formula. The extension of the formula to arbitrary elements is the obvious one.}:
$$a \star b :=Res_z(V(a,z) \frac{(z+1)^{wta}}{z} b)$$
$$a \Box b :=Res_z(V(a,z) \frac{(z+1)^{wta}}{z^2} b)$$
$$a  \Box_n b :=Res_z(V(a,z) \frac{(z+1)^{wta}}{z^{n+2}} b)$$
$$O(V):=span(a \Box_n b : a,b \in V , n \in \Nop) \quad .$$
Zhu shows in \cite{zhu} that $O(V)$ is a two sided ideal of the bilinear product $\star$. Hence this product is inherited by the quotient space\footnote{The equivalence class of $a$ is denoted by $[a]$. We only use the brackets for the equivalence classes when we are considering elements in $A(V)$, rather than 
$\tilde A(V)$ which was introduced in the introduction. In the later case we do not bother to distinguish between classes and their representants.}
$A(V):= \frac{V}{O(V)} $.
Zhu further shows, that $(A(V),\star,1)$ is an associative algebra with unit $1=[|0\!>]$.

\section{General Remarks on $A(V)$} \label{conjecture}
We recall the following theorem by Zhu \cite{zhu}, which forms the basis for our work:\\
\\
{\it Theorem:} 
If $M=\oplus_{n=0}^{\infty} M_n$ is a representation of the vertex operator algebra $V$, then the ground states $M_0$ form a representation of the associative algebra with unit $(A(V),\star,1)$. The action is given as follows:  $[a] \in A(V)$ acts on $M_0$ as $a_0$ (independent of the representant).
Conversely, if $\pi : A(V) \rightarrow End W$ is a representation of  $(A(V),\star,1)$, then there exists a reconstructible representation $M=\oplus_{n=0}^{\infty} M_n$ of $V$, such that $M_0=W$ and $a_0 v = \pi([a])v$ for every $a \in V, v \in W.$\\

A constructive proof is given in \cite{zhu}. As correlation functions are uniquely specified by the action of zero modes on the ground states, we see that Zhu's theorem provides a bijective correspondence between reconstructible representations of $(A(V),\star,1)$ and  representations of $V$.
As it preserves irreducibility, it induces a bijective correspondence of irreducible representations of $V$ and irreducible representations of $A(V)$. [Note that for reconstructible representations $M$ any subrepresentation $\hat M$ fulfils $Rad \hat M \equiv Rad_{\hat M_0'} \hat M = Rad_{M_0'} \hat M \subset Rad_{ M_0'}  M \equiv Rad M =\lbrace 0 \rbrace$ with indices denoting that in definition (\ref{def rad}) $m'$ runs over the space given by the index, the second equality being a consequence of $\hat M_0 \subset M_0$ (cf remark 2.2.1 in \cite{zhu}) and the invariance of $\hat M$. However $\hat M $ may not be generated from $\hat M_0$, so that $\hat M$ may not be reconstructible.
So the correspondence may not preserve the submodule structure.]

Further, if $V$ is rational, $A(V)$ can be decomposed into a finite number of irreducible (right) ideals. [$V$ is called {\it rational}, if it has only a finite number of irreducible representations and every finitely generated representation is a finite direct sum of irreducibles.]\\
\\

Using a few simple results about associative algebras with unit, compiled in appendix \ref{associative appendix}, we can be more precise: If $V$ is rational with inequivalent irreducible representations $M^{(i)}$, all inequivalent irreducible representations of $A(V)$ are $M_0^{(i)}$. Hence, by proposition B1, $A(V)$ can be identified with a subalgebra of $\oplus_i End M_0^{(i)}$, and hence, by proposition B5, with $\oplus_i End M_0^{(i)}$. So for rational $V$
$$  A(V) = \oplus_i  End M^{(i)}_0 .$$

In practise we often know a certain set of irreducible representations of $V$, but do not know, whether it is complete. Let us suppose that 
\begin{equation} \label{formula conj}
dim \tilde A(V) = \sum_{i \in \mbox{known irred repns}} dim End M_0^{(i)}
\end{equation}
and that the action of the basis of $\tilde A(V)$ on $\bigoplus_{i \ known} M_0^{(i)} $ is independent. This is the situation we will have  at the end of most sections.

Under these conditions $A(V)=\tilde A(V)$ (independent action) and the multiplication in $A(V)$ can be recovered from the action on the ground states. In the case of a finite dimensional $A(V)$ we can further note that the known irreducible representations are inequivalent and $A(V)$  is  isomorphic to $\bigoplus_{i \ known} End M_0^{(i)}$ as an associative algebra with unit. So, by proposition B4, all finitely generated representations of $A(V)$ decompose into finite direct sums of the known $M_0^{(i)}$. So every  reconstructible representation that is finitely generated from the ground states decomposes into a finite direct sum of the irreducible $M^{(i)}$. To prove rationality, we would need an argument (like in \cite{fz,w} for the Kac-Moody and Virasoro vertex operator algebras), that all  (finitely generated) representations decompose into reconstructible ones (i.e. are reconstructible up to shifts in the grading of components).

In the examples we will consider, $A(V)$ is often the direct sum of the polynomial algebra and a finite dimensional algebra. We discuss this case when we need it at the end of section \ref{repns pure}.

\section{Full Lattice Theory} \label{sec2}

\subsection{Heisenberg Algebra} \label{full hb alg}
As a warmup we consider the pure Heisenberg algebra\footnote{We use the normalization $\lbrack a_n , a_m \rbrack = n \delta_{n+m,0}$  $(n \in \Zop)$.} (without momenta)
$$V:=H:=span(\prod_{i=1}^N a_{-n_i} \vert 0 > : N \in \Nop, n_i \in \Nop \setminus \lbrace 0 \rbrace )$$
with the usual vertex operators (cf. (\ref{V's})). As this is a special case of the Kac-Moody vertex operator algebra ($G=U(1)$), $A(V)$ is known \cite{fz}. However we would still like to illustrate our approach in that simple case.

\subsubsection{$\tilde A(H)$}
We generate $\tilde O(V)$ by the following vectors:
\begin{eqnarray}
a_{-1} \vert 0 > \Box_n \psi &=&Res_z \sum_k a_{-k} z^{k-1} \frac{z+1}{z^{n+2}} \psi \nonumber \\ 
&=& (a_{-(n+2)}+a_{-(n+1)}) \psi \qquad n \in \Nop  \quad .\label{basicreln}
\end{eqnarray}
From this we get $\tilde A(H)=span( a_{-1}^n \vert 0 > : n \in \Nop )$.

\subsubsection{Representations} \label{cheat}
We know, that for every momentum $k \in \Cop$, we have an (irreducible) representation ($M^{(k)} =span (\prod_i a_{-n_i} \vert k > )$ as usual). The action of the $[a_{-1}^n \vert 0 >]$ is independent on the ground states $ \vert k >$ as
\begin{eqnarray}
[a_{-1}^n \vert 0\!\!>]_0 \vert k \!\!> &=& [a_{-1} \vert 0\!\!> \star \dots \star a_{-1} \vert 0 \!\!>]_0 \vert k \!\!> =  [a_{-1} \vert 0 \!\!>]_0 \dots [a_{-1} \vert 0 \!\!>]_0 \vert k \!\!> \nonumber\\
&=& a_0^n \vert k > = k^n \vert k > \nonumber
\end{eqnarray}
and the corresponding monomials are independent.  So indeed $A(V)= \tilde A(V)$, the multiplication is abelian and hence $A(V)$ is isomorphic to the full polynomial algebra with irreducible representations corresponding to the possible evaluations of the polynomials ($k \in \Cop$). So the representations $M^{(k)}$ are indeed the full set of inequivalent irreducible representations by Zhu's theorem \cite{zhu}.

\subsection{Momenta Included}
As our second example we consider the full FKS lattice theory constructed from the lattice $\Lambda = \sqrt{2} \Zop$:
$$V:=H(\sqrt{2} \Zop):=span (\prod_{a:=1}^N a_{-n_a} \vert k > : k \in \sqrt{2} \Zop, N \in \Nop, n_a \in \Nop \setminus \lbrace 0 \rbrace)$$
with corresponding vertex operators \cite{dgm} for the state 
\begin{equation} \label{state}
\psi = \prod_{a:=1}^N a_{-n_a} \vert k > 
\end{equation}
\begin{equation} \label{V's}
V(\psi,z) = e^{ikX_<(z)} e^{ikq} z^{kp} : \prod_{a=1}^{N} \sum_{l_a \in \Zop} a_{l_a} z^{-l_a-n_a} { -l_a-1 \choose n_a-1} : e^{ikX_>(z)} \sigma_k
\end{equation}
with $X_{<>}(z):=i\sum_{n<>0}\frac{a_n}{n}z^{-n}$ and $\sigma_k \vert k'> := \epsilon(k,k') \vert k'>$. \\
The $\epsilon$'s ($\in \lbrace 1,-1 \rbrace $) satisfy (cf appendix B of \cite{dgm})
\begin{eqnarray}
\epsilon(k,k') &=& (-)^{k k'} \epsilon(k',k)  \qquad \mbox{(locality)} \nonumber \\
\epsilon(k,k') \epsilon(k+k',k'') &=& \epsilon(k,k'+k'') \epsilon(k',k'') \qquad \mbox{(associativity)}\nonumber \\
 \epsilon(k,0) &=& 1  \qquad \mbox{(creation)} \nonumber \\
\epsilon(k,-k) &=& (-)^\frac{k^2}{2} \qquad \mbox{(gauge)} \nonumber \\
\epsilon(k,k') &=& \epsilon(-k,-k') \qquad \mbox{(gauge)} \quad . \label{epsilons}
\end{eqnarray}
This example is still very simple (as indeed all full lattice theories), as we have a means to promote simple objects in $O(V)$ to more complicated ones by using 
\begin{equation}
[\psi]=0 \qquad \Rightarrow \qquad 0=[\psi \star a_{-1} \vert 0 >]=[a_{-1} \psi] \qquad \Rightarrow \qquad [a_{-1}^n \psi ] =0  \quad . \label{raise}
\end{equation}
[Here we used $[a \star b] = 
[Res_z Y(b,z) \frac{(z+1)^{wtb-1}}{z} a ]$ (cf \cite{zhu}) for $b=a_{-1} \vert 0>$.]
In the $\Zop_2$-projected examples we will have no immediate equivalent of  relation (\ref{raise}) as it requires $a_{-1} \vert 0 > \in V$.
To proceed, we need a simple combinatorical identity for the exponentials appearing in the vertex operators of states with non-zero momentum:
\begin{equation} \label{combinatorics}
e^{ikX_<(z)}= \sum_{m=0}^{\infty} z^m \sum_{\sum_{i=1}^{\infty}i r_i =m} \frac{(k a_{-1})^{r_1}}{1^{r_1} r_1!} \frac{(k a_{-2})^{r_2}}{2^{r_2} r_2!} ...
 \qquad . 
\end{equation}
If we use (\ref{basicreln}) this simplifies in our context (i.e. when acting on $V/O(V)$)  to
\begin{equation}
e^{ikX_<(z)} \equiv \sum_{m=0}^{\infty} z^m \sum_{\sum_{i=1}^{\infty}i r_i =m} \frac{(k a_{-1})^{r_1+r_2+...} (-)^{r_2+r_4+r_6+...}}{1^{r_1} r_1! 2^{r_2} r_2! ...}  \quad . \label{exp}
\end{equation}

\subsubsection{$\tilde A(H({\protect\sqrt{2}} \Zop))$}
We consider 
\begin{equation} \label{interact}
\vert \sqrt{2} > \Box_n \vert m \sqrt{2} > = Res_z e^{i\sqrt{2} X_<(z)} \vert (m+1) \sqrt{2} > z^{2m} \frac{z+1}{z^{2+n}} \epsilon(\sqrt{2}, m \sqrt{2})
\end{equation}
with the exponential expanded as in (\ref{exp}). We obtain as elements of $O(V)$
\begin{eqnarray}
\vert (m+1) \sqrt{2} > & \qquad \qquad & (m \geq 1, n=2m-1) \label{a1}\\
(\sqrt{2} a_{-1} +1) \vert \sqrt{2} > && (m=n=0) \\
(2a_{-1}^3-a_{-1}) \vert 0> && (m=-1, n=0) \quad . \label{a3}
\end{eqnarray}
Similarly $\vert -(m+1) \sqrt{2} > \qquad (m \geq 1), \quad (-\sqrt{2} a_{-1} +1) \vert -\sqrt{2} > \in O(V)$. This combined with (\ref{basicreln}), (\ref{raise}) gives
\begin{equation}
\tilde A(H(\sqrt{2} \Zop))= span ( \vert 0 >, a_{-1} \vert 0>, a_{-1}^2 \vert 0> , \vert \sqrt{2} >, \vert -\sqrt{2} > )  \quad . \label{basis}
\end{equation}

\subsubsection{Representations}
Here we can offer two representations, the adjoint and shifted ones, with the shifted one being built on the ground states of weight 1/4 
$$ M_0^{(shifted)}=span ( \vert \frac{1}{\sqrt{2}} >, \vert -\frac{1}{\sqrt{2}}>  ).$$
The action of the zero modes of the five vectors in (\ref{basis}) on the ground states provides a basis of
$$ \bigoplus_{i \in \lbrace shifted,adjoint \rbrace} End M_0^{(i)} .$$
Hence $A(V)=\tilde A(V)$ and the set of reconstructible representations is complete (cf. section \ref{conjecture}). 

Note that we were able to translate the relations in $O(V)$ for zero momentum to
any momentum by using (rather symbolically) $0 \times k$. Additional relations were found by lattice interaction $k \times k'$. For $k' = -k$ we even found new relations at zero momentum, reflecting the fact, that not all representations of the Heisenberg algebra lift to representations of $V$. To conclude from (\ref{a1}-\ref{a3}) on the form of $\tilde A(V)$ we did not use all information, but only, that certain terms of higher weight can be transformed to terms of lower weight (for fixed momentum). The only valuable information is the non-zero coefficient in front of the term with high weight. In later sections we will only compute this part, to avoid lengthy expressions.

\section{$Z\!\!\!Z_2$-Projected Heisenberg Algebra} \label{pure oscillators}
In this section we analyse the following {\it subtheory} of the Heisenberg algebra with the {\it number of oscillators restricted to be even}:
$$V:=H_+:=span(\prod_{i:=1}^{2n} a_{-m_i} \vert 0 > \ : n \in \Nop ) \quad .$$
This corresponds to the subtheory invariant under the $\Zop_2$-automorphism  corresponding to the reflection twist \cite{dgm}.
Here we neither have (\ref{basicreln}) nor (\ref{raise}) at hand, as $a_{-1} \vert 0 \!> \not\in V$. However we will again try to involve as few as possible oscillators in our basic relations.

\subsection{$\tilde A(H_+)$} \label{non-tensor}
We employ the general relation
\begin{equation} \label{vir}
\psi \Box \vert 0 > = (L_0 + L_{-1}) \psi
\end{equation} 
for $\psi = a_{-n} a_{-m} \vert 0 >$ to obtain
$$((n+m)a_{-n} a_{-m} +n a_{-n-1} a_{-m} +m a_{-n} a_{-m-1}) \vert 0 > \in O(V) .$$
Looking only at top weight this reads\footnote{For the notation see below.}
\begin{equation} \label{todiagonal}
a_{-n-1}a_{-m}\vert0>\approx -\frac{m}{n} a_{-n}a_{-m-1}\vert 0 >
\end{equation}
and hence we can bring all states made of two oscillators to a normal form only involving $a_{-n}^2 \vert 0>$ or alternatively $a_{-n}a_{-1}\vert 0>$ ($n$ odd).

Warning: As we can prove that\footnote{We denote the largest number of oscillators in a term of $\psi$ as $N(\psi)$.} $N(\psi \Box \phi)=N(\psi)+N(\phi)$, it is tempting to believe that we already have all relations in $O(V)$ for $N\leq 2$. However as $\lbrace \phi \Box \psi : \phi,\psi \in V \rbrace \neq span( \phi \Box \psi : \phi, \psi \in V )$, we will see that there are additional objects in $O(V)$ with $N \leq 2$ obtained by the cancellation of the $N=4$ terms in objects like $\phi_1 \Box \psi_1 -\phi_2 \Box \psi_2$. This makes it very hard to generalize the type of proof in \cite{fz} and \cite{w} to show that a given $\tilde O(V)$ is indeed equal to $O(V)$ without using known representations explicitly. [In section \ref{z3} we show how to circumvent this problem by working out a conjecture for new representations to prove equality of $\tilde O(V)$ and $O(V)$.]

For clarity we introduce a partial ordering on states of the form $$\phi=a_{-n_1} \dots a_{-n_N} \vert k > \qquad \qquad (n_i >0, N\geq 0, k \in \Rop)$$
we say $\phi^{(1)} < \phi^{(2)}$ if  
\begin{eqnarray*}
n_1^{(1)}+...+n_{N^{(1)}}^{(1)}+\frac{{k^{(1)}}^2}{2} &\leq& n_1^{(2)}+...+n_{N^{(2)}}^{(2)}+\frac{{k^{(2)}}^2}{2} \\
\vert k^{(1)} \vert &\leq &\vert k^{(2)} \vert \\
N^{(1)} &\leq & N^{(2)}
\end{eqnarray*}
with at least one inequality being strict. As a slightly weaker notion we say  $\phi^{(1)} <_{wt} \phi^{(2)}$ if (possibly only) the first two inequalities are true, still at least one of the three being strict.
Whenever we use ``$\approx$'' (``$\approx_{wt}$'') instead of ``$=$'' we mean ``equal up to terms that are smaller than one of the terms in the expression written down (w.r.t. this partial ordering) modulo relations in $O(V)$ that have previously been pointed out''. In fact we use this notation to hide the inductions (in this ordering) tacidly performed in the sequel.

With this notation we generalize (\ref{todiagonal}) to larger $N$ by
\begin{eqnarray} 
0 &\approx & a_{-n}a_{-m} \vert 0> \Box a_{-n_1} \dots a_{-n_{N-2}} \vert 0 > \nonumber \\
&\approx &(n a_{-n-1} a_{-m} +m a_{-n} a_{-m-1}) a_{-n_1} \dots a_{-n_{N-2}} \vert 0> \quad . \label{Ntodiagonal}
\end{eqnarray}
So far we succeeded to reduce $\tilde A(V)$ to $span  ( \vert 0> , a_{-\alpha} a_{-1}^{2n+1} \vert 0 > : \alpha \ odd, n \in \Nop )$. For $n\!\geq \!1, \alpha \!\geq \!3$ we can further write $a_{-\alpha} a_{-1}^{2n\!+\!1} \vert 0 \!\!>$ in terms of $a_{-\!(\alpha \!-\!1)} a_{-\!2} a_{-\!1}^{2n} \vert 0\!\!>$ which can be reduced to lower weight ($a_{-2} a_{-1} \rightarrow a_{-1}^2$). Hence 
$$ \tilde A(V) = span  ( \vert 0> ,  a_{-1}^{2n} \vert 0 > : n \geq 2, a_{-\alpha} a_{-1} \vert 0 >: \alpha \ odd) \quad .$$
As announced before, this is still not the whole story at $N=2$. The task is to find a non-trivial relation at $N=2$ by using relations at $N=4$ with all $N=4$ parts cancelled (at least at the top weight)\footnote{Here we only show the cancellation at the top weight. However in actual fact, some more effort should enable us to show that all appearing $N=4$ objects at lower weight can be reduced to $N=2$ using only the relations that already appeared before. This has been checked using REDUCE for small values of $\alpha$.}. A number of the most obvious cancellations at $N=4$  can be shown to reduce to a trivial result at $N=2$. One way to produce a non-trivial object is to reduce $a_{-m} a_{-3} a_{-2} a_{-1} \vert 0>$ (for $m$ even) to 
$a_{-m-3} a_{-1}^3 \vert 0>$ in two different ways, (a) by first $(a_{-m} a_{-2}   \rightarrow a_{-m-1} a_{-1})$ and then $(a_{-m-1} a_{-3} \rightarrow a_{-m-3} a_{-1})$, or (b) by $(a_{-3} a_{-2} 
\rightarrow a_{-4} a_{-1})$ followed by $(a_{-m} a_{-4} \rightarrow a_{-m-3} a_{-1})$. It turns out that for $m$ even  and $m\geq 4$ this shows that we can express $a_{-m-5}a_{-1} \vert 0 > $ in terms of lower weights. A shortcut variation of this is
(for $m$ even):
\begin{eqnarray} 
0 &\approx &
 a_{-3} a_{-1} \vert 0 > \Box a_{-m} a_{-1} \vert 0 >
-a_{-m} a_{-3} \vert 0 > \Box a_{-1} a_{-1} \vert 0 >  \nonumber \\
&&-a_{-m} a_{-1} \vert 0 > \Box a_{-3} a_{-1} \vert 0 >
+m a_{-m-1} a_{-2} \vert 0 > \Box a_{-1} a_{-1} \vert 0 > \nonumber \\
&&
- \frac{m (m+1)}{2} a_{-1} a_{-1} \vert 0 > \Box a_{-m-2} a_{-1} \vert 0 >  \label{heisenberg without computer}\\
&\approx_{wt}&
\Biggl([m a_{-(m+5)} a_{-1} \lbrace { -m\!-\!1 \choose 2} + {m\!+\!4 \choose 2} \rbrace +  a_{-6} a_{-m} \lbrace { -2 \choose 2} + {5 \choose 2} \rbrace] \nonumber \\
&&
-[2 a_{-(m+5)} a_{-1} \lbrace { -2 \choose m-1}  {m+4 \choose 2} + { -2 \choose 2}  {m+4 \choose m-1} \rbrace] \nonumber \\
&&
-[3 a_{-\!(m\!+\!5)} a_{-\!1} \lbrace { -4 \choose m\!\!-\!\!1}\!\!+ \!\!{m\!\!+\!\!4 \choose m\!\!-\!\!1} \rbrace \!+ \!a_{-\!m\!-\!3} a_{-\!3} \lbrace { -2 \choose m\!\!-\!\!1} \!\!+\!\!{m\!\!+\!\!2 \choose m\!\!-\!\!1} \rbrace] \nonumber \\
&&
+m [2 a_{-(m+5)} a_{-1} \lbrace { -2 \choose m}  {m+4 \choose 1} + { -2 \choose 1}  {m+4 \choose m} \rbrace] \nonumber \\
&&
-\frac{m (m+1)}{2}[(m+2) a_{-(m+5)} a_{-1} 2 + 1 a_{-4} a_{-m-2} 2 ]\Biggr) \vert 0 >  \quad .  \label{heisenberg without computer result}
\end{eqnarray}
We collect terms not yet of the form $a_{-(m+5)} a_{-1} \vert 0 >$ and convert them using (\ref{todiagonal})
$$\Bigl(13 a_{-6} a_{-m} -(-m+\frac{(m+2) (m+1) m}{3!})a_{-m-3} a_{-3} -m (m+1) a_{-4} a_{-m-2}\Bigr) \vert 0 > \approx $$
$$\frac{(m+4) (m+3)}{2} (m- \frac{(m+2) (m+1) m}{20}) a_{-(m+5)} a_{-1} \vert 0 >.$$
Inserting this and factorizing  the coefficient finally yields
\begin{equation} \label{N=2 trick}
 0 \approx_{wt} - \frac{4}{15} m (m+2) (m+4) (m+6) (m-2)  a_{-(m+5)} a_{-1} \vert 0 >
\end{equation}
giving a non-trivial result for $m \neq 2$.
This gives
\begin{equation}
\tilde A(H_+) = span  ( a_{-1}^{2n} \vert 0 > : n \in \Nop, a_{-\alpha} a_{-1} \vert 0 >: \alpha \in \lbrace 3,5,7 \rbrace) .
\end{equation}

\subsection{Representations} \label{repns pure}

For later use we denote 
$$even(k):=\vert k > + \vert -k> \qquad \mbox{and} \qquad odd(k):=\vert k > -\vert-k> .$$
We know irreducible representations built on the following ground states with usual (untwisted) vertex operators:
 For fixed $k \in \Cop \setminus \lbrace 0 \rbrace$: $M_0^{(k,+)}\!\!\!=\!\!span(even(k))$ , $M_0^{(k,-)}=span(odd(k))$ (both have weight $k^2/2$ and are seen to be equivalent); $M_0^{(0,+)}=span( \vert 0 >)$ (adjoint), $M_0^{(0,-)}=span( a_{-1} \vert 0 >)$ (weight 1).
We also have two irreducible representations with twisted vertex operators. For later use we recall the formulae from \cite{dgm} also for non-vanishing momenta.
For the $\psi$ as in (\ref{state})
\begin{equation} \label{twistvop}
V_T(\psi,z)=V_0\left(e^{\Delta(z)}\psi,z\right)\,,
\end{equation}
where
\begin{eqnarray*}
V_0(\psi,z)&=&:\prod_{a=1}^N{i\over (n_a-1)!}{d^{n_a}\over dz^{n_a}}
R(x) e^{i k\cdot R(z)}:\gamma_k \,,\\
R(z)&=&i\sum_{r\in \Zop+{1\over 2}}{c_r\over r}z^{-r}\\
\Delta(z)&=&-{1\over 2}p^2 \ln 4z + {1\over 2}\sum_{{m,n\geq 0\atop
(m,n)\neq (0,0)}}
\left({-{1\over 2}\atop m}\right)\left({-{1\over 2}\atop n}\right)
{z^{-m-n}\over m+n} a_m\cdot a_n\,.
\end{eqnarray*}
The ground states are 
$M_0^{(T,+)}=span( \chi)$ (weight 1/16), $M_0^{(T,-)}=span( c_{-1/2} \chi)$ (weight 9/16).

To examine the action of the zero modes of our basis of $\tilde A(V)$, we note that in the untwisted case we can use the reasoning of section \ref{cheat} and obtain
\begin{eqnarray*}
(a_{-1}^{2n} \vert 0>)_0 even(k) &=& k^{2n} even(k)\\
(a_{-1}^{2n} \vert 0>)_0 odd(k) &=& k^{2n} odd(k)\\
(a_{-1}^{2n} \vert 0>)_0 \vert 0> &=& 0\\
(a_{-1} a_{-\alpha} \vert 0> )_0 \vert k>&=&k^2 \vert k> \qquad (\alpha \ odd) \quad . 
\end{eqnarray*}
Hence we see that only vectors with exclusively $N=2$ terms can possibly be candidates for being in $O(V)$. So we look at 
\begin{equation} \label{ansatz}
\sum_{i=1}^4 \lambda_i a_{-1} a_{-2 i +1} \vert 0> \quad .
\end{equation}
The condition for the zero mode of (\ref{ansatz}) to vanish on $\vert k>$ is  
$$\lambda_1+\lambda_2+\lambda_3+\lambda_4=0 \quad .$$
On the ground state $M_0^{(0,-)}$ we obtain
$(a_{-1} a_{-\alpha} \vert 0> )_0 a_{-1} \vert 0>= (\delta_{\alpha,1}+\alpha) a_{-1} \vert 0> (\alpha \ odd)
$ and hence for (\ref{ansatz}) to vanish on this ground state 
$$2 \lambda_1+ 3 \lambda_2+ 5 \lambda_3+ 7 \lambda_4=0 \quad . $$
For the twisted representation we note 
$$e^{\Delta(\!z\!)} a_{-\!1}a_{-\!\alpha} \vert 0\!\!> = a_{-\!1}a_{-\!\alpha} \vert 0\!\!> \!-\frac{\alpha}{2 (\!\alpha\!+\!1\!)}z^{\!-\!\alpha \!-\!1} {-\!\frac{1}{2} \choose \alpha} \vert 0\!\!>$$
and $(V_0(a_{-1} a_{-\alpha} \vert 0>,z))_0$ contributes only on $M_0^{(T,-)}$ with $\frac{1}{2} {-3/2 \choose \alpha -1 } +\frac{1}{2} {-1/2 \choose \alpha -1 }$. Evaluating for $\alpha=1,3,5,7$ gives the conditions for the zero mode of (\ref{ansatz}) to vanish on the twisted ground states:
$$\frac{1}{8} \lambda_1+\frac{15}{128}\lambda_2+\frac{105}{1024}\lambda_3+ \frac{3003}{32768}\lambda_4=0$$
$$\frac{9}{8}\lambda_1+\frac{159}{128}\lambda_2+\frac{1505}{1024}\lambda_3+\frac{54747}{32768}\lambda_4=0 \quad .$$
As is easily checked, the four linear equations for the $\lambda$'s are independent and hence $A(V)=\tilde A(V)$. 
Because of the independent action on the listed representations we can find a basis of $A(V)$ of the form $\lbrace F_1, F_2, F_3, E_n: n \in \Nop \rbrace$
with $F_i$ acting only in a non-trivial way on one of the ground states $M_0^{(0,-)}$, $M_0^{(T,+)}$, $M_0^{(T,-)}$ and the $E_n$ acting only on the ground states $\vert k>$ (by multiplying with $k^{2n}$).
The  algebra is commutative and the relations are 
\begin{eqnarray*}
E_n E_m &=& E_{n+m} \\
E_n F_i &=& 0 \\
F_i F_j &=& \delta_{i,j} F_i \\
1 &=& F_1 +F_2+F_3 +E_0 \qquad .
\end{eqnarray*}
The free module can (as in appendix \ref{associative appendix} proposition B3) be identified with the algebra by identifying the base vector with the unit.
The action of  $F_i$ (or $E_0$) is a projector on submodules on which only $F_i$ (or the set of $E_n$'s) is non-trivial. We see that the maximal invariant proper  submodules are spanned by {\it either} all $E_n$'s and all apart from one of the $F_i$'s, corresponding to the quotients (irreducible representations) $M_0^{(0,-)}$, $M_0^{(T,+)}$, $M_0^{(T,-)}$, {\it or} all $F_i$'s and the module generated by $k^2 E_0 -E_1$, corresponding to the quotients (irreducible representations) $\vert k>$.
In other words the algebra splits into a direct sum of three one dimensional algebras and one polynomial algebra. The irreducible representations are just the set of irreducible representations of any component.

All invariant  submodules of the free module of the polynomial algebra are given by the set of polynomials that have a given polynomial as a factor. For maximal invariant proper submodules this is an irreducible polynomial, i.e. (over $\Cop$) a linear polynomial. Polynomials with multiple zeros give rise to indecomposable representations of  dimension equal to the multiplicity. They correspond to non-unitary representations of our unitary vertex operator algebra $H_+$. All these representations are contained in the construction with ordinary vertex operators when the momentum states $\vert k >$ are replaced by such representation spaces of the polynomial algebra. In the vertex operators $(a_0)^2$ can be identified with the polynomial variable $E_1$. In later sections polynomial algebras in more variables and some of their subalgebras are appearing in $A(V)$. They are completely matched with the subalgebra of the algebra generated by $a_0^{(i)}$ that appears in the ordinary vertex operators of the given theor

y. In appendix \ref{associative appendix} (proposition B5-B9) we demonstrate how all finitely generated representation of these subalgebras of the polynomial algebra can be obtained from representations of the corresponding polynomial algebra itself. This is needed, as we indeed need to define the action of $a_0^{(i)}$ individually (and not only certain combinations).

Note that the order  presented in these notes is not the same as you would work in. Usually we do not find all relations in $O(V)$ just by staring at it, rather we get inspired by those candidates whose zero modes vanish on known representations.

\section{Including the Lattice in the $Z\!\!\!Z_2$-Projected \protect\\ Theory} \label{sec4}
Now we discuss the subtheory $H(\sqrt{2} \Zop)_+$ of the FKS lattice theory $H(\sqrt{2} \Zop)$ corresponding to the reflection twist of $\Lambda = \sqrt{2} \Zop$: 
\begin{eqnarray}
V:=H(\sqrt{2} \Zop)_+:=span(&\prod_{i:=1}^{2n} a_{-m_i} \vert 0 > &\!\!: n \in \Nop \label{vac}\\
        &\prod_{i:=1}^{2n} a_{-m_i} even(k) &\!\!: n \in \Nop , k\in \sqrt{2} \Nop \setminus \lbrace0\rbrace \label{even}\\
        &\prod_{i:=1}^{2n+1} a_{-m_i} odd(k) &\!\!: n \in \Nop , k\in \sqrt{2} \Nop \setminus \lbrace0\rbrace )  . \label{odd}
\end{eqnarray}
As before we obtain the most basic relations from (\ref{vir}). For 
$\psi \!=\!\prod_{i:=1}^{2n} a_{-\!m_i} even(\!k\!)$ we get
\begin{eqnarray} 
\prod_{i:=1}^{2n} \!a_{-\!m_i} a_{-\!1} odd(k) &\approx  -\frac{1}{k} \Bigl( &( \frac{k^2}{2}+m_1+ \!\dots \!+m_{2n}) \prod_{i:=1}^{2n} \!a_{-\!m_i} even(\!k\!) \nonumber \\
&& +\sum_{i=1}^{2n} m_i a_{-\!m_i\!-1} \!\prod_{j\neq i} a_{-\!m_j} even(\!k\!) \Bigr)  \label{get rid of}
\end{eqnarray}
and setting  
$\psi =\prod_{i:=1}^{2n+1} a_{-m_i}  odd(k)$ (with $m_{2n+1} < m_{2n} \leq \dots \leq m_1$) we get
$$\prod_{i:=1}^{2n} a_{-m_i} a_{-m_{2n+1}-1} odd(k) \approx -\frac{1}{m_{2n+1}} \Bigl( ( \frac{k^2}{2}+m_1+ \dots +m_{2n+1}) \prod_{i:=1}^{2n+1} a_{-m_i} odd(k) $$
\begin{equation}\label{get rid}
+
\sum_{i=1}^{2n} m_i a_{-m_i-1} \prod_{j\neq i} a_{-m_j} a_{-m_{2n+1}} odd(k)+ k \prod_{i:=1}^{2n+1} a_{-m_i} a_{-1} even(k)
\Bigr) \quad .
\end{equation}
Using (\ref{get rid})  repeatedly, we can write every ``odd'' vector (of the form (\ref{odd})) in terms of ``even'' vectors (of the form (\ref{even})) or odd vectors with one of the oscillators being $a_{-1}$. Such a term can then be rewritten by (\ref{get rid of}) in terms of even vectors. In the sequel this will be  frequently used. For odd terms we will take $N$ to be the corresponding $N$ after the transformation to even terms, whenever we use our induction arguments.

\subsection{Exploiting $0 \times k$} \label{without interaction}

To make a generalization to arbitrary lattices easier we first concentrate on relations of the form $0 \times k$ or $k \times 0$. We already managed to get rid of all odd vectors and hence we will translate every new relation into a relation for even vectors (using (\ref{get rid}) and (\ref{get rid of})). First 
we seek a generalization of (\ref{todiagonal}) by
\begin{eqnarray*}
0 &\approx & a_{-n} a_{-m} \vert 0> \Box even(k) \approx (n  a_{-n-1} a_{-m} + m a_{-n} a_{-m-1})even(k) \\
&&+ \Bigl( (-)^{n-1} {m+n \choose m-1} +(-)^{m-1} {m+n \choose n-1} \Bigr) k a_{-m-n-1} odd(k) \\
& \approx & (n  a_{-n-1} a_{-m} + m a_{-n} a_{-m-1})even(k) \\
&&-\frac{k^2}{n+m} a_{-m-n} a_{-1} even(k) \Bigl( (-)^{n-1} {m+n \choose m-1} +(-)^{m-1} {m+n \choose n-1} \Bigr) \quad .
\end{eqnarray*}
Hence inductively we can reduce to $a_{-\alpha} a_{-1} even(k)$ at $N=2$\footnote{This time there is no restriction for $\alpha$ to be odd.}:
\begin{equation} \label{ktodiagonal}
a_{-n} a_{-m}even(k) \approx c(m,n) a_{-n-m+1} a_{-1}even(k).
\end{equation}
Again, as in (\ref{Ntodiagonal}), this can be generalized to larger $N$. First for $N=4$:
\begin{eqnarray*}
0 &\approx& a_{-u} a_{-v} \vert 0> \star \Bigl( a_{-n} a_{-m}  - c(m,n) a_{-n-m+1} a_{-1} \Bigl) even(k) \\
&\approx& a_{-u} a_{-v}  a_{-n} a_{-m} even(k) - c(m,n) a_{-u} a_{-v} a_{-n-m+1} a_{-1}  even(k) \\
&&+ k a_{-u-v} a_{-n} a_{-m} odd(k) \Bigl( (-)^{u-1} {u+v-1 \choose v-1} +(-)^{v-1} {u+v-1 \choose u-1} \Bigr).
\end{eqnarray*}
In the last equality the (would be) second odd contribution is reduced to lower weight or lower $N$ when converted to an even term and hence neglected. The remaining odd term is reduced to terms of the form 
$$a_{-u-v-i} a_{-n-j}  a_{-m+i+j+1} a_{-1} even(k) \qquad \qquad (i,j\geq 0) .$$
We see that we can transform all terms of the form  $a_{-u} a_{-v}  a_{-n} a_{-m} even(k)$ into similar terms with $m=1$, but even more can be read from this: reordering to $v=1$ in a second step, we still increase $u$ (or alternatively produce two $a_{-1}$'s). Eventually only terms of the form $ a_{-n} a_{-m}  a_{-1} a_{-1} even(k)$ will be left. To reduce these further we use
\begin{eqnarray*}
0 &\approx& a_{-1} a_{-1} \vert 0> \star \Bigl( a_{-n} a_{-m}  - c(m,n) a_{-n-m+1} a_{-1} \Bigl) even(k) \\
&\approx& a_{-n} a_{-m}  a_{-1} a_{-1} even(k) - c(m,n) a_{-m-n+1} a_{-1}^3  even(k) \\
&&+2 k a_{-2} a_{-n} a_{-m} odd(k) \\
& \approx& (1-2 k^2) a_{-n} a_{-m}  a_{-1} a_{-1} even(k) - c(m,n) a_{-m-n+1} a_{-1}^3  even(k) \quad .
\end{eqnarray*}
For an integral lattice the coefficient $(1-2 k^2)$ is non-zero and we can reduce all $N=4$ objects to $a_{-\alpha} a_{-1}^3  even(k)$.\\
For $N>4$ we use induction:
\begin{eqnarray*}
0 &\approx& a_{-n} a_{-m} \vert 0> \star \Bigl( a_{-n_1} \dots a_{-n_{N-2}}  - \beta a_{-\alpha} a_{-1}^{N-3} \Bigr) even(k) \\
&\approx & a_{-n} a_{-m}  a_{-n_1} \dots a_{-n_{N-2}} even(k) - \beta a_{-n} a_{-m} a_{-\alpha} a_{-1}^{N-3}   even(k) \\
&& + \mbox{terms with $n$ increased} \quad .
\end{eqnarray*}
We can eventually express all in terms of the form $a_{-\!n} a_{-\!m} a_{-\!\alpha} a_{-\!1}^{N\!-\!3} even(\!k\!).$  These can be further treated by 
\begin{eqnarray*}
0 & \approx & a_{-1}^{N-4} \vert 0> \star \Bigl( a_{-n} a_{-m} a_{-\alpha} a_{-1} - \beta a_{-\gamma} a_{-1}^3 \Bigr) even(k) \\
& \approx & \Bigl( a_{-1}^{N-3}  a_{-n} a_{-m} a_{-\alpha} - \beta a_{-\gamma} a_{-1}^{N-1} \Bigr) even(k) 
\end{eqnarray*}
and hence everything is reduced to terms of the form $a_{-\alpha} a_{-1}^{N-1} even(k) $.\\
Now we want to further reduce these for $N\geq 4$ and $\alpha \geq 2$ (as in section \ref{non-tensor}). First this is achieved for $N\geq 6$  by inserting
\begin{eqnarray}
 0 &\approx& a_{-n} a_{-1} \vert 0> \Box a_{-1}^{N-2}  even(k) \nonumber \\
&\approx & \Bigl( n a_{-n-1} a_{-1} + a_{-n} a_{-2} \Bigr) a_{-1}^{N-2} even(k)
\label{insertion}
\end{eqnarray}
into
\begin{eqnarray}
0 &\!\!\!\!\approx & \!\!\!\!a_{-1}^{N-2} \vert 0> \star  \Bigl(a_{-n} a_{-1} \vert 0> \Box  even(k) \Bigr) \nonumber \\
&\!\!\!\!\approx& \!\!\!\!a_{-1}^{N\!-\!2} \vert 0\!\!> \star \Biggl( \Biggl[ n \!- \!\frac{k^2}{n\!\!+\!\!1} \Bigl\lbrace \frac{(n\!\!+\!\!1) n}{2} \!+\!(-)^{n\!-\!1} \Bigr\rbrace \Biggr] a_{-\!n\!-1} a_{-\!1} even(k) \!+\!a_{-\!n} a_{-\!2} even(k) \Biggr) \nonumber \\
&\!\!\!\!\approx & \!\!\!\!a_{-1}^{N\!-\!2} \Biggl( \Biggl[ n \!- \!\frac{k^2}{n\!\!+\!\!1}\Bigl\lbrace \frac{(n\!\!+\!\!1) n}{2} +(-)^{n\!-\!1} \Bigr\rbrace \Biggr] a_{-\!n\!-\!1} a_{-\!1}  +a_{-\!n} a_{-\!2}  \Biggr) even(k)
\label{N=6}
\end{eqnarray}
to obtain
$$0 \approx a_{-n-1}a_{-1}^{N-1} even(k) \qquad (N\geq6) \quad ,$$
as the coefficient $ - \frac{k^2}{n+1}\Bigl\lbrace \frac{(n+1) n}{2} +(-)^{n-1} \Bigr\rbrace$ is always non-zero for $k\neq 0$. \\ 
Then for $N=4$ the last line of (\ref{N=6}) has a further contribution from an odd term, which was  reduced to lower $N$ for $N\geq 6$. This contribution reads ($N=4$):
$$(N-2) (N-3) a_{-2} a_{-1}^{N-4} k a_{-n} a_{-2} odd(k) \approx -2 k^2 a_{-n} a_{-2} a_{-1}^2 even(k). $$
Hence inserting (\ref{insertion}) into this  modified version of (\ref{N=6}) we get
$$0 \approx \frac{1}{n+1} \Biggl[ 2 k^2 n (n+1)  - k^2 \Bigl\lbrace \frac{(n+1) n}{2} +(-)^{n-1} \Bigr\rbrace \Biggr] a_{-n-1} a_{-1}^3 even(k).$$
The coefficient is non-zero for $k \neq 0$ and we have achieved our reduction for $N=4$. \\
We can summarize our results as
\begin{eqnarray*}
\tilde A(V) =& \!\!\!span ( & \!\!\!a_{-1}^{2n} \vert 0 > : n \in \Nop, \qquad a_{-\alpha} a_{-1} \vert 0 >: \alpha \in \lbrace 3,5,7 \rbrace,\\
&& \!\!\!a_{-1}^{2n} even(k) : n \in \Nop, \quad a_{-\alpha} a_{-1} even(k): \alpha \in \lbrace 2,3,4,\dots \rbrace \quad ) .
\end{eqnarray*}
We can also pull the zero-momentum  relation (cf. (\ref{N=2 trick}), (\ref{todiagonal}))
$$ a_{-\alpha} a_{-1} \vert 0> \approx_{wt} 0 \qquad \mbox{for} \ \alpha \not\in \lbrace 1,3,5,7 \rbrace$$
to any momentum (unless $k^2=\alpha=2$) ($\alpha \not\in \lbrace 1,3,5,7 \rbrace$):
\begin{eqnarray}
0 &\approx_{wt}& a_{-\alpha} a_{-1} \vert 0> \star \quad even(k) \nonumber \\
&\approx& \Biggl(1-\frac{k^2}{\alpha} \Bigl(\alpha+(-)^{\alpha-1} \Bigr) \Biggr) a_{-\alpha} a_{-1} even(k) \quad . \label{pull towards momentum}
\end{eqnarray}
As the lattice is integral, the coefficient is non-zero unless $\alpha=k^2=2$. \\
So far we only found relations that had an analogue at zero-momentum. The following powerful relation breaks this rule (without yet using lattice interaction like in (\ref{a1})). For $n\geq2$ and even we obtain by expanding the exponentials $e^{ikX_<}$ as in (\ref{combinatorics}), while $e^{ikX_>}$ is just expanded to second order:
\begin{eqnarray*}
0 &\approx& even(k) \Box a_{-n} a_{-1} \vert 0> \\
&\approx& \frac{k^{n+2}}{(n+1)!} \Bigl( -1 + \frac{k^2}{n+2} \Bigr) a_{-1}^{n+2} even(k) \quad .
\end{eqnarray*}
So for $k\neq 0$ and $N\neq k^2$ we can reduce $a_{-1}^N even(k)$ ($N\geq 4$) and conclude without use of any lattice interaction:
\begin{eqnarray*}
\tilde A(V) = span  (&& a_{-1}^{2n} \vert 0 > : n \in \Nop, \qquad a_{-\alpha} a_{-1} \vert 0 >: \alpha \in \lbrace 3,5,7 \rbrace, \\
&&even(k), a_{-1}^{k^2} even(k) : k \in \sqrt{2} \Nop \setminus \lbrace 0 \rbrace, \\
&& a_{-\alpha} a_{-1} even(k): \alpha \in \lbrace 1,3,5,7 \rbrace \quad k \in \sqrt{2} \Nop \setminus \lbrace 0 \rbrace, \\
&& a_{-2} a_{-1} even(\sqrt{2}) \qquad \qquad ) .
\end{eqnarray*}

\subsection{Exploiting $k \times k'$}
Looking at the powers of $z$ in the corresponding generalization of (\ref{interact}), we see that, in $even(k) \Box_n even(k')$, the components with momenta added vanish as long as $(0 \leq ) n \leq k k' -2$. This leaves a possibly non-trivial relation for the subtracted momenta. In particular for $k=k'=\sqrt{2} m$ we obtain for $n$ odd ($0 \leq n \leq 2 m^2 -2$)
$$ 0 \approx even(m \sqrt{2}) \Box_n even(m \sqrt{2}) \approx 2 \frac{(m \sqrt{2})^{2m^2+n+1} a_{-1}^{2m^2+n+1}}{(2m^2+n+1)!} \vert 0> (-)^m \quad . $$
So we reduce  
$a_{-1}^{2l} \vert 0>$  for $l=5 \quad (n=1, m=2)$, $l=6 \quad (n=3, m=2)$, $l=7
\quad (n=5,m=2)$, $l=10 \quad (n=1, m=3)$, etc. For every value of $m$ we obtain the  whole range of $l$'s for $m^2+1 \leq l \leq 2 m^2 -1$. This range overlaps with the one for $m+1$ if $m\geq3$. Hence we only fail to find suitable $n,m$ for $l=0,1,2,3,4,8,9$. 

If we take $n=k k' -1$, we keep just one single term with momenta being added and obtain (neglecting lower momenta!)
$$0 \approx  even(k) \Box_{k k'-1} even(k') \approx_{wt} even(k+k') \epsilon(k,k') \qquad (k,k' >0) \quad . $$
So for $k\neq 0, \sqrt{2}$
\begin{equation} \label{momentum reduction}
even(k)\approx_{wt} 0
\end{equation}
and hence also
$$0 \approx_{wt} a_{-n} a_{-1} \vert 0> \star \ even(k) \approx \Bigl( 1-k^2\frac{n+(-)^{n-1}}{n} \Bigr) a_{-n} a_{-1}  even(k),$$
the coefficient being always non-zero ($k^2 \neq 2$), and similarly
$$0 \approx_{wt}  a_{-1}^{k^2} \vert 0> \star \ even(k) \approx  a_{-1}^{k^2}  even(k)$$
as $k^2 \geq 4$.
By now we managed to reduce $\tilde A(V)$ to a 16 dimensional object:
\begin{eqnarray*}
\tilde A(V) = span  (&& a_{-1}^{2n} \vert 0 > : n =0,1,2,3,4,8,9 \ ,  \qquad a_{-\alpha} a_{-1} \vert 0 >: \alpha \in \lbrace 3,5,7 \rbrace, \\
&& a_{-\alpha} a_{-1} even(\sqrt{2}): \alpha = 1,2,3,5,7 \ ,\quad even(\sqrt{2})  \qquad ) .
\end{eqnarray*}
All further arguments will rely on a somewhat more detailed comparison of terms at different momenta. So we only want to neglect terms of lower (weight,$N$) than a term of the same momentum (modulo $O(V)$). We denote this by ``$\equiv$''.

 There is one more pretty general tool that will help us to get rid of $a_{-1}^{2l} \vert 0>$ for most $l$: 
\begin{eqnarray}
0 &\equiv& a_{-1}^{2l-4} \vert 0 > \star \ \Bigl( even(\sqrt{2}) \Box_1 even(\sqrt{2}) \Bigr)  \nonumber \\
&\equiv& a_{-1}^{2l-4} \vert 0 > \star \ \Bigl( \epsilon(\sqrt{2}, \sqrt{2}) even(2 \sqrt{2}) -\Bigl( \frac{1}{3} a_{-1}^4 +\frac{1}{2} a_{-2}^2 +\frac{4}{3} a_{-3} a_{-1} \Bigr) \vert 0> \Bigr) \nonumber \\
& \equiv &  \epsilon(\sqrt{2}, \sqrt{2})  \Bigl( a_{-1}^{2l-4} \vert 0 > \star \  even(2 \sqrt{2}) \Bigr) - \frac{1}{3} a_{-1}^{2l}  \vert 0>  \quad . \label{penultimate trick}
\end{eqnarray}
Now, as we have seen in section \ref{without interaction}, we can reduce the first term (without any lattice interaction) to terms of the form  $a_{-\alpha} a_{-1} even(2 \sqrt{2})$ ($\alpha = 1,3,5,7$), 
$even(2 \sqrt{2})$, $a_{-1}^8 even(2 \sqrt{2})$. The weight of these terms is at most 12, and when we reduce them further (by lattice interaction (\ref{momentum reduction})) to zero momentum, the resulting terms can be neglected if $l \geq 7$. So
$$a_{-1}^{2l} \vert 0> \equiv 0 \qquad (l \geq 7).$$
For $l=4,5$ a slightly modified version of this argument will work. We are only 
interested in $l=4$ and present this: Here the reduction of the first term of 
(\ref{penultimate trick}) involves  only $ \beta a_{-1} a_{-3} even(2 \sqrt{2})$ at top weight. However the reduction of this to momentum zero
\begin{eqnarray} \label{reduction to momentum zero in special case}
0 &\equiv& a_{-1} a_{-3} \vert 0> \star \Bigl( even(\sqrt{2}) \Box_1 even(\sqrt{2}) \Bigr)  \nonumber \\
&\equiv& - \epsilon(\sqrt{2},\sqrt{2}) \frac{29}{3} a_{-1} a_{-3} even(2 \sqrt{2} ) - \frac{1}{3} a_{-3} a_{-1}^5 \vert 0 >
\end{eqnarray}
involves only terms with $N \leq 6$ at momentum zero, so indeed 
\begin{equation} \label{N=8}
a_{-1}^8 \vert 0> \equiv 0
\end{equation}
and we arrive at the 13 dimensional 
\begin{eqnarray}
\tilde A(V) = span  (&\!\!\!\!& a_{-1}^{2n} \vert 0 > : n =0,1,2,3 \ ,  \qquad a_{-\alpha} a_{-1} \vert 0 >: \alpha \in \lbrace 3,5,7 \rbrace, \nonumber \\
&\!\!\!\!& a_{-\alpha} a_{-1} even(\sqrt{2}): \alpha = 1,2,3,5,7 \ ,\quad even(\sqrt{2})  \quad ) . \label{13 dim}
\end{eqnarray}
As the last two steps (cf. section \ref{ultimate trick}) are a bit tricky, we prefer to present first a discussion of the representations, partly to motivate 
these steps.

\subsection{Cocycles}
Again we know the twisted and untwisted representations. However now we first want to check some details on cocycles and $\gamma$-matrix representations, as it will turn out, that there are two inequivalent choices of twisted representations corresponding to different representations of the $\gamma$-matrix algebra. We use notation and results from \cite{dgm} (in particular appendices B and C). Writing $V$ for vertex operators, $U$ for representation operators and $W$ for intertwiners, we assume to have specified the cocycles for $V$ as
\begin{equation} \label{V cocycles}
\sigma_\lambda \vert \lambda' > = \epsilon (\lambda, \lambda' )\vert \lambda' > 
\end{equation}
with $ \epsilon $ as in (\ref{epsilons}), as we obviously want to discuss $A(V)$ for a given fixed $V$. So its cocycles also have to be fixed at the start.

\subsubsection{Untwisted Representations}
We first consider untwisted representations. To specify the cocycle for $U$ we fix a base vector in the lattice of the representation $k \in \Lambda^{\star}$ and put our freedom in the definition into a function $S$:
\begin{equation} \label{U cocycles}
 \sigma_\lambda \vert \lambda' +k > = \epsilon (\lambda, \lambda' )\vert \lambda' +k> S(\lambda, \lambda') \quad .
\end{equation}
For the auxiliary intertwiners $W$ we define cocycles as
$$ \sigma_{\lambda +k} \vert \lambda'  > = \epsilon (\lambda, \lambda' ) (-)^{k \lambda'} \vert \lambda' > \sigma(\lambda, \lambda') \quad . $$
The intertwining relation $ W V =U W $ and the creation property for $W$ constrain the functions $S$ and $\sigma$ by the following two equations:
\begin{eqnarray}
\sigma(\lambda',\lambda+\lambda'') &=& S(\lambda, \lambda' +\lambda'') \sigma(\lambda', \lambda'')  \label{locality} \\
\sigma(\lambda,0) &=&1   \quad . \label{creation}
\end{eqnarray}
To derive (\ref{locality}) we let $W$ carry momentum $\lambda' +k$ and $U$ (and $V$) carry momentum $\lambda$, with the intertwining relation acting on momentum $\lambda''$. The role of the cocycles is  to provide a symmetry factor $(-)^{(k+\lambda') \lambda}$. The resulting equation contains $\sigma$, $S$ and $\epsilon$. All $\epsilon$'s can be eliminated using (\ref{epsilons}) and
we arrive at (\ref{locality}).\\
Taking $\lambda''=0$ in (\ref{locality}) and using (\ref{creation}) we get
$$\sigma(\lambda',\lambda) = S(\lambda,\lambda')  \quad . $$
Inserting this in (\ref{locality}) for $\lambda':=0$ and setting $v(\lambda):=\sigma(0,\lambda)$ we get
\begin{eqnarray}
S(\lambda, \lambda'') \quad = \quad  \sigma(\lambda'',\lambda)&=&\frac{v(\lambda+\lambda'')}{v(\lambda'')}\\
v(0)&=&1 \quad .
\end{eqnarray}
This choice always fulfils (\ref{locality}) and (\ref{creation}).
The freedom in the choice of $v$ just reflects the freedom in the choice of the relative phases for the vectors $\vert \lambda +k>$ of the representation. So all such representations are equivalent.

For $k=0$ the choice $v(\lambda) =1$, i.e. $S=1$ in (\ref{U cocycles}), keeps the reflection symmetry. For $k= \frac{1}{\sqrt{2}}$ we can restore the reflection symmetry (which seems to be broken) by demanding the action of $\sigma_\lambda $ on $\vert \lambda' + \frac{1}{\sqrt{2}} >$ to be the same as the action of $\sigma_{-\lambda}$ on $\vert -\lambda' -\frac{1}{\sqrt{2}} >$. 
Noting that (\ref{epsilons}) implies 
$$ \epsilon(\lambda, -\sqrt{2}) \epsilon( \sqrt{2} - \lambda , - \sqrt{2}) \epsilon( \sqrt{2}, -\sqrt{2}) = 1$$
and 
$$\epsilon(\lambda,\lambda') \epsilon(\lambda'+\sqrt{2},-\sqrt{2}) = 
\epsilon(-\lambda,-\lambda'-\sqrt{2}) \epsilon(\lambda+\lambda'+\sqrt{2},-\sqrt{2}) \ ,
$$ 
it is easily checked, that for this demand to be met the
 necessary and sufficient condition on $v(\lambda)$ reads (in our case)
\begin{eqnarray*}
\Bigl( v(-\sqrt{2}) \Bigr)^2 &=& -1 \\
v(-\lambda -\sqrt{2})& =& \frac{v( \lambda)}{v(-\sqrt{2})} \epsilon(\lambda+\sqrt{2}, -\sqrt{2}) \qquad (\lambda \in \sqrt{2} \Nop) \quad .
\end{eqnarray*}
We may choose $v(\lambda)=1$ ($\lambda \in \sqrt{2} \Nop$), $v(-\sqrt{2}) =i$ and $v(\lambda) = -i \epsilon(-\lambda, -\sqrt{2}) $  $(\lambda \leq - 2 \sqrt{2})$. So with this choice $S(\sqrt{2}, -\sqrt{2})=-i$ , $S(-\sqrt{2},0)=i$. This enables us to calculate all cocycles that we will need later: 
 We obtain on untwisted representations
\begin{eqnarray}
\sigma_0 &=& 1\nonumber \\
\sigma_{\pm \sqrt{2}} &=& 1 \qquad \mbox{on $\vert0>$} \nonumber \\
\sigma_{\pm \sqrt{2}} &=& -1 \qquad \mbox{on $\vert \mp \sqrt{2}>$} \nonumber \\
\sigma_{\pm \sqrt{2}} &=& i \qquad \mbox{on $\vert \mp \frac{1}{\sqrt{2}} >$} \quad .
\end{eqnarray}

\subsubsection{Twisted Representations}
In the case of the twisted representations we are meant to give $\gamma$-matrix representations that correspond to the  cocycle relations (\ref{V cocycles}). First we note, that a change in the phase for the basic $\gamma$-matrices corresponding to a basis of the lattice does not influence the algebra (i.e. the symmetry factors) (cf. appendix B,C in \cite{dgm}). However we have to keep the gauge $\gamma_\lambda =\gamma_{-\lambda} $, as otherwise even sectors (weight $1/16 + \Zop$) and odd sectors (weight $9/16 + \Zop$) are linked, corresponding to non-meromorphic representations. This forces us to a fixed sign of $\gamma_{e_i}^2$ for the basis vectors $e_i$. In fact our only freedom is to choose different signs for the centre of the algebra of basic $\gamma$-matrices, corresponding to different irreducible representations. Here the freedom amounts to the choice of sign
\begin{equation} \label{gamma sign}
\gamma_{\sqrt{2}} = \pm i \quad .
\end{equation} 
Hence we consider the following representations and their ground states, with $T_+$ referring to the choice $+$ in (\ref{gamma sign}) for the $\gamma$-matrix and similarly for $T_-$:
\begin{eqnarray*}
M_0^{(T_+,+)}=span( \chi) & \qquad&
M_0^{(T_-,+)}=span( \chi) \nonumber \\
M_0^{(T_+,-)}=span( c_{-1/2} \chi) &\qquad&
M_0^{(T_-,-)}=span( c_{-1/2} \chi)\nonumber \\
M_0^{(0,+)}=span( \vert 0 >) &&\nonumber \\
M_0^{(1/\sqrt{2},+)}=span(even(1/\sqrt{2})) &\qquad& M_0^{(1/\sqrt{2},-)}=span(odd(1/\sqrt{2})) 
\end{eqnarray*}
\begin{equation}\label{basis of repns}
M_0^{(0,-)}=span( a_{-1} \vert 0 >, odd(\sqrt{2})) \quad .
\end{equation}

\subsection{Action on Representations}
 We calculate the action of the zero modes of the 13 vectors in (\ref{13 dim})
on the space of ground states $\bigoplus_i M_0^{(i)}$ with respect to our basis (in the order of (\ref{basis of repns})), noting that only those components of the zero mode contribute with at most one lowering oscillator $c_{\frac{1}{2}}$ or $a_{1}$.\\
On untwisted ground states
\begin{eqnarray*}
(a_{-1}a_{-\alpha}\vert 0>)_0 &\!\!=& \!\!a_0^2 \!+\!a_{-\!1} a_{1} (\delta_{\alpha,1} \!\!+\!\!\alpha) = (0,\frac{1}{2},\frac{1}{2}, 
\left(\begin{array}{cc}
 \delta_{\alpha,1} \!\!+\!\!\alpha  &   0        \\
     0        &    2         
      \end{array} \right))
\quad (\alpha odd) \\
(a_{-1}^{2n} \vert0> )_0 &\!\!=& \!\!a_0^{2n}= (0,\frac{1}{2^n},\frac{1}{2^n}, \left(\begin{array}{cc}
    0         &     0      \\
    0         &     2^n        
      \end{array} \right)
)  \qquad (n\geq2)\\
(a_{-\!1} a_{-\!\alpha} even(\sqrt{2}))_0 &\!\!=& \!\!(0,\frac{i}{2} (-)^{\alpha-1}, \frac{i}{2} (-)^{\alpha},  (1-\delta_{\alpha,1}) (-)^{\alpha-1} 2 \sqrt{2} \left(\begin{array}{cc}
    0         &     1      \\
    0         &     0        
      \end{array} \right)) \\
(even(\sqrt{2}))_0 &\!\!=& \!\!(0,i,-i, 
\left(\begin{array}{cc}
    0         & 2 \sqrt{2}  \\
 -\sqrt{2}    &     0        
      \end{array} \right)) \\
(\vert 0>)_0 &\!\!=&\!\!(1,1,1,
\left(\begin{array}{cc}
    1         &     0      \\
    0         &     1        
      \end{array} \right)) \qquad .
\end{eqnarray*}
On  twisted ground states 
\begin{eqnarray*}
(a_{-1} a_{-\alpha} \vert 0> )_0^T&=& c_{-\frac{1}{2}}c_{\frac{1}{2}} \lbrace {-\frac{3}{2} \choose \alpha-1 } +{-\frac{1}{2} \choose \alpha-1 } \rbrace -\frac{\alpha}{2 (\alpha +1)} {-\frac{1}{2} \choose \alpha }  \\
(a_{-1}^{2n} \vert 0>)_0^T &=&(e^{\Delta(z)} a_{-1}^{2n} \vert 0>)_0^0=
( \sum_{k=0}^{n} a_{-1}^{2n-2k} \vert 0> \frac{(2 k)!}{k! 2^{4k}} 
{2n \choose 2k})_0^0 \\
&=&c_{-\frac{1}{2}}c_{\frac{1}{2}} \frac{(2 n)!}{(n-1)!} 2^{-4 n +4} +   \frac{(2 n)!}{(n)!} 2^{-4 n } \\
(even(\sqrt{2}))_0^T&=&\frac{1}{2} (1-8 c_{-\frac{1}{2}}c_{\frac{1}{2}}) \gamma_{\sqrt{2}} \\
(a_{-1}a_{-\alpha} even(\sqrt{2}))_0^T &=& (e^{\Delta (z)} a_{-1}a_{-\alpha} even(\sqrt{2}))_0^0 
\end{eqnarray*}
\begin{eqnarray*}
&=& \frac{1}{4} \Biggl( -\frac{3 \alpha +2}{2 (\alpha +1)} {-\frac{1}{2} \choose \alpha} even(\sqrt{2}) -\frac{1}{2} \sqrt{2} a_{-\alpha} odd(\sqrt{2}) \\
&& +{-\frac{1}{2} \choose \alpha} \sqrt{2} a_{-1} odd(\sqrt{2}) +a_{-1} a_{-\alpha} even(\sqrt{2})   \Biggr)_0^0 \\
&=& \frac{1}{4} \Biggl( -\frac{3 \alpha \!\!+\!\!2}{2 (\alpha \!\!+\!\!1)} {-\frac{1}{2} \choose \alpha} 2 (1\!\!-\!\!8 c_{-\frac{1}{2}}c_{\frac{1}{2}})
 \!-\!\frac{1}{2} \sqrt{2} (1\!\!-\!\!\delta_{\alpha,1}) (-4 \sqrt{2}) {-\frac{3}{2} \choose \alpha \!\!-\!\!2} c_{-\!\frac{1}{2}}c_{\frac{1}{2}} \\
&& + 2 \lbrace {-\frac{3}{2} \choose \alpha-1 } +{-\frac{1}{2} \choose \alpha-1 }  \rbrace c_{-\frac{1}{2}}c_{\frac{1}{2}} \Biggr) \gamma_{\sqrt{2}} \qquad .
\end{eqnarray*}
The aim of all this is to see whether these actions are  indeed spanning the 11 dimensional space $\bigoplus_i End M_0^{(i)}$ and (if so) to find the 2 dimensional subspace of $\tilde A (V)$ with trivial action on the representations. We observe that 
$(\vert 0> )_0$, $(even(\sqrt{2}))_0$ provide unique non-trivial components for 
$End M_0^{(0,+)}$ and (in matrix notation) $(End M_0^{(0,-)})_{2,1}$.
Among the remaining 11 vectors only the five $ a_{-1}a_{-\alpha} even(\sqrt{2})$
have non-vanishing components of type\footnote{In this notation algebraic operations are performed on the components with respect to our basis.} 
$$\lbrace  End M_0^{(1/\sqrt{2},+)} - End M_0^{(1/\sqrt{2},-)}, (End M_0^{(0,-)})_{1,2} $$
$$End M_0^{(T_+,-)} - End M_0^{(T_-,-)}, End M_0^{(T_+,+)} - End M_0^{(T_-,+)} \rbrace .$$ 
A certain combination of the first two components $End M_0^{(1\!/\!\sqrt{2},+)} \!\!- \!End M_0^{(1\!/\!\sqrt{2},-)} \!\!- \frac{i}{2 \sqrt{2}}  (End M_0^{(0,-)})_{1,2}$ is only non-zero on $ a_{-1}^2 even(\sqrt{2})$ leaving us with four vectors on a 3 dimensional space.
 
Similarly only the six (out of 11) vectors of type $a_{-1} a_{-\alpha} \vert 0>$ and $ a_{-1}^{2n} \vert 0>$ are unique in having components of type 
$$\lbrace End M_0^{(1/\sqrt{2},+)} + End M_0^{(1/\sqrt{2},-)}, (End M_0^{(0,-)})_{1,1}, (End M_0^{(0,-)})_{2,2} $$
$$End M_0^{(T_+,-)} + End M_0^{(T_-,-)},End M_0^{(T_+,+)} + End M_0^{(T_-,+)} \rbrace .$$

So we split our problem of 13 vectors in an 11 dimensional space into 3 independent vectors in  a 3 dimensional space, 4 vectors in a 3 dimensional space, 6 vectors in a 5 dimensional space. Plugging in the different values for $n$ and $\alpha$, we find that indeed the maximal dimension is attained and the two vectors vanishing on all representations are in general position\footnote{We call a vector in general position, if none of its components is vanishing in the given basis.} in the 6 (respectively 4) dimensional subspaces of $\tilde A(V)$ corresponding of the splitting of the problem. For our inductive reasoning it is not important to know the precise coefficients of these two vectors. Only the fact that  they lie in general position is important to notice.

\subsection{The Final Two Relations} \label{ultimate trick}
Now we can go ahead to prove that these two vectors (rather than hinting on further representations) are indeed in $O(V)$ .  If we can prove that 
\begin{eqnarray} 
a_{-1}a_{-7} \vert 0> + \beta_1 a_{-1}a_{-5} \vert 0>
 + \beta_2 a_{-1}a_{-3} \vert 0>
 + \beta_3 a_{-1}^2 \vert 0> &&\nonumber \\
 + \beta_4 a_{-1}^4 \vert 0>
 + \beta_5 a_{-1}^6 \vert 0>
&\in& O(V) \label{final kick}
\end{eqnarray}
for some $\beta_i$'s, we can write 
\begin{equation}
a_{-1}a_{-7} \vert 0> \approx_{wt} 0 \quad .
\end{equation}
This in turn would imply (cf. (\ref{pull towards momentum}))
$$a_{-1}a_{-7} even( \sqrt{2})  \approx_{wt} 0 $$
and we could conclude
\begin{eqnarray}
 A(H(\sqrt{2})) = span  (&& a_{-1}^{2n} \vert 0 > : n =0,1,2,3 \ ,  \qquad a_{-\alpha} a_{-1} \vert 0 >: \alpha \in \lbrace 3,5 \rbrace, \nonumber \\
&& a_{-\alpha} a_{-1} even(\sqrt{2}): \alpha = 1,2,3,5 \ ,\quad even(\sqrt{2})  \quad ) . 
\end{eqnarray}
As our proof of (\ref{final kick}) seems rather lengthy, we first discuss, why a few simpler attempts would fail. The first idea would be to eliminate $a_{-1}^6 \vert0>$ along the lines of (\ref{penultimate trick}) for $l=3$. This can only fail as $a_{-1}a_{-7} \vert 0>$ is not included (induction in weight), but (\ref{final kick}) needs to be in general position. We also know from section \ref{pure oscillators}, that $0 \times 0$ cannot produce (\ref{final kick}). So we somehow need lattice interaction at weight 8. This was however already used for (\ref{N=8}). Yet we might try to introduce two inequivalent lattice interactions  at weight 8 and cancel the contributions of $a_{-1}^8\vert0>$, so that the only remaining contribution at weight 8 is $a_{-1} a_{-7} \vert 0>$ with hopefully a non-vanishing coefficient.
This works as follows\footnote{The equivalence notion used here only discards terms of lower weight or in $O(V)$. Terms of same weight and lower $N$ are kept.}:
\begin{equation} \label{details of final kick}
0 \equiv_{wt} a_{-1}^{4} \vert 0 > \star \ \Bigl( even(\sqrt{2}) \Box_1 even(\sqrt{2}) \Bigr) +\frac{7 \ 5! }{2 \sqrt{2}} (a_{-5} odd(\sqrt{2})) \Box (even(\sqrt{2})) \, .
\end{equation}
The (weight 8) parts of (\ref{details of final kick}) carrying momentum  can be reduced  by 
\begin{eqnarray*}
0 &\equiv_{wt}& even(2 \sqrt{2}) \Box a_{-2} a_{-1} \vert0> \\
&\equiv_{wt}& \frac{32}{3} a_{-1}^4 even(2 \sqrt{2}) - 5 a_{-2}^2 even(2 \sqrt{2}) -\frac{14}{3} a_{-3}a_{-1} even(2 \sqrt{2})
\end{eqnarray*}
and
\begin{eqnarray*}
0 &\equiv_{wt}& a_{-2} a_{-1} \vert0> \Box even(2 \sqrt{2})  \\
&\equiv_{wt}& a_{-2}^2 even(2 \sqrt{2}) -\frac{10}{3} a_{-3}a_{-1} even(2 \sqrt{2})
\end{eqnarray*}
to  $\epsilon(\sqrt{2},\sqrt{2}) \frac{536}{3} a_{-3}a_{-1} even(2 \sqrt{2})$. This is converted to momentum zero by
a more precise version of  
(\ref{reduction to momentum zero in special case})
\begin{eqnarray*} 
0 &\equiv_{wt}& a_{-1} a_{-3} \vert 0> \star \Bigl( even(\sqrt{2}) \Box_1 even(\sqrt{2}) \Bigr)  \nonumber \\
&\equiv_{wt}& - \epsilon(\sqrt{2},\sqrt{2}) \frac{29}{3} a_{-1} a_{-3} even(2 \sqrt{2} ) - \Bigl( \frac{1}{3} a_{-3} a_{-1}^5 +\frac{1}{2} a_{-3}a_{-2}^2a_{-1} \\
&&+\frac{4}{3} a_{-1}^2a_{-3}^2 +12 a_{-5} a_{-1}^3 +32 a_{-6}a_{-2} +100a_{-1}a_{-7} +12 a_{-5}a_{3}
\Bigr) \vert 0 > \ .
\end{eqnarray*}
Having converted all of (\ref{details of final kick}) to zero momentum we write, using only $0 \times 0$, all weight 8 parts in terms of $a_{-1}^8 \vert0>$ and $a_{-1}a_{-7}\vert0>$ and find indeed (after some trivial but tedious steps) that the coefficient of the former vanishes while the one of the later is non-zero ($-\frac{64}{3}$).

\section{ $Z\!\!\!Z_2$-Projected Heisenberg Algebra \protect\\ in d Dimensions} \label{sec6}
Denoting with $H$ the Heisenberg algebra discussed in section \ref{full hb alg} and $H_+$ its subtheory discussed in section \ref{pure oscillators}, we want to study the $d$ dimensional versions $H^d$, $(H_+)^d$ and $(H^d)_+$.

In general we know Zhu's algebra for tensor product theories as $A(V_1 \bigotimes V_2) = A(V_1) \bigotimes A(V_2)$, and the irreducible representations are of course the tensor products of the irreducible representations of the components. A proof of this can be found in  appendix \ref{tensor appendix}. So the cases $H^d$ and $(H_+)^d$ are trivial.

To obtain $A((H^d)_+)$ we mainly use results from the one dimensional case together with versions of (\ref{todiagonal}) and (\ref{heisenberg without computer}) that mix different types of modes corresponding to the different components of $H^d$. To avoid messy notation we first discuss $d=2$ before looking at general $d$.

\subsection{$(H^2)_+$} \label{2dim}
For the commuting sets of oscillators $a_m$, $b_n$ we write
$$ V:=(H^2)_+ := span(\prod_{i:=1}^{M_a} \prod_{j:=1}^{M_b} a_{-m_i}b_{-n_j} \vert 0 > : M_a + M_b \ \mbox{even}) .$$
Rewriting 
\begin{eqnarray*}
\phi \in H^{(a)}_+ , \quad \psi \in O(H^{(b)}_+) &\Rightarrow& \phi \otimes \psi \in O((H^2)_+) \\
\phi_1 \!\otimes \!\psi_1 \in (H^{2})_+ , \ \phi_2 \!\otimes \!\psi_2 \in O((H^{2})_+) &\Rightarrow& (\phi_1 \!\otimes \!\psi_1)  \star  (\phi_2 \!\otimes \!\psi_2) \in O((H^2)_+) 
\end{eqnarray*}
into our ``inductive notation''
\begin{eqnarray*}
 \psi \approx_{(wt)} 0 &\Rightarrow& \phi \otimes \psi \approx_{(wt)} 0\\
 \phi_2 \otimes \psi_2 \approx_{(wt)} 0 &\Rightarrow& (\phi_1 \otimes \psi_1) \ \star \ (\phi_2 \otimes \psi_2) \approx_{(wt)} 0 ,
\end{eqnarray*}
we immediately realise that the results of section \ref{pure oscillators} imply
\begin{eqnarray}
\tilde A(V) &=& span  \Bigl( a_{-1}^{n} b_{-1}^{m}\vert 0 > : n+m \ \mbox{even},\\ &&a_{-\alpha} a_{-1} \vert 0 >, \ b_{-\beta} b_{-1} \vert 0 >, \\
&& a_{-\alpha} a_{-1} b_{-\beta} b_{-1} \vert 0 >, \ a_{-\alpha} a_{-1}^2 b_{-\beta} b_{-1}^2 \vert 0 >, a_{-\alpha} a_{-1}^2  b_{-\delta} \vert 0 >,\nonumber \\
&& a_{-\gamma} \ b_{-\beta} b_{-1}^2 \vert 0 >, \label{multiple 2dim} \\
&& a_{-\gamma} b_{-\delta} \vert 0> \label{single 2dim} \\
&&\mbox{for} \ \alpha, \beta \in \lbrace 1,3,5,7 \rbrace \quad \gamma, \delta \in \lbrace 1,2,3,4,\dots \rbrace  \Bigr) \nonumber .
\end{eqnarray}
Now 
$$ 0 \approx (L_0+L_{-1}) a_{-n} b_{-m} \vert 0> \approx n a_{-n-1} b_{-m} + m a_{-n} b_{-m-1} \vert 0>$$
and we can write
$$a_{-n} b_{-m} \vert 0> \approx C(m,n) a_{-n-m+1} b_{-1} \vert 0> \approx \tilde C(m,n) a_{-n-m+2} b_{-2} \vert 0> .$$
By this tool we can rewrite $a_{-\!\alpha} a_{-\!1} b_{-\!\beta} b_{-\!1} \vert 0 \!\!>$ first in terms of $a_{-\!\alpha\!-\!\beta \!+\!2} a_{-\!1} b_{-\!2} b_{-\!1} \vert 0 \!\!>$ and then reduce this to lower weight, because $b_{-2} b_{-1} \vert 0 > \approx 0$. So we can discard (\ref{multiple 2dim}) from $\tilde A(V)$ and specialise to $\delta =1 $ in (\ref{single 2dim}). 

The following  version of (\ref{heisenberg without computer}) provides $a_{-m-5} b_{-1} \vert 0> \approx 0$ for $m \geq 1$:
 \begin{eqnarray} 
0 &\approx &
 a_{-3} a_{-1} \vert 0 > \Box b_{-m} a_{-1} \vert 0 >
-b_{-m} a_{-3} \vert 0 > \Box a_{-1} a_{-1} \vert 0 >  \nonumber \\
&&-b_{-m} a_{-1} \vert 0 > \Box a_{-3} a_{-1} \vert 0 >
+m b_{-m-1} a_{-2} \vert 0 > \Box a_{-1} a_{-1} \vert 0 > \nonumber \\
&&
- \frac{m (m+1)}{2} a_{-1} a_{-1} \vert 0 > \Box b_{-m-2} a_{-1} \vert 0 > \nonumber \\
&\approx_{wt}&
\Biggl([ 1 a_{-6} b_{-m} \lbrace { -2 \choose 2} + {5 \choose 2} \rbrace] 
-[2 b_{-(m+5)} a_{-1}  { -2 \choose 2}  {m+4 \choose m-1} ] \nonumber \\
&&
-[3 b_{-(m+5)} a_{-1}  {m+4 \choose m-1}  + 1 b_{-m-3} a_{-3} {m+2 \choose m-1}] \nonumber \\
&&
+m [2 b_{-(m+5)} a_{-1} { -2 \choose 1}  {m+4 \choose m} ] 
-\frac{m (m+1)}{2}[ 1 a_{-4} b_{-m-2} 2 ]\Biggr) \vert 0 > \nonumber \\
& \approx & -\frac{4}{15} m (m+1) (m+2) (m+3) (m+4) b_{-(m+5)} a_{-1} \vert 0 >  \quad . \nonumber 
\end{eqnarray}
As we could have picked any set of two orthogonal directions in $\Rop^{d=2}$ as axes for the moding, we can equally conclude (here for $m=2$)
$$(a+b)_{-7} (a-b)_{-1} \vert 0 > \approx 0 \quad , \quad a_{-7} b_{-1} \vert 0 > \approx 0 \quad \mbox{and} \quad b_{-7} a_{-1} \vert 0 > \approx 0 $$
and hence
$$ b_{-7} b_{-1} \vert 0 > \approx a_{-7} a_{-1} \vert 0 > \quad . $$
So we arrive at 
\begin{eqnarray*}
\tilde A((H^2)_+) &\!\!\!\!=& \!\!\!\!span  \Bigl( a_{-1}^{n} b_{-1}^{m}\vert 0 > : n+m \ \mbox{even},\\
 &&\!\!a_{-\!\alpha} a_{-\!1} \vert 0 \!\!>,  b_{-\!\beta} b_{-\!1} \vert 0 \!\!>, a_{-\!\gamma} b_{-\!1} \vert 0\!\!>  : \alpha\!=\!3,5,7 \ , \ \beta\!=\!3,5 \ , \ \gamma\!=\!2,3,4,5  \Bigr)  .
\end{eqnarray*}
The action of this on the known representations is independent, as we will see below in the more general case and we have $A(V)=\tilde A(V)$.

\subsection{$(H^d)_+$}
We denote the commuting sets of oscillators by $a_{-m}^{(i)}$ and write
$$ (H^d)_+ := span(\prod_{i_1:=1}^{M_1} a_{-m_{i_1,1}}^{(1)} \dots \prod_{i_d:=1}^{M_d} a_{-m_{i_d,d}}^{(d)}  \vert 0 > : M_1 + \dots + M_d \ \mbox{even}) .$$
Using only the results for $H_+$ and $(H^2)_+$ , and noting that the argument to get rid of (\ref{multiple 2dim}) in $\tilde A((H^2)_+)$ in section \ref{2dim} does not use the passive $a_{-1}$, which may be replaced by any oscillator $a_{-n}^{(j)}$, we see
\begin{eqnarray*}
\lefteqn{\tilde A(V) =span  \Bigl( {a_{-1}^{(1)}}^{n_1} \dots {a_{-1}^{(d)}}^{n_d}\vert 0 > : n_1 + \dots + n_d \ \mbox{even},}\\ &&\!\!a_{-\alpha}^{(i)} a_{-1}^{(i)} \vert 0 >: i=1,\dots,d \ , \  \alpha=3,5 \  \  \ , a_{-7}^{(1)} a_{-1}^{(1)} \vert 0 > \ , \  \\
&&\!\!a_{-\!\gamma}^{(\!i_1\!)} a_{-\!1}^{(\!i_2\!)} \!\dots \!a_{-\!1}^{(\!i_l\!)} \vert 0 \!\!> : l \ \mbox{even}  ,  \lbrace i_1,\!\dots\!,i_l \rbrace \mbox{any ordered subset of} \ \lbrace 1,\!\dots\!,d \rbrace  , \\
&& \qquad \qquad \qquad \gamma\!=\!2,3,4,5  \Bigr)  .
\end{eqnarray*}
We can restrict to $l=2$ by the following argument (for $l=4$, $i_j=j$):
\begin{eqnarray*}
0 & \approx & (L_0+L_{-1})  a_{-m}^{(1)} a_{-1}^{(2)}  a_{-1}^{(3)} a_{-1}^{(4)} \vert 0 > \\
&  \approx & m a_{-\!m\!-\!1}^{(1)} a_{-\!1}^{(2)}  a_{-\!1}^{(3)} a_{-\!1}^{(4)} \vert 0 \!\!> 
\!\!+ a_{-\!m}^{(1)} a_{-\!2}^{(2)}  a_{-\!1}^{(3)} a_{-\!1}^{(4)} \vert 0 \!\!> \\ &&+ a_{-\!m}^{(1)} a_{-\!1}^{(2)}  a_{-\!2}^{(3)} a_{-\!1}^{(4)} \vert 0 \!\!> \!+ a_{-\!m}^{(1)} a_{-\!1}^{(2)}  a_{-\!1}^{(3)} a_{-\!2}^{(4)} \vert 0 \!\!> \\
& \approx & - 2 m a_{-m-1}^{(1)} a_{-1}^{(2)}  a_{-1}^{(3)} a_{-1}^{(4)} \vert 0 > \quad .
\end{eqnarray*}
So independence on the representations (see below) confirms
\begin{eqnarray}
A((H^d)_+) &=& span  \Bigl( {a_{-1}^{(1)}}^{n_1} \dots {a_{-1}^{(d)}}^{n_d}\vert 0 > : n_1 + \dots + n_d \ \mbox{even}, \nonumber \\ 
&&a_{-\alpha}^{(i)} a_{-1}^{(i)} \vert 0 >: i=1,\dots,d \ , \  \alpha=3,5 \ , \quad a_{-7}^{(1)} a_{-1}^{(1)} \vert 0 > \ , \ \nonumber \\
&&a_{-\gamma}^{(i)} a_{-1}^{(j)} \vert 0 > : i < j \ , \ \gamma=2,3,4,5  \Bigr)  . \label{A(V) ddim}
\end{eqnarray}

\subsection{Representations}
Again we have twisted and untwisted types of vertex operators \cite{dgm}. We list the ground states:
\begin{eqnarray*}
M_0^{(k)}=span(\vert k>) :k \in \Cop^d \quad &,& \quad M_0^{(0,-)}=span( a_{-1}^{(i)} \vert 0 > :i=1,\dots,d) \\
M_0^{(T,+)}=span( \chi) \quad &,& \quad M_0^{(T,-)}=span( c_{-1/2}^{(i)} \chi :i=1,\dots,d ) \quad .
\end{eqnarray*}
As in section \ref{pure oscillators} we only need to consider independence of the $N=2$ parts, as all the others are separated by the action on $M_0^{(k)} : k \in \Cop^d$. So we look at conditions for
\begin{equation}
\psi:=(\sum_{i<j} \sum_{l=1}^5 \lambda_{ijl} a_{-l}^{(i)} a_{-1}^{(j)} + \sum_i \sum_{\tiny l=1,3,5} \mu_{il} a_{-l}^{(i)}  a_{-1}^{(i)} + \mu_{1 7} a_{-7}^{(1)} a_{-1}^{(1)} ) \vert 0>
\end{equation}
to have a vanishing zero mode on these ground states.
On $M_0^{(k)}$ we see
\begin{equation} \label{k constraint}
\sum_l \lambda_{ijl} =0 \quad (i<j) \ , \ \sum_{\tiny l=1,3,5} \mu_{il} = 0 \quad (i \neq 1) \ , \ \sum_{\tiny l=1,3,5,7} \mu_{1 l} = 0 \quad .
\end{equation}
Noting that $e^{\Delta (z)} a_{-\alpha}^{(i)} a_{-1}^{(j)} \vert 0> = a_{-\alpha}^{(i)} a_{-1}^{(j)} \vert 0>$ $(i \neq j)$, we calculate the constraint in the off-diagonal components ($a_{-1}^{(i)} a_{1}^{(j)}, a_{-1}^{(j)} a_{1}^{(i)}, c_{-\frac{1}{2}}^{(i)} c_{\frac{1}{2}}^{(j)},  c_{-\frac{1}{2}}^{(j)} c_{\frac{1}{2}}^{(i)} $) on $M_0^{(0,-)}$, $ M_0^{(T,-)}$ as
\begin{eqnarray*}
\sum_{l=1}^5 \lambda_{ijl} \delta_{l,1} =0 &,&
\sum_{l=1}^5 \lambda_{ijl} {-2 \choose l-1} =0 \\
\sum_{l=1}^5 \lambda_{ijl} {-\frac{1}{2} \choose l-1} =0 &,&
\sum_{l=1}^5 \lambda_{ijl} {-\frac{3}{2} \choose l-1} =0 \quad .
\end{eqnarray*}
These, combined with the first equation of (\ref{k constraint}), form an independent set of equations, as is easily checked, and we obtain $\lambda_{ijk}=0$. 

To constrain the $\mu_{i,l}$'s, we look at the diagonal components $id$, $c_{-\frac{1}{2}}^{(i)} c_{\frac{1}{2}}^{(i)}$, $a_{-1}^{(i)} a_{1}^{(i)}$ on $M_0^{(T,+)}$, ``$M_0^{(T,-)}-M_0^{(T,+)}$'', $M_0^{(0,-)}$, noting that $e^{\Delta(z)} a_{-\alpha}^{(i)} a_{-1}^{(i)} \vert 0 > = (a_{-\alpha}^{(i)} a_{-1}^{(i)} + \beta_{\alpha} ) \vert 0>$ with $\beta_{\alpha}$ being indepenent
of $i$:
\begin{eqnarray}
-\sum_{i,l} \mu_{i,l}  \frac{l}{2 (l+1)} {-1/2 \choose l} &=& 0 \label{eqn1}\\
\sum_{l} \mu_{i,l} \lbrace {-\frac{3}{2} \choose l-1 } +{-\frac{1}{2} \choose l-1 } \rbrace &=& 0 \qquad (i \quad \mbox{fixed}) \label{eqn2} \\
\sum_{l} (\delta_{l,1} + l) \mu_{i,l} &=& 0 \qquad (i \quad \mbox{fixed}) 
\quad .  \label{eqn3}
\end{eqnarray}
For fixed $i \neq 1$ equations (\ref{eqn2}), (\ref{eqn3}) and the second equation in (\ref{k constraint}) form an independent system (cf. section \ref{repns pure}) and $\mu_{i,l} = 0$ ($i \neq 1$, $l=1,3,5$). Knowing this, we obtain four independent  (cf. section \ref{repns pure}) equations for $i=1$ ((\ref{eqn1}), (\ref{eqn2}), (\ref{eqn3}), last equation in (\ref{k constraint})), and  we conclude $\mu_{1,l}=0$ ($l=1,3,5,7$), i.e. the action of our basis is independent, proving the claimed form (\ref{A(V) ddim}) of $A(V)$.

\subsection{Conclusion}
We studied the embedding 
$$ (H_+)^d \subset (H^d)_+ \subset H^d .$$
We saw that $A(H) \subset A(H_+)$ with the  quotient\footnote{According to our presentation, we are only looking at vector space embeddings. However as $A(H_+)$ decomposes as an algebra (as in (\ref{blabla})), we can even prove algebraic embeddings, but then we have to be more careful in picking  the suitable 3-dimensional subspace of the space spanned by $a_{- \alpha} a_{-1} \vert 0 > : \alpha = 1,3,5,7$. Compare also footnote \ref{fussnote10} and section \ref{z3conj}.}
 being spanned by $a_{- \alpha} a_{-1} \vert 0 > : \alpha = 3,5,7$. These three vectors give rise to three new representations, namely those built on $a_{-1} \vert 0 >$, $\chi$ and $c_{-\frac{1}{2}} \chi$. One of these (corresponding to $a_{-1} \vert 0>$) was trivially expected, as the adjoint representation of $H$ has to decompose into more than one irreducible representation of the subtheory. For an automorphism of order $n$ we expect $n$ sectors.
The two twisted representations are genuinely new. 

Similarly $A(H^d) \subset A((H^d)_+) $ with the quotient splitting into a $d^2$ dimensional part corresponding to $span( a_{-1}^i \vert 0> : i =1,\dots,d)$, a $d^2$ dimensional part corresponding to $span( c_{-\frac{1}{2}}^i \chi : i =1,\dots,d)$ and a one dimensional part for $span(\chi)$.

The difference between $A((H_+)^d)=(A(H_+))^d$ and $A((H^d)_+)$ corresponds to two possibilities:\\
\\
(a) objects in $(A(H_+))^d$ missing in $A((H^d)_+)$:\\
\\
$a_{-\alpha} a_{-1} b_{-1}^{2m} \vert 0>$  : standard ground states $even(k)$, $odd(k)$ do not tensor consistently with the three special representations when objects like $a_{-1} b_{-1} \vert 0> $ have to be represented (cf section 5.5 in \cite{montague});\\ 
$a_{-\alpha} a_{-1} b_{-\alpha} b_{-1} \vert 0>$: similar inconsistencies in tensoring\footnote{\label{fussnote10} The argument is a bit sloppy, as we did not identify the correct one dimensional subspaces of $A(V)$ with the representations, but rather faked some sort of counting argument.  Hence this picture confuses $\chi \otimes \vert 0>$ with $\chi \otimes \chi$ and   $c_{-\frac{1}{2}} \chi \otimes \vert 0>$ with $c_{-\frac{1}{2}}\chi \otimes \chi$. Obviously the argument can be made more precise.};\\
$b_{-7}b_{-1} \vert 0>$: this was traded for $a_{-7}a_{-1} \vert 0>$ reflecting the fact that (in our faulty picture) ``$\chi \otimes \vert 0>$'' and ``$\vert 0> \otimes \chi$'' both correspond to $\chi \otimes \chi$.\\
\\
(b) new objects in $A((H^d)_+)$:\\
\\
$a_{-\alpha}^{(i)} a_{-1}^{(j)} \vert 0> : i <j , \ \alpha=2,3,4,5$ correspond to glueing together  $c_{-\frac{1}{2}}^{(i)} \chi \otimes \chi$ and $\chi \otimes c_{-\frac{1}{2}}^{(j)} \chi$ (or $a_{-1}^{(i)} \vert 0> $ and $a_{-1}^{(j)} \vert 0> $)  into a single irreducible representation with $d$ dimensional ground state.

\section{$Z\!\!\!Z_3$-Projected Heisenberg Algebra \protect\\ (in Two Dimensions) $(H^2)_0$ } \label{z3}

\subsection{The Algebra and its known Representations} \label{z3intro}
This section summarizes some results from \cite{z3montague}. As we see shortly, $H^2$ allows for a $\Zop_3$-automorphism $\Theta$. We will consider the invariant subspace $(H^2)_0$ and its representation theory. For this it is convenient to describe $H^2$ in a new basis: Let $a_n^{(1)}, a_m^{(2)}$ be the original sets of oscillators used so far. Define 
$$ b_n:=\frac{1}{\sqrt{2}} ( a_n^{(1)}+ i a_n^{(2)} ) \quad, \qquad 
\bar b_n:=\frac{1}{\sqrt{2}} ( a_n^{(1)}- i a_n^{(2)} ) \ . $$
Then $[b_m,\bar b_n]=m \delta_{m,-n}$ and $[b_m,b_n]=[\bar b_m,\bar b_n]=0$.
For
\begin{equation}
\psi=\prod_{a=1}^Mb_{-m_a}\prod_{b=1}^N\bar b_{-n_b}|0>   \ ,   \label{states1}
\end{equation}
which we call of type $(M,N)$, we have
\begin{equation}
V(\psi,z)=:\prod_{a=1}^M (\sum_{k_a \in \Zop} {-k_a-1 \choose m_a-1} z^{-k_a-m_a} b_{k_a} ) \prod_{b=1}^N (\sum_{l_b \in \Zop} {-l_b-1 \choose n_b-1} z^{-l_b-n_b} \bar b_{l_b}) :  \ . \label{vop}
\end{equation}
The automorphism is defined as 
$$ \Theta \vert 0 > := \vert 0> \ , \qquad \Theta b_n \Theta^{-1} = e^{2 \pi i /3} b_n \ , \qquad \Theta \bar b_n \Theta^{-1} = e^{4 \pi i /3} \bar b_n \ .$$
The invariant space $(H^2)_0$ is hence spanned by vectors of the form (\ref{states1}) with $M-N \equiv 0 \  mod \  3$ $( M,N \in \Nop)$ .

Obvious irreducible representations of $(H^2)_0$ are those (cf. section \ref{cheat}) corresponding to ``usual'' vertex operators of a full lattice theory: Ground states are 
$ M_0^{(k)}:= span( \vert k > )$ ($k \in \Cop \setminus \lbrace 0 \rbrace$), $M_0^{(0,0)}:=span( \vert 0 >)$,
$M_0^{(0,1)}:=span( b_{-1} \vert 0 >)$, $M_0^{(0,2)}:=span( \bar b_{-1} \vert 0>)$.

There is a twisted representation $T_1$, built from $\Zop \pm \frac{1}{3}$ moded oscillators $c_r$, with $[c_r,c_s]:=r \delta_{r,-s}$, acting on a ground state $\chi$, which satisfies $c_r \chi =0$ $(r>0)$.
The corresponding vertex operator of the state (\ref{states1}) is  (cf. (\ref{twistvop}))
\begin{equation} 
V_{T_1}(\psi,z)=V_1\left(e^{\Delta_1(z)}\psi,z\right)\,,
\end{equation}
where
$$
V_1(\psi,z)=:\prod_{a=1}^M (\sum_{r_a \in \Zop+\frac{1}{3}} {-r_a\!\!-\!\!1 \choose m_a\!\!-\!\!1} z^{-\!r_a\!-\!m_a} c_{r_a} ) \prod_{b=1}^N (\sum_{s_b \in \Zop-\frac{1}{3}} {-s_b\!\!-\!\!1 \choose n_b\!\!-\!\!1} z^{-\!s_b\!-\!n_b}  c_{s_b}) : \ ,
$$
$$\Delta_1(z)= \sum_{m,n > 0}
\left({-{2\over 3}\atop n}\right)\left({-{1\over 3}\atop m}\right)
{z^{-m-n}\over m+n} b_n \bar b_m\,.
$$
This representation can be decomposed into three irreducible representations of 
$(H^2)_0$ with the following ground states:
$M_0^{(T_1,0)}:=span(\chi)$,  $M_0^{(T_1,1)}:=span(c_{-{1 \over 3}} \chi)$, $M_0^{(T_1,2)}:=span(c_{-{1\over 3}} c_{-{1 \over 3}} \chi \ , c_{-{2 \over 3}} \chi )$.

In a second twisted representation $T_2$ the roles of $b$ and $\bar b$ are interchanged. There we denote the oscillators by $\bar c_r$ and the ground states are $M_0^{(T_2,0)}:=span(\bar \chi)$,  $M_0^{(T_2,1)}:=span(\bar c_{-{1 \over 3}} \bar \chi)$, $M_0^{(T_2,2)}:=span(\bar c_{-{1\over 3}} \bar c_{-{1 \over 3}} \bar \chi \ , \bar c_{-{2 \over 3}} \bar \chi )$.

\subsection{$\tilde A( (H^2)_0 )$}
As in the other examples, we first reduce $\tilde A(V)$ to a tractable size:
We use our usual inductions in weight and number of oscillators. Like in (\ref{Ntodiagonal}) we have
$$ 0 \approx b_{-n} \bar b_{-m} \vert 0> \Box \tilde \psi \approx (n b_{-n-1} \bar b_{-m} +m b_{-n} \bar b_{-m-1}) \tilde \psi \ . $$
Hence, as long as there are at least one $b$-oscillator and one $\bar b$-oscillator, we can reduce $\psi$ to terms of the form $b_{-\alpha} b_{-1}^{M-1} \bar b_{-1}^{N} \vert 0 >$. In the case $\alpha \geq 2$, we can (for $N\geq 3$) rewrite this in terms of 
$$ b_{-(\alpha-1)} b_{-1}^{M-1} \bar b_{-2} \bar b_{-1} \bar b_{-1} \bar b_{-1}^{N-3} \vert 0 \!\!> \approx \frac{1}{3}  \bar b_{-1} \bar b_{-1} \bar b_{-1}\vert 0 \!\!> \Box b_{-(\alpha-1)} b_{-1}^{M-1} \bar b_{-1}^{N-3}\vert 0 \!\!> \approx 0 \ .$$
[For $N=1,2$ and $M\geq 3$ we instead use  $ b_{-2} b_{-1}  b_{-1}  b_{-1}^{M-3} \bar b_{-(\alpha-1)} \bar b_{-1}^{N-1} \vert 0 \!\!>\approx \dots \approx 0$.]

So apart from objects of the form 
\begin{equation} \label{infset}
\phi = b_{-1}^M \bar b_{-1}^N \vert 0>
\end{equation}
we are only left with vectors of types\footnote{Remember that $M-N \equiv 0 \ mod \ 3$.} $(3k,0)$, $(0,3l)$, $(1,1)$, $(2,2)$.

Now we want to reduce type $(3k,0)$: First note that, by (\ref{vir}),
\begin{eqnarray}
0 &\approx& L_{-1}b_{-n}b_{-m}b_{-(l-1)} \vert 0> \nonumber \\
 &=& n b_{-n-1}b_{-m}b_{-(l-1)} \vert 0>+m b_{-n}b_{-m-1}b_{-(l-1)} \vert 0> \nonumber \\
&&+ (l-1) b_{-n}b_{-m}b_{-l} \vert 0> \ . \label{al}
\end{eqnarray}
So for $n\geq m \geq l \geq 2$ we can rewrite $ b_{-n}b_{-m}b_{-l} \vert 0>$ in terms of vectors with smaller $l$, and finally $l=1$.

Now we can use the same argument in the presence of more oscillators, by starring (\ref{al}) with the other oscillators, so that we always reduce any vector of type (3k,0) to objects of the form $b_{-n} b_{-m} b_{-1}^{3k-2} \vert 0> $ ($n \geq m$).
If $k>1$ and $n>1$, (\ref{al}) implies
$$b_{-n} b_{-1} b_{-1} \vert 0> \approx -\frac{2}{n-1} b_{-(n-1)} b_{-2} b_{-1} \vert 0> \ ,$$
so 
\begin{eqnarray*}
b_{-n} b_{-m} b_{-1}^{3k-2} \vert 0> & \approx & b_{-n} b_{-1}^2 \vert 0> \star b_{-m} b_{-1}^{3k-4} \vert 0> \\
& \approx & - \frac{2}{n-1} b_{-(n-1)} b_{-2} b_{-1} \vert 0> \star b_{-m} b_{-1}^{3k-4} \vert 0> \\
& \approx & - \frac{2}{n-1} b_{-2} b_{-1} b_{-1} \vert 0> \star b_{-(n-1)}  b_{-m} b_{-1}^{3k-5} \vert 0> \\
& \approx & - \frac{2}{3 (n-1)} b_{-1} b_{-1} b_{-1} \vert 0> \Box b_{-(n-1)}  b_{-m} b_{-1}^{3k-5} \vert 0> \\
& \approx & 0 \ .
\end{eqnarray*}
The same analysis holds for type $(0,3l)$, so we are left with (\ref{infset}),  
type $(1,1)$, $(2,2)$, $(3,0)$ and $(0,3)$. Now we turn to the analysis of these remaining types:

\subsubsection{$(3,0)$}
We try a cancellation at type $(4,1)$, similar in spirit to (\ref{heisenberg without computer result}). This will have two free parameters to have a chance to be enough to reduce type $(3,0)$ to a finite number of terms. The cancellation can be understood by looking at the diagram
\begin{displaymath} \begin{array}{ccc}
\bar b_{-m} b_{-n-1} b_{-2} b_{-1} b_{-1} &\!\!\!\rightarrow\!\!\! &  \bar b_{-m-1} b_{-n} b_{-2} b_{-1} b_{-1} \\
\downarrow && \downarrow\\
\bar b_{-m-1} b_{-n-1} b_{-1} b_{-1} b_{-1} &\!\!\!\rightarrow\!\!\! & \bar b_{-m-2} b_{-n} b_{-1} b_{-1} b_{-1} \quad ,
\end{array} \end{displaymath}
the only purpose of which is to eluminate the following equation (for $n \geq 1$, $m\geq 1$):
\begin{displaymath} \begin{array}{cl}
\!\!0\approx & \!\!\!(n-1) \bar b_{-m} b_{-n} \vert 0 > \Box b_{-2} b_{-1} b_{-1} \vert 0 > 
- n (n\!-\!1) \bar b_{-\!m} b_{-\!1} \vert 0 \!\!> \Box b_{-\!n\!-\!1} b_{-\!1} b_{-\!1} \vert 0 \!\!> \\
& \!\!\!+ m (n\!-\!1) \bar b_{-\!m\!-\!1} b_{-\!n} \vert 0 \!\!> \Box b_{-\!1} b_{-\!1} b_{-\!1} \vert 0 \!\!>
- m (n\!-\!1) \bar b_{-\!m\!-\!1} b_{-\!1} \vert 0 \!\!> \Box b_{-\!n} b_{-\!1} b_{-\!1} \vert 0 \!\!>\\
\!\!\approx_{wt} & \!\!\!(n\!-\!1) [ 2 {-3 \choose m-1} {m+n+2 \choose n-1} b_{-m-n-3} b_{-1} b_{-1} +2 {-2 \choose m-1} {m+n+1 \choose n-1} b_{-m-n-2} b_{-2} b_{-1}] \vert 0 \!\!> \\
& \!\!\!-n (n-1) [ (n+1) {-n-2 \choose m-1}  b_{-m-n-3} b_{-1} b_{-1} +2 {-2 \choose m-1}  b_{-m-3} b_{-n-1} b_{-1}] \vert 0 > \\
& \!\!\!+m (n-1) [ 3 {-2 \choose m} {m+n+2 \choose n-1} b_{-m-n-3} b_{-1} b_{-1}] \vert 0 > \\
& \!\!\!-m (n-1) [ n {-n-1 \choose m}  b_{-m-n-3} b_{-1} b_{-1} 
+2 {-2 \choose m}  b_{-m-4} b_{-n} b_{-1}] \vert 0 > \quad .
\end{array} \end{displaymath}
With a little bit of trivial algebra and the use of
$$ b_{-m-n-2} b_{-2} b_{-1} \vert 0 > \approx -\frac{m+n+2}{2} b_{-m-n-3} b_{-1} b_{-1} \vert 0 >$$
we obtain for $N(=m+n)\geq 3, n=1,\dots, N-1$
\begin{eqnarray} 
0  &\approx_{wt}& (-)^{N-n} (N-n) (n-1) \Bigl( a( n) b_{-\!N\!-\!3} b_{-\!1} b_{-\!1} +b(n) b_{-\!N\!+\!n \!-\!4} b_{-\!n } b_{-\!1} \nonumber \\
&& + c(n) b_{-\!N\!+\!n \!-\!3} b_{-\!n \!-\!1} b_{-\!1} \Bigr) \vert 0\!\!>
\label{30system}
\end{eqnarray}
with
\begin{eqnarray*}
a(n)& = &{N+2 \choose  n-1 } (3(N-n) +5 ) \\
b(n) &=& -2 (N-n+1)\\
c(n) &=& 2 n \ .
\end{eqnarray*}
As the result for $n=1$ is trivial, we only take $n=2,\dots,N-1$.\\
All equations are combinations of the vectors 
\begin{equation}
 b_{-N-3}b_{-1}b_{-1}\vert0\!\!>  , \ b_{-N-2}b_{-2}b_{-1}\vert0\!\!>  ,  \dots  ,  b_{-N+[{N\over 2}]-2}b_{-[{N\over 2}]-2}b_{-1}\vert0\!\!> \ .
\end{equation}
As for $N\geq 7$ the number of equations $(N-2)$ is greater or equal to the number of vectors $([{N\over 2}]+2)$, one might like to believe that a suitable combination could reduce all the vectors to lower weight . Unfortunately this is not the case (as can be seen with quite some effort). We will include  two more equations (\ref{30spec1},\ref{30spec2}) and combine  them with the first two equations of (\ref{30system}) (for $ n =2,3$) (assuming $N\geq 4$, so that $ n=3 \leq N-1$). Combining these four equations will yield  
\begin{eqnarray}
 b_{-N-3}b_{-1}b_{-1}\vert0> \approx 0 && b_{-N-2}b_{-2}b_{-1}\vert0>  \approx 0
\nonumber \\
b_{-N-1}b_{-3}b_{-1}\vert0> \approx 0 && b_{-N}b_{-4}b_{-1}\vert0>  \approx 0 \ . \label{30special}
\end{eqnarray}
Using this, we can simply use (\ref{30system}) for $ n = 4,\dots, [{N\over 2}]+1$ (in this order!) to show that 
$b_{-N+1}b_{-5}b_{-1}\vert0> \approx 0$,$\dots$, $ b_{-N+[{N\over 2}]-2}b_{-[{N\over 2}]-2}b_{-1}\vert0> \approx 0$. Combining this with (\ref{al}), we can conclude, that  for $weight \geq 9$ ($N\geq 4$)  type $(3,0)$ can be reduced to lower weight.

Let us fill in the details to prove (\ref{30special}): The two additional equations are from (\ref{al}) for $m=l-1=1,2$ (with each vector reduced by the standard procedure using (\ref{al}) with $n\geq m\geq l \geq2$):
\begin{eqnarray}
0 &\approx& L_{-1}b_{-N-2}b_{-1}b_{-1} \vert 0> \nonumber \\
 &\approx& (N+2) b_{-N-3}b_{-1}b_{-1} \vert 0>+2 b_{-N-2}b_{-2}b_{-1} \vert 0>  \label{30spec1}
\end{eqnarray}
\begin{eqnarray}
0 \!\!\!\!&\approx& \!\!\!\! L_{-1}b_{-N}b_{-2}b_{-2} \vert 0> \nonumber \\
 &\approx& \!\!\!\! N b_{-N-1}b_{-2}b_{-2} \vert 0>+4 b_{-N}b_{-3}b_{-2} \vert 0> \nonumber \\
 &\approx& \!\!\!\!-N (N\!\!+\!\!1) b_{-\!N\!-\!2}b_{-\!2}b_{-\!1} \vert 0\!\!> \!-2 N  b_{-\!N\!-\!1}b_{-\!3}b_{-\!1} \vert 0\!\!> \! - 4 N b_{-\!N\!-1}b_{-\!3}b_{-\!1} \vert 0\!\!> \nonumber \\
&&\!\!\!\!-12 b_{-\!N}b_{-\!4}b_{-\!1} \vert 0\!\!> \nonumber \\
&\approx& \!\!\!\!-N (N\!\!+\!\!1) b_{-\!N\!-\!2}b_{-\!2}b_{-\!1} \vert 0\!\!> \!-6 N  b_{-\!N\!-\!1}b_{-3}b_{-1} \vert 0\!\!> \! -12 b_{-\!N}b_{-\!4}b_{-\!1} \vert 0\!\!> \, .  \label{30spec2}
\end{eqnarray}
The set of our four equations can be written as
\begin{displaymath} 
\left( \begin{array}{cccc} 
a(2)&b(2)&c(2)&0\\
a(3)&0&b(3)&c(3)\\
N+2&2&0&0\\
0&-N (N+1) &-6 N & -12
\end{array} \right) 
\left( \begin{array}{c}
b_{-N-3}b_{-1}b_{-1}\vert0> \\
b_{-N-2}b_{-2}b_{-1}\vert0> \\
b_{-N-1}b_{-3}b_{-1}\vert0> \\
b_{-N}b_{-4}b_{-1}\vert0>  
\end{array} \right)
\approx 0 \ .
\end{displaymath}
The determinant of the matrix can be evaluated as 
$$ -648 N (N+2) (N-1) \neq 0 \qquad (\mbox{for } \ N\geq 4) \ .$$
So (\ref{30special}) holds. 

Having reduced all objects of type  $(3,0)$ of $weight \geq 9$, we turn to the special case of lower weight:
For $N=3$ (weight 8) we have to drop the equation for $ n =3$ and instead of four equations we get (noting that $b_{-4}b_{-3}b_{-1}\vert0> =
b_{-3}b_{-4}b_{-1}\vert0>$)  
\begin{displaymath} 
\left( \begin{array}{ccc} 
a(2)&b(2)&c(2)\\
N+2&2&0\\
0&-N (N+1) &-6 N  -12
\end{array} \right) 
\left( \begin{array}{c}
b_{-6}b_{-1}b_{-1}\vert0> \\
b_{-5}b_{-2}b_{-1}\vert0> \\
b_{-4}b_{-3}b_{-1}\vert0>   
\end{array} \right)
\approx 0 \ .
\end{displaymath}
As the determinant is $-3240 \neq 0$, weight 8 can equally be reduced.

For $weight \leq 7$ we proceed differently:
We exploit {\it all} information from (\ref{al}), instead of just only the cases $n\geq m \geq l \geq 2$:\\
{\it Claim:} Using (\ref{al}) we can exactly reduce to $b_{-n} b_{-n} b_{-m} \vert 0>$ ($n \geq m$) \\
{\it Proof:} We first introduce the total lexicographic ordering of $(n+m+k,n,m,k)$ on all objects of the form  $b_{-n} b_{-m} b_{-k} \vert 0>$ with $n \geq m \geq k$. In this proof ``$\approx$'' refers to equality up to smaller terms in this ordering (and up to $O(V)$).

Now for $n \geq m \geq k$:
$$0 \approx L_{-1} b_{-n} b_{-m} b_{-k} \vert 0> \approx const \ b_{-n-1} b_{-m} b_{-k} \vert 0> \ .$$
So for every $(n,m,k)$ there is one new reduction and all objects $(n+1,m,k)$ with $n \geq m \geq k$ are reduced. So precisely the objects $(n,n,m)$ ($n \geq m$) are left out by this.\\
{\it qed .}\\
\\
\\
So we can reduce all vectors of type $(3,0)$ to a combination of
\begin{equation}
b_{-1}b_{-1}b_{-1}\vert0>,
b_{-2}b_{-2}b_{-1}\vert0>,
b_{-2}b_{-2}b_{-2}\vert0>,
b_{-3}b_{-3}b_{-1}\vert0> \ .
\end{equation}
In a similar way we can reduce all vectors of type $(0,3)$ to
\begin{equation}
\bar b_{-1}\bar b_{-1}\bar b_{-1}\vert0>,
\bar b_{-2}\bar b_{-2}\bar b_{-1}\vert0>,
\bar b_{-2}\bar b_{-2}\bar b_{-2}\vert0>,
\bar b_{-3}\bar b_{-3}\bar b_{-1}\vert0> \ .
\end{equation}

\subsubsection{$(2,2)$ and $(1,1)$}
The strategy is to get cancellations at type $(3,3)$, resulting in objects of type $(2,2)$ and $(1,1)$ (up to lower weight). These are then reduced to standard form $b_{-n-5}b_{-1}\bar b_{-1}\bar b_{-1}\vert0\!>$ and $b_{-n-7}\bar b_{-1}\vert0\!>$. The resulting (reducing) equation for these is what we are interested in. At every weight we propose two such cancellations and show that the resulting equations are independent (for $n$ sufficiently large).

We use the following new notation:
$$[n_1,n_2,\dots,n_N;m_1,m_2,\dots,m_M]:=\prod_{a=1}^{N} b_{-n_a} \prod_{b=1}^{M} \bar b_{-m_b} \vert 0\!> \ .$$
The diagram
\begin{eqnarray*}
0 &\rightarrow &[2,1,1;k,2,l] \rightarrow [k+1,1,1;1,2,l] \rightarrow
[k+2,1,1;,1,1,l] \\
&\rightarrow& [k+l+1,1,1;1,1,1] \rightarrow [k+l,1,1;2,1,1] \rightarrow 0
\end{eqnarray*}
illustrates the cancellation for type $(3,3)$
\begin{eqnarray}
\lefteqn{0 \approx 
[1,1,1;] \Box [;k,2,l]} \label{eq1} \\ 
&&-  3 \Bigl([2;k]+(-)^k k [k+1;1]\Bigr) \star [1,1;2,l] \label{eq2} \\ 
&& -  3 k (-)^{k-1} \Bigl([k+1;2]+(k+1) [k+2;1]\Bigr) \star [1,1;1,l] \label{eq3}\\
&&-3 k (k+1) (-)^k \Bigl([k+2;l]+(-)^l {l+k \choose l-1} [k+l+1;1]\Bigr) \star [1,1;1,1] \label{eq4} \\
&& -3 k (k\!+\!1) (-)^{l+k-1} {l\!+\!k \choose l\!-\!1} \Bigl([k\!+\!l\!+\!1;1]+{1 \over k\!+\!l} [k\!+\!l;2] \Bigr)\star [1,1;1,1] \label{eq5} \\
&&+\frac{3 k (k+1)}{k+l}(-)^{k+l-1}{k+l \choose l-1} [;2,1,1] \star [k+l,1,1;] \ . \label{eq6}
\end{eqnarray}
We are interested in two versions of this expression, namely $(k,l)=(n,2)$ and $(k,l)=(n+1,1)$. We discuss the contributions at top weight: type $(3,3)$ cancels; all terms contribute to type $(2,2)$; at first only the terms (\ref{eq1}) and (\ref{eq6}) contribute to type $(1,1)$, however, whilst reducing terms of type $(2,2)$ to standard form, we encounter objects of type $(1,1)$ also in all other terms. We are not planning to give {\it all} details, but list a number of important steps:
Bringing the general vector of type $(2,2)$, $[n,m;k,l]$, to standard form via
$$ [n,m;k,l] \rightarrow [n,1;k+m-1,l] \rightarrow [n+k+m-2,1;1,l] \rightarrow [n+k+m+l-3,1;1,1] $$
is summarized by
\begin{eqnarray}
\lefteqn{[n,m;k,l] \approx (-)^{k+l} \frac{(n+m+k+l-4)!}{(n-1)! (m-1)! (k-1)! (l-1)!} [n+m+k+l-3,1;1,1]} \nonumber \\
&& + (-)^{k+l} \frac{(n+m+k+l-2)!}{(n-1)! (m-1)! (k-1)! (l-1)!} \Bigl( 1+\frac{2}{n+m+k+l-2}  -\frac{1}{n+l} \nonumber \\
&& - \frac{1}{n+k} -\frac{1}{m+k} -\frac{1}{m+l} \Bigr) [n+m+k+l-1;1] 
 \ . \label{alphan}
\end{eqnarray} 
The top weight contribution of $[m;n] \star [1,1;t,u]$ (appearing in (\ref{eq2},\ref{eq3},\ref{eq4},\ref{eq5})), apart from the bits of type $(3,3)$ 
(which anyway cancel with the other terms), is
$$ [m;n] \star [1,1;t,u] - [m,1,1;n,t,u] \approx  2 {-2 \choose n-1} {m+n \choose m-1} [m+n+1,1;t,u] $$
\begin{eqnarray}
&& + t {-t-1 \choose m-1} {m+n+t-1 \choose n-1} [1,1;m+n+t,u] \nonumber \\
&& + u {-u-1 \choose m-1} {m+n+u-1 \choose n-1} [1,1;t,m+n+u] \ . \label{betan}
\end{eqnarray}
Using the two formulae (\ref{alphan},\ref{betan}) we can rewrite (\ref{eq2},\ref{eq3},\ref{eq4},\ref{eq5}) as a finite number (independent of $(k,l)$ or $n$) of terms. The precise result is rather lengthy and not very eluminating and hence not shown.

The top weight part (without type $(3,3)$) of (\ref{eq1}) is
\begin{eqnarray}
\lefteqn{[1,1,1;] \Box [;k,2,l] - [2,1,1;k,2,l] \approx 3 k \sum_{r=1}^{3+k} [r,4+k-r;2,l]} \label{eq11} \\
&& + 6 \sum_{r=1}^{5} [r,6-r;k,l]   +3 l \sum_{r=1}^{3+l} [r,4+l-r;k,2]  \label{eq13} \\
&& + \Bigl( 12 k (-)^{l-1} {k+l+4 \choose l-1} - 6 k l (k+l+4) +12 l (-)^{k-1} {k+l+4 \choose k-1} \Bigr)  \nonumber \\
&& * [k+l+5;1] \ . \label{eq14}
\end{eqnarray}
Inserting (\ref{alphan}) in (\ref{eq13},\ref{eq14}) gives a finite number of terms as we eventually will fix $l=1$ or $l=2$ (independent of $n$!). However (\ref{eq11})  involves a summation depending on $k$ (or $n$). To compute this sum, we have to know the following formulae:
\begin{equation}
\sum_{r=1}^{k+3} \frac{1}{(r-1)! (k+3-r)! } = \frac{1}{(k+2)!} 2^{k+2} \label{gamman}
\end{equation}
\begin{eqnarray}
\lefteqn{\sum_{r=1}^{k+3} \frac{1}{(r-1)! (k+3-r)!} \frac{1}{r+(\alpha-1)} = \sum_{r=0}^{k+2} \frac{1}{r! (k+2-r)!} \int_0^1 x^{r+\alpha-1} dx } \nonumber \\ && =
\frac{1}{(k+2)!} \int_0^1 (1+x)^{k+2} x^{\alpha-1} dx =: \frac{1}{(k+2)!} S(k+2,\alpha) \ . \label{deltan}
\end{eqnarray}
Using successive integration by parts, we can express $S(k+2,\alpha)$ (which we need for $\alpha = 2,3$) in terms of 
\begin{equation}
S(\tilde k+2,1)=\frac{2^{\tilde k+3}-1}{\tilde k+3}  \label{epsilonn} 
\end{equation}
with $\tilde k = k+\alpha-1$. 
Making use of (\ref{alphan},\ref{gamman},\ref{deltan},\ref{epsilonn}), (\ref{eq11}) can be written as a finite number (independent of $n$) of terms, the detailed forms of which are again not terribly eluminating.

The last term to discuss is (\ref{eq6}). Its top weight part (without (3,3) objects) is (dropping the coefficient)
\begin{eqnarray}
\lefteqn{[;2,1,1] \star [k+l,1,1;] - [k+l,1,1;2,1,1] \approx} \nonumber \\
&& 2 (k+l) \sum_{r=1}^{k+l+3} (r-1) [1,1;r,k+l+4-r]  \label{eq1n} \\
&& + 4 \sum_{r=1}^{4} (r-1) [k+l,1;r,5-r] -4 \sum_{r=1}^{4}  [k+l,1;r,5-r] \label{eq2n} \\
&&+(k+l) (-1-k-l) \sum_{r=1}^{k+l+3}  [1,1;r,k+l+4-r] \label{eq3n} \\
&&+ \Bigl( 10 (-)^5 {k+l+4 \choose 5}  +4 (k+l) (k+l+4) (-)^{k+l+4}
                 -8 (-)^5 {k+l+4 \choose 5} \nonumber \\
&&               +4 (k+l) (-k-l-1)  (-1)^{k+l+4} 
                 -8 (k+l)  (-)^{k+l+4} \Bigr) [k+l+5;1] \ .
\end{eqnarray}
Apart from (\ref{eq1n},\ref{eq3n}) these are a finite number of terms, but (\ref{eq1n},\ref{eq3n}) can be summed using (\ref{alphan},\ref{gamman},\ref{deltan},\ref{epsilonn}) and 
$$
\sum_{r=1}^{k+3} \frac{r-1}{(r-1)! (k+3-r)! } = \frac{1}{(k+1)!} 2^{k+1} \ .
$$

Hence substituting all this into (\ref{eq1}-\ref{eq6}) for $(k,l)=(n,2)$ and $(n+1,1)$ we obtain
\begin{eqnarray*}
a_1(n) [n+5,1;1,1] + a_2(n) [n+7;1] &\approx& 0 \\
a_3(n) [n+5,1;1,1] + a_4(n) [n+7;1] &\approx& 0
\end{eqnarray*}
with the coefficients $a_1(n),\dots,a_4(n)$ explicitly given by a finite number of terms. We can conclude
\begin{eqnarray}
[n+5,1;1,1] &\approx& 0 \nonumber \\
\lbrack n+7;1 \rbrack &\approx& 0 \label{22result}
\end{eqnarray}
provided we can prove that $d(n):=det \left( \begin{array}{cc} a_1(n) & a_2(n)\\
a_3(n) & a_4(n) \end{array} \right) \neq 0$. Explicitly calculating $d(n)$ shows that $d(n)=\frac{(n+1) P(n)}{(n+2)^4 (n+3)^3 (n+4) (n+5) (n+6)}$ with $P(n)$ a linear combination of terms of the form $2^{n s} n^r$ with $s=0,1,2$ and $r=0,\dots,20$. Comparing coefficients, we can safely estimate that the terms of the same sign dominating for $n \!\rightarrow \!\infty$ already ``dominate'' for $n \geq 90$. Finally we just calculate $P(n)$ for $n < 90$ and find, that the only (integer) zero appears for $n=1$. So (\ref{22result}) holds for $n \geq 2$.
For $n=1$ the equations are dependent, and we only have
$ [6,1;1,1] \approx - \frac{27}{2}  [8;1]  $ at weight 9.

\subsubsection{Preliminary $\tilde A((H^2)_0)$}
Summarizing the results so far, we can reduce $A(V)$ to\footnote{By abuse of notation, we use the same symbol $\tilde A(V)$ for the subspace of $V$ defined by the rhs of (\ref{tentative basis}) and by the space $V/\tilde O(V)$ with a basis of cosets represented by the vectors in (\ref{tentative basis}). We hope it is clear from the context, whichever is used. Most of the time both interpretations are valid.} 
\begin{eqnarray}
\tilde A((H^2)_0) = &span \Bigl(& \phi_{M,N}:= b_{-1}^M \bar b_{-1}^N \vert 0 \!\!> : M,N \in \Nop (M-N \equiv 0 \ mod \ 3), \nonumber \\
&& 
b_{-2}b_{-2}b_{-1}\vert0\!\!>,
b_{-2}b_{-2}b_{-2}\vert0\!\!>,
b_{-3}b_{-3}b_{-1}\vert0\!\!>, \nonumber \\
&&
\bar b_{-2}\bar b_{-2}\bar b_{-1}\vert0\!\!>,
\bar b_{-2}\bar b_{-2}\bar b_{-2}\vert0\!\!>,
\bar b_{-3}\bar b_{-3}\bar b_{-1}\vert0\!\!>, \nonumber \\
&&
b_{-\alpha} \bar b_{-1} \vert 0 \!\!> : \alpha=2,\dots,8 \quad , \nonumber \\
&&b_{-\beta}b_{-1} \bar b_{-1} \bar b_{-1}\vert0\!\!> : \beta = 2,\dots,5 \quad \Bigr) \quad . \label{tentative basis}
\end{eqnarray}
Looking only at top weight in candidate relations to decrease $A(V)$ further will not prove fruitful, as has been checked by brute force, using the computer to systematically list all reductions of $a \Box b$ for $wt a \!+ \!wt b \!\leq \!8$ at top weight. Before we turn to the results of a more detailed computer search for more relations in $O(V)$, we look at the action of this basis (\ref{tentative basis}) of $\tilde A(V)$ on the known representations:

We determine the subspace $A_1 \subset \tilde A((H^2)_0)$ with 
\begin{eqnarray*}
A_1&\!\!\!\!:=\!\!\!\!&\biggl\lbrace \phi \in \tilde A((H^2)_0): (\phi)_0 \vert_{M_0^{(i)}} = 0 \ \mbox{for all representations listed in section \ref{z3intro}} \\
&& \mbox{i.e.} \quad (i)\in \Bigl\lbrace (k): k\in \Cop^2, (0,1), (0,2), (T_1,0), (T_1,1), \\
&& (T_1,2), (T_2,0), (T_2,1), (T_2,2) \Bigr\rbrace \biggr\rbrace \quad .
\end{eqnarray*}
We find:
\begin{equation}
A_1 = span( \varphi_1,\varphi_2, \varphi_3)
\end{equation}
with
\begin{eqnarray}
\varphi_1&:=&-8/9 [1,1,1;] + 29/9 [2,2,1;] + [2,2,2;] -4/3 [3,3,1;] \\
\varphi_2&:=& 11/54 [2;1] \!+\!38/27 [3;1] \!+\! 173/54 [4;1] \!+\!3 [5;1] \!+\! [6;1] \!+\! 1/27 [1,1;1,1] \nonumber \\
&& +31/108 [2,1;1,1] +1/2 [3,1;1,1] +1/4 [4,1;1,1] \\
\varphi_3&:=& -8/9 [;1,1,1] +29/9 [;2,2,1] +[;2,2,2] -4/3 [;3,3,1] \ .
\end{eqnarray}

Now the task is to determine the subspace of $A_1$ which is in $O(V)$.
The idea behind the computer search is to calculate all products $a \Box b$ with $wt a + wt b \leq max$, for some given limitation $max$ in the search, and then reduce the result along the lines of this section until it is represented by a vector  in  $\tilde A((H^2)_0)$. More explicitly, when we have a reduction sequence, as outlined before, like
$$a \Box b \approx \alpha_{1,1} \approx \alpha_{2,1}
\approx \dots \approx  \alpha_{n_1,1} \in \tilde A((H^2)_0)$$
and the objects in $O(V)$ corresponding to ``$\approx$'' are $d_{i,1}$, we calculate 
$$\tilde \alpha_{n_1,1}:=d_{1,1}+\dots+d_{n_1,1}+ a \Box b \ .$$
Now $ \alpha_{0,2}:= \tilde \alpha_{n_1,1}- \alpha_{n_1,1}$ is of lower weight (or has less oscillators) than $ \alpha_{n_1,1}$, and we can reduce it in a similar way to $\alpha_{n_2,2} \in  \tilde A((H^2)_0)$. We continue like this until we reach weight $0$ and get
$$a \Box b + \sum_{i,j} d_{i,j} = \sum_i \alpha_{n_i,i} \in  \tilde A((H^2)_0) \cap  O((H^2)_0) \ .$$
In this way we can systematically explore those parts of  $O(V) \cap A_1$ that we missed so far. A complete computer search conducted for $max =11$ (and a partial search for max=12,13,14) obtained only
\begin{equation}
\varphi_2 \Cop \subset O((H^2)_0) \cap A_1
\end{equation}
(by eg. using $a=[1,1,1;]$ and $b=[;3,3,1]$). Using this, we can reduce $[6,1]$ in terms of the remaining basis elements. By symmetry we can argue that if any nonzero linear combination of $\varphi_1,\varphi_3$ were in $O(V)$, so would $\varphi_1$ and $\varphi_3$ be. So we are left with only two possibilities:
\begin{description}
\item[(a) $\varphi_1,\varphi_3 \in O(V)$:] 
  This would mean that $A(V)$ is reduced to the size expected from the known representations and we would conclude that $A(V)=P_0^2 \oplus (\oplus_{(i) \neq (k)} End M_0^{(i)})  $, with $P_0^2$ the algebra of polynomials in two variables, restricted to combinations of monomials with the difference of the degrees in the two variables being a multiple of 3.
The set of (reconstructible) representations would be complete.
\item[(b) $\varphi_1,\varphi_3 \not\in O(V)$:] 
  $A(V) $ would be larger than expected, and to learn about the representation theory, we would have to explicitly calculate the product structure on $(A(V),\star)$
\end{description}

\subsection{A Conjecture} \label{z3conj}
Possibility (a) would seem fairly unusual, as we know that (after reduction) products $a \Box b $ with $wt a +wt b \leq 11 $ do  not  produce a component of $\varphi_1$. Without a hidden symmetry (a new representation) this miraculous cancellation of $wt12, wt11, \dots, wt7=wt(\varphi_1)$ contributions of all such products seems fairly unlikely to me. This suggests to conjecture that in fact there is a new representation lurking behind the cancellation, which necessitates $A(V)$ to be bigger than previously expected. So we conjecture that option (b) is in fact true. For the moment we assume that this holds\footnote{A proof of the conjecture is given in section \ref{pf of repnconj}. However the approach towards the proof is best motivated by looking first at conclusions we can draw from the conjecture.}.

To calculate the product structure on $(A(V),\star)$, it is useful to make a choice of basis reflecting the structure of $\frac{A(V)}{span(\varphi_1,\varphi_3)}=  P_0^2 \bigoplus \Bigl(\oplus_{(i) \neq (k)} End M_0^{(i)}\Bigr)$.  Given our basis of $M_0^{(i)}$, this choice is essentially unique up to adding arbitrary combinations of  $\varphi_1,\varphi_3 $  to basis vectors. This freedom can be removed by asking for a multiplicative structure ``as simple as possible''.

In appendix \ref{app C} we list this choice of basis, labeled in a suggestive way:
\begin{eqnarray}
& \varphi_1, \qquad \varphi_3, \qquad \varphi^{k^n \bar k^{\bar n}}, \qquad \varphi^{(0,1)},\qquad  \varphi^{(0,2)}, & \nonumber \\
&\varphi^{T_1,c_{\!-\!{1\over 3}}c_{\!-\!{1\over 3}}c_{{\!1\over 3}} c_{{\!1\over 3}}} , \varphi^{T_1,c_{\!-\!{1\over 3}} c_{{\!1\over 3}}}, \varphi^{T_1,c_{\!-\!{2\over 3}} c_{{\!2\over 3}}}, \varphi^{T_1,id},  \varphi^{T_1,c_{\!-\!{1\over 3}}c_{\!-\!{1\over 3}}c_{{\!2\over 3}}}, \varphi^{T_1,c_{\!-\!{2\over 3}}c_{{\!1\over 3}} c_{{\!1\over 3}}}, & \nonumber \\
&\varphi^{T_2,\bar c_{\!-\!{1\over 3}}\bar c_{\!-\!{1\over 3}}\bar c_{{\!1\over 3}} \bar c_{{\!1\over 3}}} , \varphi^{T_2,\bar c_{\!-\!{1\over 3}} \bar c_{{\!1\over 3}}}, \varphi^{T_2,\bar c_{\!-\!{2\over 3}} \bar c_{{\!2\over 3}}}, \varphi^{T_2,\bar {id}},  \varphi^{T_2,\bar c_{\!-\!{1\over 3}}\bar c_{\!-\!{1\over 3}}\bar c_{{\!2\over 3}}}, \varphi^{T_2,\bar c_{\!-\!{2\over 3}}\bar c_{{\!1\over 3}} \bar c_{{\!1\over 3}}} \ . & \label{z3 basis choice}
\end{eqnarray}
We denote  elements of the basis of the form  $\varphi^{k^n \bar k^{\bar n}}$ by  $\varphi_{poly}$ and all others by $\varphi_{rest}$. Using the associativity and (\ref{app11}), which suggests the definition of $\varphi^{k^n \bar k^{\bar n}}$ in terms of products of $\varphi^{k^1 \bar k^{1}}$,$\varphi^{k^0 \bar k^{3}}$,$\varphi^{k^3 \bar k^{0}}$ (cf appendix \ref{app C}), we only need to calculate a finite number of products. As we need to reduce the result of each product to  a linear combination of vectors in (\ref{z3 basis choice}), this computation is again done using the computer (Maple). We obtain:
$$ \varphi^{k^0 \bar k^0} \star \varphi_{rest}=\varphi_{rest} \star \varphi^{k^0 \bar k^0} = 0 $$
$$ \varphi^{k^3 \bar k^0} \star \varphi_{rest}=\varphi_{rest} \star \varphi^{k^3 \bar k^0} = 0 $$
$$ \varphi^{k^0 \bar k^3} \star \varphi_{rest}=\varphi_{rest} \star \varphi^{k^0 \bar k^3} = 0 $$
$$ \varphi^{k^1 \bar k^1} \star \varphi_{rest}=\varphi_{rest} \star \varphi^{k^1 \bar k^1} = 0 \ . $$
We can conclude (using the definition in  appendix \ref{app C}) 
$$ \varphi_{poly} \star \varphi_{rest}=\varphi_{rest} \star \varphi_{poly} = 0$$
$$\varphi^{k^n \bar k^{\bar n}} \star \varphi^{k^m \bar k^{\bar m}}= \varphi^{k^{m+n} \bar k^{\bar n +\bar m}} \ .$$
Finally, we calculate all products $\varphi_{rest} \star \varphi_{rest}$. The result is:
\begin{equation} \label{direct sum}
(A(V),*)=  P_0^2 \bigoplus \Bigl({\oplus \atop  {\tiny \begin{array}{c} j\!=\!1,2 \\
k\!=\!0,1,2 \end{array}  }} End M_0^{(T_j,k)}  \Bigr) \bigoplus Alg
\end{equation}
with $Alg$ being the 4-dimensional algebra spanned by $\varphi_1,\varphi_3,\varphi^{(0,1)},\varphi^{(0,2)}$
with the product structure
\begin{eqnarray}
\varphi_1 \star \varphi^{(0,1)} = 0 &\qquad& \varphi^{(0,1)} \star \varphi_1 = \varphi_1 \nonumber \\
\varphi_3 \star \varphi^{(0,1)} = \varphi_3 &\qquad& \varphi^{(0,1)} \star \varphi_3 = 0 \nonumber \\
\varphi_1 \star \varphi^{(0,2)} = \varphi_1
 &\qquad& \varphi^{(0,2)} \star \varphi_1 = 0 \nonumber \\
\varphi_3 \star \varphi^{(0,2)} = 0
 &\qquad& \varphi^{(0,2)} \star \varphi_3 = \varphi_3 \nonumber \\
 \varphi^{(0,1)}  \star \varphi^{(0,1)} = \varphi^{(0,1)} &\qquad& \varphi^{(0,2)}  \star \varphi^{(0,2)} = \varphi^{(0,2)} \label{prod structure}
\end{eqnarray}
and all other products being zero.

To classify representations of (\ref{direct sum}), which are generated by one vector, we note that $A(V)$ contains the projectors $\varphi^{k^0 \bar k^{\bar 0}}$, $\varphi^{T_1, id} +\varphi^{T_2,\bar {id}}$ and  $ \varphi^{(0,1)}+ \varphi^{(0,2)}$ onto $P_0^2$, $\oplus_{j,k} End M_0^{(T_j,k)}$ and $Alg$ respectively.
Hence all invariant subspaces can be decomposed into the invariant subspaces of the components of the direct sum. So all representations generated by one vector (i.e. quotients of the free module and an invariant subspace) can be decomposed into  representations of the components, as the quotient respects the direct sum. Hence we only need to look for representations of the components, of which only $Alg$ is new to us.

\subsection{Representations of $Alg$} \label{sec 7.4}
To find invariant subspaces of the free module $Alg_f$ of $Alg$, we identify the unit $\varphi^{(0,1)} \!+ \!\varphi^{(0,2)}$ with the generating vector as usual and hence $Alg_f$ with $Alg$ as vector spaces. The free module is 4-dimensional. It is a simple exercise to compute all 3,2,1-dimensional invariant subspaces, simply by looking at the possible orbits of an arbitrary vector in $Alg_f$. As demonstrated in appendix \ref{subspace classification}, we obtain:\\
\\
{\it Proposition:} $\qquad $ The only {\it 3-dimensional} submodules are $span(\varphi^{(0,2)}, \varphi_1,\varphi_3)$ and $span(\varphi^{(0,1)}, \varphi_1,\varphi_3)$. The only {\it 2-dimensional} submodules are $span(\varphi_1,\varphi_3)$, $V_{\lambda}:=span(\varphi_1,\varphi^{(0,2)}+\lambda \varphi_3)$ and $V^{\lambda}:=span(\varphi_3,\varphi^{(0,1)}+\lambda \varphi_1)$. The only {\it 1-dimensional} submodules are $span( \varphi_1)$ and $span( \varphi_3)$.\\
\\
Taking the quotient of $Alg_f$ with these subspaces, we obtain {\it all} representations of $Alg$ generated by a single vector:\\
\\
{\it Corollary:} $\quad$ The {\it 1-dimensional} representations are just the two known irreducible representations $M_0^{(0,1)}$ and  $M_0^{(0,2)}$. The {\it 2-dimensional} representations are $Alg_f/span(\varphi_1,\varphi_3) = M_0^{(0,1)} \bigoplus M_0^{(0,2)}$, $Alg_f/V_{\lambda}$ and  $Alg_f/V^{\lambda}$.
The {\it 3-dimensional } representations are  $Alg_f/span( \varphi_1) \cong Alg_f/V_{0} \bigoplus M_0^{(0,2)}$ and $Alg_f/span( \varphi_3) \cong Alg_f/V^{0} \bigoplus M_0^{(0,1)}$. The {\it 4-dimensional } representation  $Alg_f$ decomposes into the two invariant subspaces $V_0$, $V^0$ and $Alg_f \cong V_0 \bigoplus V^0 \cong Alg_f/V^0 \bigoplus   Alg_f/V_0$.

Note that this decomposition of $Alg_f$ is not respected by some invariant subspaces (e.g. $V_\lambda$). This is because the projection operators corresponding to this decomposition are missing in $Alg$.

$Alg_f/V^{\lambda}$ arises from $Alg_f/V_{\lambda}$ if we interchange the role of $b$ and $\bar b$, so we do not need to discuss it seperately. Further we  see in appendix \ref{pf of equivalence} by explicit computation that $Alg_f/V_{\lambda}$ and   $Alg_f/V_0$ are equivalent representations. So we only have one genuinely new representation, namely $Alg_f/V_0$.

\subsection{$Alg_f/V_0$}
As a vector space, $Alg_f/V_0 = \Cop^2$. In matrix notation, the action of $Alg$ on the space (with choice of basis as in appendix \ref{pf of equivalence}) is 
\begin{eqnarray} \varphi^{(0,2)} = \left( \begin{array}{cc} 0&0\\0&1 \end{array} \right) &\qquad& \varphi^{(0,1)} = \left( \begin{array}{cc} 1&0\\0&0 \end{array} \right) \nonumber \\
 \varphi_1 = \left( \begin{array}{cc} 0&0\\0&0 \end{array} \right) &\qquad& \varphi_3 = \left( \begin{array}{cc} 0&0\\1&0 \end{array} \right) \quad . \label{alg structure}
\end{eqnarray}
By Zhu's Theorem, there is a representation $M^{new}$ of $(H^2)_0$ corresponding to this representation of $Alg$ (and hence $A((H^2)_0)$). To reconstruct this representation, we need to reconstruct the correlation functions. We look at  correlation functions involving fields of type $(n_1,m_1),\dots,(n_k,m_k)$ with $\Delta:=n_1+\dots+n_k-m_1-\dots-m_k$ ($\Delta = 0 \ mod \ 3$). Looking at (\ref{alg structure}) and the general construction in \cite{zhu}, we already see that correlation functions with $\Delta \geq 3$ or $\Delta \leq -6$ vanish, and those with $\Delta = 0$ are the same as in   $M^{(0,1)} \bigoplus M^{(0,2)}$. However those with $\Delta = -3$ are in general  non-zero and the new feature of this representation $M^{new}$.

At this point of the argument  the existence of this representation is only a conjecture, however we know that this (and its partner obtained by interchanging $b$ and $\bar b$) are the {\it only} possible candidates for new reconstructible representations. We would like to emphasize the unusual feature of $M^{new}$, that it is reducible (with single irreducible component being $M^{(0,2)}$) but indecomposable. This is only possible for a non-unitary representation\footnote{We already had (somewhat trivial) non-unitary representations associated with the polynomial algebra. Compared to these, our  new representations are somehow unusual, as they arise from a finite dimensional factor of $A(V)$ and are still indecomposable.}. {\it So $M^{new}$ (if it exists) is a non-unitary representation of a unitary vertex operator algebra.} Before we prove the existence of this representation we have to get some feeling for the correlation functions.

By the proof of Zhu's theorem, we only need  complete knowledge of the 1-point-functions for $\Delta=-3$ to construct the representation (at least in principle). At first this seems to be again a tedious exercise on the computer, as for each $\psi \in V$ we have to know to which representant in $A(V)$ it will be reduced, to calculate the action on $Alg_f/V_0$. In general we can only hope to spot a nice pattern from such an exercise. Here we are indeed lucky and the results for low weights (and number of oscillators) are compatible with
\begin{equation} \label{computer and pi}
\Biggl\langle {0 \choose 1}, \Bigl(b_{-n_1} \dots b_{-n_N} \bar b_{-m_1} \dots \bar b_{-m_M} \vert 0 \!\!> \Bigr)_0 {1 \choose 0}  \Biggr\rangle=\pi ( b_{-n_1} \dots b_{-n_N} \bar b_{-m_1} \dots \bar b_{-m_M} \vert 0 \!\!>)
\end{equation}
where the linear functional $\pi : V \rightarrow \Cop$ is defined by linear extension of
$$ \pi ( b_{-n_1} \dots b_{-n_N} \bar b_{-m_1} \dots \bar b_{-m_M} \vert 0 \!\!>) := \frac{9}{4} \delta_{N,0} \delta_{M,3} \Biggl( {-\!1\!-\!1 \choose m_1\!\!+\!\!m_2\!\!-\!\!1} {-\!1\!+\!1 \choose m_3\!-\!1} $$
\begin{equation}
+ {-\!1\!-\!1 \choose m_2\!\!+\!\!m_3\!\!-\!\!1} {-\!1\!+\!1 \choose m_1\!-\!1} + {-\!1\!-\!1 \choose m_3\!\!+\!\!m_1\!\!-\!\!1} {-\!1\!+\!1 \choose m_2\!-\!1} \Biggr) \label{defn of pi}
\end{equation}
and $\langle , \rangle$ denotes the bilinear form (correlation function).
We have written (\ref{defn of pi}) to exhibit the similarity of $\frac{4}{9} \pi(\bar b_{-m_1} \bar b_{-m_2} \bar b_{-m_3} \vert 0 \!\!>)$ to the following correlation function obtained from ordinary vertex operators:
\begin{equation}
\Biggl\langle \bar b_{-1} \vert 0 \!\!> , \Bigl(b_{-\!m_1\!-\!m_2} \bar b_{-m_3} \vert 0 \!\!> \!\!+b_{-\!m_2\!-\!m_3} \bar b_{-m_1} \vert 0 \!\!> \!\!+b_{-\!m_3\!-\!m_1} \bar b_{-m_2} \vert 0 \!\!> \Bigr)_0 b_{-1} \vert 0 \!\!> \Biggr\rangle \ . \label{suggestive vops}
\end{equation}
However, so far we are only able to point out such a correspondence for 1-point-functions.

\subsection{Proof of the Conjecture} \label{pf of repnconj}
Now we have developed enough intuition to prove that $\varphi_3 \not\in O(V)$. First (\ref{computer and pi}) suggests, that $\pi$ is just a measure of the $\varphi_3$-component. So if $\varphi_3 \not\in O(V)$, we expect $\pi$ to vanish on all of $O(V)$. This is what we show now:\\
\\
{\it Theorem:} $\pi (a \Box b) = 0 $ for $a,b \in V$. \\
{\it Proof:} Because of the Kronecker symbols in (\ref{defn of pi}), the only chance for $\pi(a \Box b)$ to be non-zero occurs for $[type(a),type(b)]$ being
$[(N,M+3),(M,N)]$, $[(N,M+2),(M,N+1)]$, $[(N,M+1),(M,N+2)]$ or $[(N,M),(M,N+3)]$ with $N,M \in \Nop$. For each of these we only need to look at terms involving $N+M$ contractions. 

We demonstrate the case of
\begin{eqnarray*}
a&=& b_{-n_1} \dots b_{-n_N} \bar b_{-\bar m_1} \dots \bar b_{-\bar m_M} \bar b_{-r} \bar b_{-s} \bar b_{-t} \vert 0 \!\!> \\
b&=& b_{-m_1} \dots b_{-m_M} \bar b_{-\bar n_1} \dots \bar b_{-\bar n_N} \vert 0 \!\!> \qquad (N \equiv M \, mod \ 3) \ .
\end{eqnarray*}
The other cases are similar and discussed in  appendix \ref{other cases of pf}.

Now $\pi(a \Box b)$ is a sum of a large number of terms, each corresponding to a different way the contractions appear. All of these terms vanish individually, and we show this for one of the equivalent ways to contract: We assume the parts of $V(a,z)$ stemming from $b_{-n_i}$, $\bar b_{-\bar m_j}$ to be contracted with $\bar b_{-\bar n_i}$, $b_{-m_j}$ respectively and denote this by $\Box_{con}$ instead of $\Box$. Similarly, evaluating $\pi$ on the result gives three (equivalent) ways of combining the terms from two of $\bar b_{-r},\bar b_{-s}, \bar b_{-t}$ to ${-2 \choose m_1+m_2-1}$ and one to ${0 \choose m_3-1}$. We only look at the term coming from evaluating the $\bar b_{-t}$ term as ${0 \choose m_3-1}$. We denote this way of picking a term by $\tilde \pi$ instead of $\pi$. We obtain
\begin{eqnarray} 
\tilde \pi(a \Box_{con} b)&=&\alpha \sum_{k=0}^{wt\,a} {wt\,a \choose k} \sum_{u=r}^{wta\!+\!wtb\!-\!s\!-\!k} {u\!\!-\!\!1 \choose r\!\!-\!\!1} {wta\!\!+\!\!wtb\!\!-\!\!k\!\!-\!\!u\!\!-\!\!1 \choose s\!\!-\!\!1} {0 \choose t\!\!-\!\!1} * \nonumber \\
&&*\, \tilde \pi(\bar b_{-u}\bar b_{-wt\,a-wt\,b+k+u}\bar b_{-1} \vert0 \!\!>) \label{monster expression} 
\end{eqnarray}
with 
$$\alpha=\bar n_1 {-\!\bar n_1\!\!-\!\!1 \choose n_1\!\!-\!\!1} \dots \bar n_N {-\!\bar n_N\!\!-\!\!1 \choose n_N\!\!-\!\!1}\, m_1 {-\!m_1\!\!-\!\!1 \choose \bar m_1 \!\!-\!\!1 } \dots  m_M {-\!m_M\!\!-\!\!1 \choose \bar m_M \!\!-\!\!1 } \ .$$
Using the two combinatorial identities (\ref{prud36},\ref{prud47}) from appendix \ref{binomial identities}, we can rewrite (\ref{monster expression}) as 
\begin{eqnarray} 
\tilde \pi(a \Box_{con} b)&=&\frac{9}{4}\alpha \sum_{k=0}^{wt\,a} {wt\,a \choose k} 
\delta_{t,1} (-)^{wta\!+\!wtb\!-\!k\!-\!1}(wta\!+\!wtb\!-\!k) * \nonumber\\
&& * \, \sum_{u=0}^{wta\!+\!wtb\!-\!s\!-\!k\!-\!r} {u\!\!+\!\!r\!\!-\!\!1 \choose r\!\!-\!\!1} {wta\!\!+\!\!wtb\!\!-\!\!s\!\!-\!\!k\!\!-\!\!r\!\!-\!\!u\!\!+\!\!s\!\!-\!\!1 \choose s\!\!-\!\!1} \nonumber \\
&=&\frac{9}{4}\alpha \sum_{k=0}^{wt\,a} {wt\,a \choose k} 
\delta_{t,1} (-)^{wta\!+\!wtb\!-\!k\!-\!1}(wta\!+\!wtb\!-\!k) * \nonumber \\
&& * \, {wta\!+\!wtb\!-\!k\!-\!1 \choose r+s-1} \nonumber \\
&=&\frac{9}{4}\delta_{t,1} (-)^{wtb\!-\!1} (r\!+\!s) \alpha \sum_{k=0}^{wt\,a} {wt\,a \choose k} (-)^k {k\!+\!wtb \choose r+s} \nonumber \\
 &=&\frac{9}{4}\delta_{t,1} (-)^{wtb\!-\!1} (r\!+\!s) \alpha (-)^{wta} \delta_{wta,r\!+\!s} \nonumber \\
&=& 0 \qquad \qquad \qquad \qquad\mbox{as} \qquad r\!+\!s < wt\,a \quad .  \label{first case5}
\end{eqnarray}
{\it qed.}\\
\\
As  $\pi(\varphi_3) =1$, we can conclude  $\varphi_3 \not\in O(V)$. Hence our conjectures are proven to  hold and $M^{new}$ is indeed a representation of $V$.
[Note that the basis (\ref{z3 basis choice}) is such that $\pi$ vanishes on all vectors of this basis apart from $\varphi_3$, as is easily checked.] Now it is also simple to see that (\ref{computer and pi}) holds in general: we know that  (\ref{computer and pi}) is true for all vectors of (\ref{z3 basis choice}) and $O(V)$. As (by the reduction procedure) every vector can be written as a linear combination of vectors from (\ref{z3 basis choice}) and a part in $O(V)$, we can conclude that (\ref{computer and pi}) holds on all of $V$. So we have:\\
\\
{\it Theorem:}
\begin{itemize}
\item $(A(V),*)=  P_0^2 \bigoplus \Bigl({\oplus \atop  {\tiny \begin{array}{c} j\!=\!1,2 \\
k\!=\!0,1,2 \end{array}  }} End M_0^{(T_j,k)}  \Bigr) \bigoplus Alg$\\
 with $dim \, Alg = 4$ and products as in (\ref{prod structure}). 
\item $(H^2)_0$ has an indecomposable reducible representation $M^{new}$ with two dimensional space of ground states $span(e_1,e_2)$. Its 1-point-functions are given by
\begin{eqnarray*}
\lefteqn{ \left( \begin{array}{cc} \langle e_1',Y(\psi,z) e_1 \rangle & \langle e_1',Y(\psi,z) e_2\rangle \\  \langle e_2',Y(\psi,z) e_1 \rangle  & \langle e_2',Y(\psi,z) e_2 \rangle  \end{array} \right)= }\\ 
&& \left( \begin{array}{cc} \langle (b_{-1} \vert 0 \!\!>)',Y(\psi,z) b_{-1} \vert 0 \!\!> \rangle_{M^{(0,1)}}  & 0 \\  z^{-wt\psi } \pi(\psi) & \langle (\bar b_{-1} \vert 0 \!\!>)',Y(\psi,z) \bar b_{-1} \vert 0 \!\!> \rangle_{M^{(0,2)}} \end{array}\right)
\end{eqnarray*}
with primes denoting corresponding dual basis vectors and with $\pi$ as in (\ref{defn of pi}). $\langle.,.\rangle_{M^{(0,i)}}$ denotes the correlation functions in $M^{(0,i)}$.
\item
All  reconstructible representations that are finitely generated from the ground states are direct sums of $M^{new}$, $\bar M^{new}$ (obtained from $M^{new}$ by exchanging the role of $b$ and $\bar b$), $M^{(T_i,j)}$, $M^{(0,j)}$ and $M^{(k)}$ (with (k) labeling a finitely generated representation of the polynomial algebra $P^2_0$).
\end{itemize}

\section{Conclusion and Outlook}
We demonstrated how to explicitly calculate Zhu's algebra  for $H$, $H(\sqrt{2} \Zop)$, $H_+$, $H(\sqrt{2} \Zop)_+$, $(H^d)_+$ and $(H^2)_0$.
In all but the last case, our result confirms that the known representations by
vertex operators  constructed from integer or half integer moded oscillators comprise the complete set of reconstructible representations.

For the $\Zop_3$ projection $(H^2)_0$ our result for Zhu's algebra is rather surprising. We proved the existence of two new non-unitary reducible representations. These and the previously known representations, which use $\frac{1}{3} \Zop$ moded oscillators, provide all components of finitely generated reconstructible representations.

When calculating $A(V)$ for  FKS lattice theories and their $\Zop_2$ or $\Zop_3$ projections, we can, so far, only treat one lattice at a time. However the simplicity of the expected result suggests, that there should be a simple argument treating all lattices simultaneously. 

Further we would like to get a better understanding for the appearance of the new non-unitary representation of $(H^2)_0$. In particular an understanding in terms of vertex operators based on (integer moded) oscillators, as suggested in 
(\ref{computer and pi},\ref{defn of pi},\ref{suggestive vops}), is desired. This is currently pursued by the author.
\\
\\
\\
{\it Acknowledgements:} \\
I would like to thank Peter Goddard for many discussions and encouragement and Matthias Gaberdiel for comments about non-reconstructible representations and representations of polynomial algebras. I am grateful to the Studienstiftung des Deutschen Volkes and the German Academic Exchange Service (HSPII) for funding this work via PhD-studentships.
Explicit calculations were performed on computers purchased on EPSRC
grant GR/J73322.

\appendix

\section{Tensor Products} \label{tensor appendix}
In this appendix we show that $A(V_1 \bigotimes \dots \bigotimes V_n) = A(V_1) \bigotimes \dots \bigotimes A(V_n)$ as associative algebras. As vector spaces this holds by\\
\\
{\it Lemma:}  
\begin{eqnarray*}
O(V_1 \otimes \dots \otimes V_n) & = span \Bigl(&  O(V_1) \otimes V_2 \otimes \dots \otimes V_n \ , \ V_1 \otimes O(V_2) \otimes \dots \otimes V_n \ , \\ && \dots, \ V_1 \otimes \dots  \otimes V_{n-1} \otimes O(V_n) \Bigr) 
\end{eqnarray*}
\\
{\it Proof:}\\
rhs $\subset$ lhs by
$$(w_1 \Box v_1) \otimes v_2 \otimes \dots \otimes v_n = (w_1 \otimes \vert 0> \otimes \dots \otimes \vert 0> ) \Box ( v_1 \otimes v_2 \otimes \dots \otimes v_n) $$
Conversely for 
$$v=v_1^{(1)} \otimes  \dots \otimes v_n^{(1)} +\dots + v_1^{(k)}  \otimes \dots \otimes v_n^{(k)}$$
$$w=w_1^{(1)} \otimes  \dots \otimes w_n^{(1)} + \dots + w_1^{(l)}  \otimes \dots \otimes w_n^{(l)}$$
we have
\begin{eqnarray*}
v \Box w &\!\!\!\!=& \!\!\sum_{i,j} (v_1^{(i)} \otimes  \dots \otimes v_n^{(i)}) \Box (w_1^{(j)} \otimes  \dots \otimes w_n^{(j)}) \\
&\!\!\!\!=& \!\!Res_z (V(v_1,z) w_1) \otimes (V(v_2,z) w_2) \otimes \dots \otimes (V(v_n,z) w_n) \frac{(z\!\!+\!\!1)^{wt  v_1 +\dots + wt  v_n}}{z^2} \\
&\!\!\!\!=&\!\!\!\!\!\!\!\!\!\!\!\!\!\!\!\!\!\!\!\! \sum_{\tiny \begin{array}{c} k_1,\dots,k_n \in \bbbz \\
                          k_1+\dots +k_n=-1
           \end{array}}\!\!\!\!\!\!\!\!\!\!\!\!\!\!\!\!\!
Res_{z_1}\Bigl(V(v_1^{(i)},z_1) w_1^{(j)} \frac{(z_1\!\!+\!\!1)^{wt  v_1^{(j)}}}{z_1} z_1^{k_1} \Bigr) \otimes \dots \otimes Res_{z_n}\Bigl(V(v_n^{(i)},z_n) w_n^{(j)} \frac{(z_n\!\!+\!\!1)^{wt  v_n^{(j)}}}{z_n} z_n^{k_n} \Bigr) . 
\end{eqnarray*}
Now any component with $k_l < 0$ is in $V_1 \otimes \dots \otimes O(V_l) \otimes \dots \otimes V_n$ and as at least one of the $k$'s must be negative to obtain a negative sum, we have established that lhs $\subset$ rhs.\\
{\it qed}\\
\\
We finally show that the product structure is respected:\\
\\
\leftline{$(v_1 \otimes  \dots \otimes v_n) \star (w_1 \otimes  \dots \otimes w_n) $}
\begin{eqnarray*}
&=& \!\!\!\!\!\!\!\!\!\!\!\!\!\!\!\!\! \sum_{\tiny \begin{array}{c} k_1,\dots,k_n \in \bbbz \\
                          k_1+\dots +k_n=0
           \end{array}}\!\!\!\!\!\!\!\!\!\!\!\!\!\!
Res_{z_1}\Bigl(V(v_1,z_1) w_1 \frac{(z_1\!\!+\!\!1)^{wt  v_1}}{z_1} z_1^{k_1} \Bigr) \otimes \dots \otimes Res_{z_n}\Bigl(V(v_n,z_n) w_n \frac{(z_n\!\!+\!\!1)^{wt  v_n}}{z_n} z_n^{k_n} \Bigr) \\
& \equiv & \!\!\!\!\!\!\!\!\!\!\!\!\!\!\!\!\! \sum_{\tiny \begin{array}{c} k_1,\dots,k_n \geq 0 \\
                          k_1+\dots +k_n=0
           \end{array}}\!\!\!\!\!\!\!\!\!\!\!\!\!\!
Res_{z_1}\Bigl(V(v_1,z_1) w_1 \frac{(z_1\!\!+\!\!1)^{wt  v_1}}{z_1} z_1^{k_1} \Bigr) \otimes \dots \otimes Res_{z_n}\Bigl(V(v_n,z_n) w_n \frac{(z_n\!\!+\!\!1)^{wt  v_n}}{z_n} z_n^{k_n} \Bigr) \\
&=& (v_1 \star w_1) \otimes \dots \otimes (v_n \star w_n) .
\end{eqnarray*}

\section{Some Lemmas on Associative Algebras} \label{associative appendix}
In this appendix we list a few simple results on representations of associative algebras, that are relevant for understanding the general properties of Zhu's algebra.\\
Let ($A,\star,1$) be an associative algebra with unit.\\
\\
{\it Proposition B1:}
Let $A_1$ be the set of all elements of $A$ that are vanishing on {\it all} representations of $A$. Then $A_1=\lbrace 0 \rbrace$.\\
\\
{\it Proof:} Look at the adjoint representation of $A$. Any non-zero $a_1 \in A_1$ acts on the unit $1$ as $a_1 \star 1 = a_1$, hence non-trivially, which contradicts the assumption that it acts trivially on all representations. So $A_1=\lbrace 0 \rbrace$.\\
{\it qed.}\\
\\
The statement of this was suggested to me by M. Gaberdiel.\\
\\
\\
{\it Proposition B2:} Let $A_1 \neq \lbrace 0 \rbrace$ be the set of all elements of $A$ that act trivially on all {\it known} representations of $(A,\star,1)$. Then $A_1$ is an ideal by the representation property. If $A_1$ has a non-trivial finite  dimensional irreducible representation $M$, then this can be extended to a genuinely new representation  of $(A,\star,1)$.\\
\\
{\it Proof:}
We define $\bar A_1$ to be the ideal with trivial action on $M$. $\tilde A_1 := \frac{A_1}{\bar A_1}$ is isomorphic to $End M$ by the irreducibility of $M$. [This holds as irreducibility delivers all $\vert e_i >< e_j \vert $ and in the finite dimensional case these span $End M$.] So there is an element $u \in A_1$ acting as the identity on $M$. We denote the action of $a_1 \in A_1$ on $M$ as $(a_1)_{\tilde 0}$. For $a \in A$ we define the action $(a)_0$ on $M$ as 
$$(a)_0 := (u \star a)_{\tilde 0} \quad .$$
The rhs is defined, as the action of $A_1$ on $M$ is defined. For $a \in A_1$ this is consistent as $(u \star a)_{\tilde 0} = (u)_{\tilde 0} (a)_{\tilde 0} = (a)_{\tilde 0} $.
To show the representation property, we observe for $a_1 , a_2 \in A$
$$ (a_1)_0 (a_2)_0 = (u \star a_1)_{\tilde 0} (u \star a_2)_{\tilde 0} = ( (u \star a_1) \star (u \star a_2) )_{\tilde 0} = ( ( (u \star a_1) \star u ) \star a_2)_{\tilde 0} $$
As $u \star a_1 \in A_1$, we have $(u \star a_1) \star u = u \star a_1 \ mod \ \bar A_1$ and hence $((u \star a_1) \star u) \star a_2) = (u \star a_1) \star a_2 \ mod \ \bar A_1$. So we obtain
$$(a_1)_0 (a_2)_0 =(  (u \star a_1)  \star a_2)_{\tilde 0} = ( u \star (a_1 \star a_2) )_{\tilde 0} = ( a_1 \star a_2 )_0 \ .$$
This is the representation property.\\
{\it qed.}\\
\\
\\
\\
{\it Proposition B3:} Let $V_i$ ($i \in I$)  be a finite collection of finite dimensional vector spaces.  The set of inequivalent irreducible representations of the associative algebra with unit $\bigoplus End  V_i$ is $\lbrace V_i :i \in I \rbrace$. \\
\\ 
{\it Proof:} The free module is constructed by the free action of the algebra on a base vector $\vert 1>$. We can identify the free module with  $\bigoplus End  V_i $ itself by identifying  the base vector with the unit element. Irreducible modules are the quotients of the free module with maximal invariant proper submodules. These quotients are just $End V_i$ divided by a maximal invariant proper submodule of
 $End V_i$, as the projectors $id_i$ are contained in the algebra. Going in the standard basis of $End V_i$ (in the algebra and the representation space) $\lbrace E_{k,l}^{(i)} \rbrace$, we observe
\begin{equation} \label{matrix algebra}
E_{m,n}^{(j)} E_{k,l}^{(i)}= \delta_{n,k} \delta_{i,j} E_{m,l}^{(i)}
\end{equation}
So we have invariant subspaces if we fix $l$. Maximal invariant proper submodules contain all $l$'s apart from one\footnote{This $l$ does not need to correspond to our choice of basis. It  can refer to any linear combination $\sum_{j} \alpha_j e_j^{(i)}$ ($\lbrace e_j^{(i)} \rbrace$ the basis of $V_i$). }, so the quotient modules $V_{i,l}$ are isomorphic to
$$ span ( E_{m,l}^{(i)} : m \ \mbox{varies})$$
Now the actions of  $\bigoplus End  V_i$ on $V_{i,l}$, $V_{i,l'}$ are equivalent (by (\ref{matrix algebra})) and indeed equivalent to the action on $V_i$. This proves the proposition. \\
{\it qed} \\
\\
If $V_i$ is allowed to be infinite dimensional, we can keep the statement, but have to comment on the proof:
The basis of $ End  V_i$ is no longer just $\lbrace E_{k,l}^{(i)} \rbrace$, as infinite linear combinations $\sum_{j,k} \alpha_{j,k} E_{j,k}^{(i)}$, with $\alpha_{j,k} \neq 0$ only for a finite number of $j$'s for each $k$, are clearly also contained in $ End  V_i$. In an irreducible representation we can however still use the argument, as we can first apply $E_{j,j}^{(i)}$ to fix $j$ and then start to create a representation on that. We again obtain only equivalent sets of irreducible representations $V_{i,{\bf \alpha}} = span( (\sum_{l} \alpha_{l} E_{m,l}^{(i)}) : m \quad \mbox{varies}) $.\\
\\
If there are an infinite number of representations $\bigoplus_i End  V_i$ ceases to contain a unit. So the free module can only almost be identified with the algebra. It contains one more basis vector, namely the base $\vert 1>$. However irreducible representations are still classified by the irreducible representations of the components, as we can project any representation on the parts corresponding to $End V_i$.\\
\\
\\
{\it Proposition B4:} Let $V_i$ ($i \in I$) be a finite collection of finite dimensional vector spaces. Any representation of $\bigoplus_i End V_i$ that is finitely generated is a finite direct sum of the irreducibles.\\
\\
{\it Proof:}
Let $M$ be generated by  $a_1,\dots,a_n$. 
By the proof of proposition B3, the submodule $<a_1>$, which can be identified with the quotient of the free module by  an invariant subspace, is a finite direct sum of the irreducible modules  $V_{j,l}$ ($l \in V_j$). For a given $j$, at most {\it one} $l$ appears in the decomposition, as $V_{j,l} \cong V_{j,l'}$ and $<a_1>$ is generated by one vector. So every $<a_i>$ is {\it uniquely} decomposed into $\oplus_{some \, j} \ V_{j,l_j^{(i)}}$.

Hence it is enough to give a proof for $I=\lbrace 1 \rbrace$ (we call $V:=V_1$, $V_{l^{(i)}}:= V_{l_1^{(i)}}$): now $<a_i>=V_{l^{(i)}}$ (some $l^{(i)} \in V$). Define $M_1:=V_{l^{(1)}}$. We define $M_k$ by induction: if $V_{l^{(k)}} \subset M_{k-1}$, then define $M_k:=M_{k-1}$. Otherwise $V_{l^{(k)}} \cap M_k = \lbrace 0 \rbrace$ (by irrducibility of $V_{l^{(k)}}$) and we define $M_k:= M_{k-1} \oplus V_{l^{(k)}}$. So by construction $M_n$ is the desired decomposition of $M$ into irreducibles.\\
{\it qed.}\\ 
\\
\\
{\it Lemma:} 
$End \ \Cop^n $ is simple as an associative algebra.\\
\\
{\it Proof:}
Let $A \neq \lbrace 0 \rbrace$ be a two-sided ideal. We pick a non-zero element $a \in A$, which has a non-trivial  component $\lambda E_{i,j}$ for some $i,j$.  
We can project out this component as $E_{i,j}=\frac{1}{\lambda} E_{i,i} \star a \star E_{j,j} \in A$. From this we can generate the whole basis of $End \ \Cop^n$ by  $E_{m,n}= E_{m,i} \star E_{i,j}  \star E_{j,n} \in A$. So $A= End \  \Cop^n$. So $End \ \Cop^n $ does not contain any non-trivial two-sided ideal.\\
{\it qed.}\\
\\
{\it Proposition B5:}
Let $V_1,\dots, V_n$ be a finite collection of finite dimensional vector spaces.
Let $A$ be a  subalgebra of $\oplus_{i=1}^{n} End V_i$ acting irreducibly and inequivalent on the $n$ components of   $\oplus_{i=1}^{n} V_i$.
Then $A =\oplus_{i=1}^{n} End V_i$, the full algebra.\\
\\
{\it Proof:} We prove this by induction in $n$. 
For $n=1$ we note that, by the irreducibility, $A$ necessarily contains all $E_{i,j}$ and hence $A$ is the full algebra of endomorphisms as claimed. Now assume the statement is proved for ``$n-1$'':

We can project\footnote{The projections we introduce are the natural ones, defined  to respect the direct sums.}  $A$ on $ \oplus_{i=1}^{n-1} End V_i$. The projection is itself a subalgebra of $ \oplus_{i=1}^{n-1} End V_i$ which acts irreducibly and inequivalent on the components of $\oplus_{i=1}^{n-1} V_i$ and must (by induction) be the whole of $ \oplus_{i=1}^{n-1} End V_i$. We can pick a set of vectors in $A$, whose projection forms a basis of  $\oplus_{i=1}^{n-1} End V_i$. We can extend this to a basis of $A$ by vectors, whose projection is vanishing.
So we split $A$ into $A_n \oplus A^{(n-1)}$, with $A_n \subset End V_n$ and $A^{(n-1)}$ projecting on the whole of $ \oplus_{i=1}^{n-1} End V_i$ without decreasing the dimension. Hence any vector with a trivial projection on $ \oplus_{i=1}^{n-1} End V_i$ must be in $A_n$. Hence $A_n$ is a two-sided ideal of $A$.

By the irreducibility  of $V_n$ (cf. $n=1$) a projection of $A$ on $End V_n$ has to yield the full algebra  $End V_n$. Now $A_n$ is a two-sided ideal of $A$ and hence of the projection of $A$ on $End V_n$. So $A_n$ is a two-sided ideal of $End V_n$. By the previous Lemma there are only two cases:\\
(a) $A_n = End V_n$:
This implies that we can remove the projection on $End V_n$ from $A^{(n-1)}$.
We obtain $A=End V_n \oplus (\oplus_{i=1}^{n-1} End V_i) =\oplus_{i=1}^{n} End V_i$ as claimed. \\
(b) $A_n = \lbrace 0 \rbrace$:
Now $A^{(n-1)}$ is closed under the multiplication and hence the multiplication rule for the projection on $End V_n$ is determined by the multiplication rule of $\oplus_{i=1}^{n-1} End V_i$. So $A$ is isomorphic to $\oplus_{i=1}^{n-1} End V_i$ and hence has only $V_1,\dots, V_{n-1}$ as  inequivalent irreducible representations. Hence $V_n$ must be equivalent to one of these, which was excluded by assumption.\\
{\it qed.}\\
\\
\\
Now we discuss some properties of polynomial algebras and their representations. We denote the polynomial algebra in $d$ variables by $P^d$, and its restriction to linear combinations of monomials of even degrees by $(P^d)_+$ ($(P^2)_+$ is generated by $1,x^2,y^2,xy$, if $x,y$ are the variables).  More generally we study  subalgebras  $(P^d)_s$ of $P^d$ with degrees being multiples of $s \in \Nop$. 

Our goal is to show, that the representations of $(P^d)_s$ can be understood entirely by studying representations of $P^d$. This implies that all representations of our vertex operator algebras that arise from the factor $(P^d)_s$ of $A(V)$ are given by the untwisted construction of vertex operators. It is instructive to start with the case $d=1$ (denote $P:=P^1$):\\
\\
{\it Proposition B6:}
The representations of $P$ which are generated by one vector are quotients  of $P$ by  the space of polynomials that have a given polynomial $g$ as a common factor. $g$ determines the representation uniquely. \\
\\
{\it Proof:} As in the proof of proposition B3, we  identify the  free module $P_f$ with $P$ by identifying the unit with the base vector. A representation is again the quotient of $P_f$ by a submodule $U$. By the Euclidean algorithm, if $h_1,h_2 \in U$, then their largest common factor is also in $U$. So the largest common factor $g$ of {\it  all} elements of $U$ is contained in $U$ (and generates $U=Pg$). \\
{\it qed.}\\
\\ 
\\
We can explicitly give the representation matrix of the generator $x$ on $M:=P_f/Pg$ for $g=(x-a_1)^{\mu_1} \dots (x-a_n)^{\mu_n}$ $\quad (a_i \neq a_j)$: Denoting cosets by $[.]$, we choose the standard basis of $M$ as 
$$\begin{array}{rcl} \lbrace \phi_{1,1}&:=&[(x-a_1)^{\mu_1-1} (x-a_2)^{\mu_2} \dots (x-a_n)^{\mu_n}] ,\\
\phi_{2,1}&:=&[(x-a_1)^{\mu_1-2} (x-a_2)^{\mu_2} \dots (x-a_n)^{\mu_n}] ,\\ &\vdots&\\
\phi_{\mu_1,1}&:=&[\qquad \qquad \qquad (x-a_2)^{\mu_2} \dots (x-a_n)^{\mu_n}] ,\\
& \vdots&
\end{array}$$
$$\begin{array}{rcl}
& \vdots&\\
 \phi_{1,n}&:=&[(x-a_1)^{\mu_1}\dots (x-a_{n-1})^{\mu_{n-1}}  (x-a_n)^{\mu_n-1}]\\
 &\vdots&\\
\phi_{\mu_n,n}&:=&[(x-a_1)^{\mu_1}  \dots (x-a_{n-1})^{\mu_{n-1}} \qquad \qquad \qquad ] \rbrace .
\end{array}$$ 
In this basis $x$ is represented by a matrix in Jordan normal form with $n$ Jordan blocks of size $\mu_1,\dots,\mu_n$ and corresponding eigenvalues $a_1, \dots, a_n$. Irreducible representations are obtained for linear $g=x-a$. These are the evaluations of polynomials at $a \in \Cop$
\\ 
\\
{\it Proposition B7:} Every representation $M$ of $P_+$, which is generated from a single (base) vector, originates from a representation of $P$ (generated from  one vector) on $M$, or on a space containing $M$. \\
\\
{\it Proof:} $P_+$  is itself a polynomial algebra in the variable $x^2$. Hence $M$ can be written as $(P_+)_f/P_+ g $ with $g$ of the form $g=(x^2-a_1)^{\mu_1} (x^2-a_2)^{\mu_2} \dots (x^2-a_n)^{\mu_n}$. As the representation decomposes into Jordan blocks, we only need to consider the case of one block, i.e. $g=(x^2-a)^\mu$.

For $a \neq 0$, we can write g as $(x-\sqrt{a})^\mu (x+\sqrt{a})^\mu$ when regarded as an element of $P$. This defines a representation of $P$  with $x$ represented by two Jordan blocks of size $\mu$. We pick the first block corresponding to $\tilde g = (x-\sqrt{a})^\mu$ and look how the generator $x^2$ of the subalgebra  $P_+$ is represented on this $\mu$ dimensional space. We can change basis so that the matrix of $x^2$ is in Jordan normal form and see that it coincides with the matrix of $x^2$ on $(P_+)_f/P_+g$ in  the standard basis.
So the representation of $P_+$ (given by $g$)  arises by restriction of the representation of $P$ on $M$ (given by $\tilde g$). 

For $a=0$  $x$ is represented as a single Jordan block of size $2 \mu$ (and eigenvalue 0) on $P_f/gP$. Denoting the standard basis vectors by $\phi_1,\dots,\phi_{2 \mu}$, $x^2$ leaves $span(\phi_1, \phi_3,\dots ,\phi_{2 \mu -1})$ and  $span(\phi_2, \phi_4,\dots ,\phi_{2 \mu})$ invariant. Each of these subspaces may be identified with $M$ and the action of $P_+$ on them is accordingly.\\
{\it qed.}\\
\\
Now we generalize\footnote{This generalization has been suggested to me by M. Gaberdiel.} to arbitrary $d$:\\
\\
{\it Proposition B8:} Every representation $M$ of $(P^d)_s$, which is generated from a single (base) vector, originates from a representation of $P^d$ on a space containing $M$.\\
\\
{\it Proof:} $M$ is the quotient $((P^d)_s)_f/U_M$ with $U_M$ an ideal of $(P^d)_s$. Regarding $U_M$ as a subspace of $P^d$, we see that $P^d U_M$ splits into $s$ parts corresponding  to different gradings modulo $s$. 
$$P^dU_M=U^{(0)}_M \oplus \dots \oplus U^{(s-1)}_M \qquad \mbox{with} \ U^0_M=U_M . $$
In the same way $P^d$ splits as 
$$P^d={P^d}^{(0)} \oplus \dots \oplus {P^d}^{(s-1)} \qquad \mbox{with} \ {P^d}^{(0)} = (P^d)_s .$$
We have ${P^d}^{(i)}={P^d}^{(i)}{P^d}^{(0)}$, $U_M^{(i)}={P^d}^{(i)}U_M^{(0)}$. 
We can decompose the representation $ (P^d)_f/P^dU_M$ of $P^d$ as
\begin{eqnarray}
(P^d)_f/P^dU_M&=&{P^d}^{(0)}/U^{(0)}_M \oplus \dots \oplus {P^d}^{(s-1)}/U^{(s-1)}_M \nonumber \\
&=& M \oplus \dots \oplus {P^d}^{(s-1)}/U^{(s-1)}_M . \label{mgquotient}
\end{eqnarray}
Restricting the action on this space to $(P^d)_s$, we see that each component is invariant and the action on the first component is as desired.\\
{\it qed.}\\
\\
Note that the components in (\ref{mgquotient}) are not invariant under $P^d$. Comparing this construction with the construction in proposition B7, we see that, even though they are equivalent as representations of $P_+$, the constructions are not equivalent with respect to the larger algebra $P$. So there are several {\it seemingly} different untwisted vertex operator constructions, which are isomorphic as representations of $V$. They only seem different, as the incarnations of $a_0$, they are based on, are inequivalent. 

The construction (\ref{mgquotient}) is yet too redundant to allow a generalization to finitely generated representations. We introduce the following equivalence relation: for $p \in {P^d}^{(i)} \quad (i \neq 0)$ we say $p \sim 0$ iff $p'p \in U_M$ for all $p' \in {P^d}^{(s-i)}$. We obtain the desired representation of $P^d$ if we take the quotient of (\ref{mgquotient}) by this equivalence relation:
\begin{equation}
\frac{(P^d)_f}{P^dU_M,\sim} = M \oplus \frac{{P^d}^{(1)}}{U^{(1)}_M,\sim} \oplus \dots \oplus \frac{{P^d}^{(s-1)}}{U^{(s-1)}_M,\sim} \ . \label{myquotient}
\end{equation}
Suppose $M$ is a representation of $(P^d)_s$ generated from two vectors $1_1, 1_2$ and the spaces they generate are $M_1, M_2$ (with associated $U_{M_1}, U_{M_2}$) with overlap $S:=M_1 \cap M_2$. 
We define $\tilde M_1:=\frac{(P^d)_f}{P^dU_{M_1},\sim}$ and  $\tilde M_2:=\frac{(P^d)_f}{P^dU_{M_2},\sim}$.  The task is now to join these two spaces consistently, so that the overlap region $S$ is identified and the glued space remains a representation of $P^d$. This is done as follows:
First we define $S_1^{(i)}:= {P^d}^{(i)} S \subset 
\frac{{P^d}^{(i)}}{U^{(i)}_{M_1},\sim}$ and $S_1:=S \oplus S_1^{(1)} \oplus \dots \oplus S_1^{(s-1)} $. Similarly we define $S_2^{(i)}, S_2$.
Now $S_1, S_2$ are invariant under the action of $P^d$. To show that they are isomorphic, we simply observe that for $v_1,\dots,v_n \in S$ and $p_1,\dots,p_n \in {P^d}^{(i)}$: 
\begin{eqnarray}
\sum_{j=1}^n p_j v_j = 0 && \mbox{ as an element of $\tilde M_1$} \qquad \mbox{iff} \nonumber \\
q \sum_{j=1}^n p_j v_j = 0 \quad \mbox{for all} \quad  q \in {P^d}^{(i)} && \mbox{regarded as an equation in  $S \subset \tilde M_1$} \qquad \mbox{iff} \nonumber \\
q \sum_{j=1}^n p_j v_j = 0 \quad \mbox{for all} \quad  q \in {P^d}^{(i)} && \mbox{regarded as an equation in $S \subset \tilde M_2$} \qquad \mbox{iff} \nonumber \\
\sum_{j=1}^n p_j v_j = 0 && \mbox{ as an element of $\tilde M_2$} \ . \label{reasoning using quotient}
\end{eqnarray}
The first and last ``iff'' hold because we regard quotients by the relation $\sim$. This proves that $S_1, S_2$ are isomorphic. So we can idendify them consistently as in $\tilde {\tilde M} :=\tilde M_1 \oplus_{ _{S_1 \equiv S_2}} \tilde M_2$. $\tilde {\tilde M}$ carries a representation of $P^d$ and decomposes into $s$ subspaces according to the grading, the first of which is $M$.

To be able to use induction in the case that $M$ is generated from more vectors, we have to take the quotient of $\tilde {\tilde M}$ by a relation analogous to $\sim$ to enable a reasoning analogous to (\ref{reasoning using quotient}). We call this quotient $\tilde M$. Summarizing the result we obtain:\\
\\
{\it Proposition B9:} Any finitely generated representation $M$ of $(P^d)_s$ can be obtained as the restricted action on a subspace of the representation $\tilde M$ of $P^d$.

\section{A basis for $A((H^2)_0)$} \label{app C}
In this appendix we give the basis used in section \ref{z3conj} for $A((H^2)_0)$. The basis is chosen in a way, so that the zero modes act in a simple way on certain representations ($T_1,T_2,(k),(0,1),(0,2)$) (with which the basis vector is labeled) and vanish on others;  in the case of ``momentum representations'' $(k)$ we demand the action to be a certain monomial in the components of the momentum. In the case of $T_1, T_2$ we label the vectors with the operator corresponding to the action. All the remaining freedom is removed by demanding a simple product structure, i.e. the choice of basis should respect the decomposition into direct sums, which is possible (cf section \ref{z3conj}). Our basis vectors are ($\varphi_1, \varphi_3$ and):

\begin{eqnarray*}
\varphi^{T_1,c_{\!-\!{1\over 3}}c_{\!-\!{1\over 3}}c_{{\!1\over 3}} c_{{\!1\over 3}}}      &:=&
   10425/4   [   2   ;   1   ]
+   18564   [   3   ;   1   ]
+   42573   [   4   ;   1   ]
+   71415/2   [   5   ;   1   ]\\
&&   -14013   [   7   ;   1   ]
   -19683/4   [   8   ;   1   ] 
+   447   [   1,1   ;   1,1   ]\\
&&+   29409/8   [   2,1   ;   1,1   ]
+   28347/4   [   3,1   ;   1,1   ] 
+   1377/2   [   5,1   ;   1,1   ]\\
&&+   36369/8   [   4,1   ;   1,1   ] \\
        \varphi^{T_1,c_{\!-\!{1\over 3}} c_{{\!1\over 3}}}   &:=&
   22599   [   2   ;   1   ]
+   156411   [   3   ;   1   ]
+   692955/2   [   4   ;   1   ]
+   278154   [   5   ;   1   ]\\
&&   -95013   [   7   ;   1   ] 
   -59049/2   [   8   ;   1   ] 
+   8055/2   [   1,1   ;   1,1   ]\\
&&+   64395/2   [   2,1   ;   1,1   ]
\!+   \!118305/2   [   3,1   ;   1,1   ]
\!+   \!6075/2   [   5,1   ;   1,1   ]\\
&&+   34020   [   4,1   ;   1,1   ]
\\
\varphi^{T_1,c_{\!-\!{2\over 3}} c_{{\!2\over 3}}}
                         &:=&
   94689/4   [   2   ;   1   ]
\!+   \!158112   [   3   ;   1   ]
\!+   \!669141/2   [   4   ;   1   ]
\!+   \!504387/2   [   5   ;   1   ]\\
&&   -66825   [   7   ;   1   ] 
   -59049/4   [   8   ;   1   ] 
+   4437   [   1,1   ;   1,1   ]\\
&&+   272889/8   [   2,1   ;   1,1   ]
+   234243/4   [   3,1   ;   1,1   ] 
   -1701/2   [   5,1   ;   1,1   ]\\
&&+   224289/8   [   4,1   ;   1,1   ]
\\
\varphi^{T_1,id}
                                  & :=&
   1919457/8   [   2   ;   1   ]
+   1619838   [   3   ;   1   ]
+   6953931/2   [   4   ;   1   ]\\
&&+   10700991/4   [   5   ;   1   ] 
   -1554957/2   [   7   ;   1   ] 
   -1594323/8   [   8   ;   1   ] \\
&&+   88209/2   [   1,1   ;   1,1   ]
+   5494473/16   [   2,1   ;   1,1   ] 
+   4825251/8   [   3,1   ;   1,1   ] \\
&&+   19683/4   [   5,1   ;   1,1   ]
+   4940433/16   [   4,1   ;   1,1   ]
\\
       \varphi^{T_2,\bar c_{\!-\!{1\over 3}}\bar c_{\!-\!{1\over 3}}\bar c_{{\!2\over 3}}}             &:=&
   27/2   [   2  , 2,1   ;   ]
+   9/2   [   2  , 2  , 2   ;   ]
   -9   [   3 ,  3,1   ;   ] \\
          \varphi^{T_1,c_{\!-\!{2\over 3}}c_{{\!1\over 3}} c_{{\!1\over 3}}}                         &:=&
   -9   [   1,1 ,  1   ;   ]
+   27   [   2  , 2,1   ;   ]
+   9   [   2 ,  2  , 2   ;   ]
   -9   [   3,   3,1   ;   ]
\\
                \varphi^{T_2,\bar c_{\!-\!{1\over 3}}\bar c_{\!-\!{1\over 3}}\bar c_{{\!1\over 3}} \bar c_{{\!1\over 3}}}                   &:=&
   -39417/4   [   2   ;   1   ]
   -65406   [   3   ;   1   ]
   -137193   [   4   ;   1   ]
   -204147/2   [   5   ;   1   ]\\
&&+   25353   [   7   ;   1   ] 
+   19683/4   [   8   ;   1   ] 
   -1857   [   1,1   ;   1,1   ]\\
&&   -113601/8   [   2,1   ;   1,1   ] 
   -96231/4   [   3,1   ;   1,1   ]
+   1377/2   [   5,1   ;   1,1   ]\\
&&   -88209/8   [   4,1   ;   1,1   ]
\\
 \varphi^{T_2,\bar c_{\!-\!{1\over 3}} \bar c_{{\!1\over 3}}}
                                 & :=&
   -104391/2   [   2   ;   1   ]
   -347472   [   3   ;   1   ]
   -732150   [   4   ;   1   ]
   -548532   [   5   ;   1   ]\\
&&+   141183   [   7   ;   1   ] \
+   59049/2   [   8   ;   1   ] 
   -19575/2   [   1,1   ;   1,1   ]\\
&&   -299475/4   [   2,1   ;   1,1   ] 
   -127350   [   3,1   ;   1,1   ]
+   6075/2   [   5,1   ;   1,1   ]\\
&&   -236925/4   [   4,1   ;   1,1   ]
\\
\varphi^{T_2,\bar c_{\!-\!{2\over 3}} \bar c_{{\!2\over 3}}}
                                   &:=&
   -54873/4   [   2   ;   1   ]
   -93816   [   3   ;   1   ]
   -409473/2   [   4   ;   1   ]\\
&&   -322299/2   [   5   ;   1   ]
+   51273   [   7   ;   1   ] 
+   59049/4   [   8   ;   1   ] \\
&&   -2484   [   1,1   ;   1,1   ]
   -156213/8   [   2,1   ;   1,1   ] 
   -139491/4   [   3,1   ;   1,1   ]\\
&&   -1701/2   [   5,1   ;   1,1   ]
   -149445/8   [   4,1   ;   1,1   ]
\\
\varphi^{T_2,\bar {id}}
                                   &:=&
   -2117745/8   [   2   ;   1   ]
   \!-\!1780947   [   3   ;   1   ]
   \!-\!7607115/2   [   4   ;   1   ]
   \!-\!11619531/4   [   5   ;   1   ] \\
&&
+   1633689/2   [   7   ;   1   ] 
+   1594323/8   [   8   ;   1   ] 
   -48843   [   1,1   ;   1,1   ]\\
&&   -6031017/16   [   2,1   ;   1,1   ] 
   -5213079/8   [   3,1   ;   1,1   ]
+   19683/4   [   5,1   ;   1,1   ]\\
&&   -5097897/16   [   4,1   ;   1,1   ]
\\
  \varphi^{T_1,c_{\!-\!{1\over 3}}c_{\!-\!{1\over 3}}c_{{\!2\over 3}}}                &:=&
   27/2   [   ;2  , 2,1      ]
+   9/2   [  ; 2  , 2  , 2     ]
   -9   [  ; 3 ,  3,1     ] \\
         \varphi^{T_2,\bar c_{\!-\!{2\over 3}}\bar c_{{\!1\over 3}} \bar c_{{\!1\over 3}}}                        &:=&
   -9   [   ;1,1 ,  1      ]
+   27   [   ;2  , 2,1     ]
+   9   [   ;2 ,  2  , 2     ]
   -9   [   ;3,   3,1      ]
\\
 \varphi^{k^0 \bar k^3}
                                   &:=&
   -16   [   ;1,1,   1      ]
+   78   [  ; 2 ,  2,1      ] 
+   27   [  ; 2  , 2  , 2      ]
   -36   [  ; 3 ,  3,1      ] 
\\
                       \varphi^{k^3 \bar k^0}     &:=&
   -16   [   1,1,   1   ;   ]
+   78   [   2 ,  2,1   ;   ]
+   27   [   2  , 2  , 2   ;   ]
   -36   [   3 ,  3,1   ;   ] 
\\
                       \varphi^{k^1 \bar k^1}       &:=&
   23835   [   2   ;   1   ]
+   154000   [   3   ;   1   ]
+   311010   [   4   ;   1   ]
+   218106   [   5   ;   1   ]\\
&&   -37260   [   7   ;   1   ] 
+   4620   [   1,1   ;   1,1   ]
+   67935/2   [   2,1   ;   1,1   ] \\
&&
+   53445   [   3,1   ;   1,1   ]
   -5670   [   5,1   ;   1,1   ]
+   36855/2   [   4,1   ;   1,1   ]
\\
        \varphi^{(0,1)}                         &:=&
      [   1   ;   1   ]
   -30653/8   [   2   ;   1   ]
   -25655   [   3   ;   1   ]
   -54381   [   4   ;   1   ]\\
&&   -164295/4   [   5   ;   1   ] 
+   21951/2   [   7   ;   1   ]
+   19683/8   [   8   ;   1   ] \\
&&   -1443/2   [   1,1   ;   1,1   ]
   -88941/16   [   2,1   ;   1,1   ]
  -76647/8   [   3,1   ;   1,1   ]\\
&&+   405/4   [   5,1   ;   1,1   ]
   -74277/16   [   4,1   ;   1,1   ]
\\
                \varphi^{(0,2)}                  &:=&
   19145/8   [   2   ;   1   ]
+   16321   [   3   ;   1   ]
+   70995/2   [   4   ;   1   ]
+   111267/4   [   5   ;   1   ]\\
&&   -17415/2   [   7   ;   1   ] 
   -19683/8   [   8   ;   1   ] 
+   435   [   1,1   ;   1,1   ]\\
&&+   54465/16   [   2,1   ;   1,1   ] 
\!+\!   48255/8   [   3,1   ;   1,1   ]
\!+\!   405/4   [   5,1   ;   1,1   ]\\
&&+   50625/16   [   4,1   ;   1,1   ] \\
\varphi^{k^0 \bar k^0}       &:=&
[;]-\varphi^{(0,1)} - \varphi^{(0,2)} - \varphi^{T_1, id}-
\varphi^{T_2,\bar {id}} \ .
\end{eqnarray*}
To define $\varphi^{k^n \bar k^{\bar n}}$, we first have to know the following crucial result (obtained by reducing the product in the familiar way on the 
computer)\footnote{The non-trivial information in this is, that the two sides of the equations do not contain different amounts of $\varphi_1 ,\varphi_3$.}:
\begin{eqnarray}
 \varphi^{k^1 \bar k^1} \star  \varphi^{k^0 \bar k^3} &=& \varphi^{k^0 \bar k^3} \star  \varphi^{k^1 \bar k^1}
\nonumber \\
 \varphi^{k^1 \bar k^1} \star  \varphi^{k^3 \bar k^0} &=& \varphi^{k^3 \bar k^0} \star  \varphi^{k^1 \bar k^1}
\nonumber \\
\varphi^{k^0 \bar k^3} \star \varphi^{k^3 \bar k^0} =\varphi^{k^3 \bar k^0} \star \varphi^{k^0 \bar k^3} &=& \varphi^{k^1 \bar k^1} \star  \varphi^{k^1 \bar k^1} \star  \varphi^{k^1 \bar k^1} \ . \label{app11}
\end{eqnarray}
In addition we need to know that $\varphi^{k^0 \bar k^0}$ multiplies $\varphi^{k^0 \bar k^3}$, $ \varphi^{k^3 \bar k^0}$ and $ \varphi^{k^1 \bar k^1}$ like a unit, which is an equally non-trivial fact.
This allows us to define  $\varphi^{k^n \bar k^{\bar n}}$ as the corresponding product of $\varphi^{k^1 \bar k^1}$'s, $\varphi^{k^3 \bar k^0}$'s and $\varphi^{k^0 \bar k^3}$'s (independent of the order) so that 
\begin{equation}
\varphi^{k^n \bar k^{\bar n}} \star \varphi^{k^m \bar k^{\bar m}}= \varphi^{k^{m+n} \bar k^{\bar n +\bar m}}
\end{equation}
for all $n,\bar n,m,\bar m$.

\section{Classification of Invariant Subspaces of $Alg_f$} \label{subspace classification}
Here we explicitly demonstrate the classification of invariant subspaces $U$ of the free module $Alg_f$, as outlined in section \ref{sec 7.4}. We assume that
$$ \varphi := \alpha_1 \varphi_1 +\alpha_2 \varphi_3 +\alpha_3 \varphi^{(0,1)}
+\alpha_4 \varphi^{(0,2)}$$ is an element of $U$, and we look at its orbit under the action of the algebra:
\begin{description}
\item[Case 1:] $\alpha_3,\alpha_4 \neq 0$ \\
Then 
\begin{eqnarray}
\varphi_1 \star \varphi &=& \alpha_4 \ \varphi_1 \\
\varphi_3 \star \varphi &=& \alpha_3 \ \varphi_3 \label{phi3 projection}\\
\Bigl( \varphi^{(0,2)} - \frac{\alpha_2}{\alpha_3} \varphi_3 \Bigr) \star \varphi &=& \alpha_4 \ \varphi^{(0,2)} . 
\end{eqnarray}
Hence $\varphi_1,\varphi_3,\varphi^{(0,2)},\varphi \in U$ and hence $U=Alg_f$.
\item[Case 2:] $\alpha_3 \neq 0$ and $U \neq Alg_f$\\
This implies, that for all $\tilde \varphi \in U$ the corresponding $\tilde \alpha_4$ vanishes. Using (\ref{phi3 projection}), we see that $\varphi_3 \in U$. Hence we are left with the following two possibilities:
Either $U=span(\varphi_1,\varphi_3,\varphi^{(0,1)}) $ or $U=V^\lambda:=span(\varphi_3,\varphi^{(0,1)}+\lambda \varphi_1) $ (for some fixed $\lambda$). Both are indeed invariant subspaces.
\item[Case 3:]  $\alpha_4 \neq 0$ and $U \neq Alg_f$\\
By symmetry we get $U=span(\varphi_1,\varphi_3,\varphi^{(0,2)}) $ or $U=V_\lambda:=span(\varphi_1,\varphi^{(0,2)}+\lambda \varphi_3) $.
\item[Case 4:] $\alpha_3 =\alpha_4 =0$ (for all $\varphi \in U$)\\
\begin{eqnarray*}
\varphi^{(0,1)} \star \varphi &=& \alpha_1 \ \varphi_1 \\
\varphi^{(0,2)} \star \varphi &=& \alpha_2 \ \varphi_3 .
\end{eqnarray*}
Hence we may have $U=span(\varphi_1,\varphi_3)$,  $U=span(\varphi_1)$ or $U=span(\varphi_3)$. All three  are indeed invariant subspaces.
\end{description}
This completes the classification of invariant subspaces of the free module.

\section{$ Alg_f/ V_{\lambda} \protect\cong Alg_f/ V_0$} \label{pf of equivalence}
Here we show that $Alg_f/V_\lambda$ and $ Alg_f/V_0$ are equivalent representations of $Alg$: We pick the basis of $Alg_f/V_\lambda$ for $\lambda \neq 0$ as  $e_1:= - \lambda [\varphi^{(0,1)}], e_2:=[\varphi^{(0,2)}]$. Recalling that $V_{\lambda}=span(\varphi_1,\varphi^{(0,2)}+\lambda \varphi_3)$, we can calculate the action of $Alg$ as 
$$\varphi_3 \ e_1 = - \lambda [ \varphi_3 \star \varphi^{(0,1)} ] = - \lambda [ \varphi_3 ] = [ \varphi^{(0,2)} ] = e_2$$
$$\varphi_3 \ e_2 = 0 \ ,$$
hence $\varphi_3 = \left( \begin{array}{cc} 0&0\\1&0 \end{array} \right) $ and similarly $\varphi_1 = \left( \begin{array}{cc} 0&0\\0&0 \end{array} \right) $, $\varphi^{(0,1)} = \left( \begin{array}{cc} 1&0\\0&0 \end{array} \right) $ and  $\varphi^{(0,2)} = \left( \begin{array}{cc} 0&0\\0&1 \end{array} \right) $. These are the same representation matrices as on $Alg_f/V_0$ with choice of basis $e_1:= [\varphi^{(0,1)}], e_2:=[\varphi_3]$.

\section{Two Binomial Identities} \label{binomial identities}
In section \ref{pf of repnconj}  we needed the following identities \cite{prudnikov}:
\begin{eqnarray}
\sum_{k=0}^n {a\!+\!k \choose a} {b\!+\!n\!-\!k \choose b} &=& {a+b+n+1 \choose a+b+1} \quad a,b,n \in \Nop  \label{prud36} \\
\sum_{k=0}^n (-1)^k {n \choose k} {a\!+\!k \choose m}&=& (-1)^n \delta_{n,m} \quad a \in \Cop, \  n,m \in \Nop, \ n\geq m \label{prud47}
\end{eqnarray}

\section{Remaining Cases in the Proof of the \protect\\ Conjecture}  \label{other cases of pf}
Here we outline the other three cases in the proof that $\varphi_3 \not\in O(V)$:\\
\\
First take 
\begin{eqnarray*}
a&=& b_{-n_1} \dots b_{-n_N} \bar b_{-\bar m_1} \dots \bar b_{-\bar m_M} \bar b_{-r} \bar b_{-s}  \vert 0 \!\!> \\
b&=& b_{-m_1} \dots b_{-m_M} \bar b_{-\bar n_1} \dots \bar b_{-\bar n_N}  \bar b_{-t} \vert 0 \!\!>  \qquad (N\!+\!1 \equiv M \, mod \ 3) \ .
\end{eqnarray*}
We again take only the terms corresponding to the contraction $\Box_{con}$. However, this time  $\pi$ has two types of inequivalent contributions: $\tilde \pi $ treats the term from $\bar b_{-t}$ to result in ${0 \choose t-1}$ as before, while in $\tilde {\tilde \pi}$ this role is taken by $\bar b_{-r}$. The calculation of $\tilde \pi(a \Box_{con} b)$ is identical to (\ref{monster expression},\ref{first case5}) in section \ref{pf of repnconj}. Here, however, the reasoning in the final line, that $r+s < wt\, a$, relies on the fact that $N+M>0$ here. Note that, hence, $\pi$ does {\it not} vanish on $O(H^2)$, the full two dimensional Heisenberg algebra. This explains, why our new representation did not appear as a representation of the full algebra in section \ref{sec6}. 

The reasoning for $\tilde {\tilde \pi}$ is slightly simpler (only using (\ref{prud47})):
\begin{eqnarray}
\tilde {\tilde \pi}(a \Box_{con} b)&=&\alpha \sum_{k=0}^{wt\,a} {wt\,a \choose k} \delta_{r,1} {wta\!\!+\!\!wtb\!\!-\!\!k\!\!-\!\!t\!\!-\!\!1 \choose s\!\!-\!\!1}  * \nonumber \\
&&*\, \tilde {\tilde \pi}(\bar b_{-1}\bar b_{-wt\,a-wt\,b+k+t}\bar b_{-t} \vert0 \!\!>) \nonumber\\
&=&\frac{9}{4}\alpha \sum_{k=0}^{wt\,a} {wt\,a \choose k} 
\delta_{r,1} (-)^{wta\!+\!wtb\!-\!k\!-\!1} s
 {wta\!+\!wtb\!-\!k\!-\!t \choose s} \nonumber \\
&=&\frac{9}{4}\alpha \delta_{r,1} (-)^{wtb\!-\!1} s \sum_{k=0}^{wt\,a} {wt\,a \choose k} (-)^k {k+\!wtb\!-\!t \choose s} \nonumber \\
&=& 0 \qquad \qquad \qquad \qquad\mbox{as} \qquad s < wt\,a \quad . \label{2nd case6}
\end{eqnarray}
\\
Secondly take
\begin{eqnarray*}
a&=& b_{-n_1} \dots b_{-n_N} \bar b_{-\bar m_1} \dots \bar b_{-\bar m_M} \bar b_{-r}  \vert 0 \!\!> \\
b&=& b_{-m_1} \dots b_{-m_M} \bar b_{-\bar n_1} \dots \bar b_{-\bar n_N}  \bar b_{-s}  \bar b_{-t} \vert 0 \!\!>  \qquad (N\!+\!2 \equiv M \, mod  \ 3) \ .
\end{eqnarray*}
Again there are two inequivalent types of contributions $\tilde \pi$ and $\tilde {\tilde \pi}$.
This time there is no contribution for $\tilde {\tilde \pi}$ as our usual summation index $k$ would need to become $k=wt\,a+wt\,b-s-t > wt\,a \ $ ($wt\,b-s-t > 0$ since $N+M>0$), which is outside the range.

For $\tilde \pi$ the calculation is very similar to (\ref{2nd case6}):
\begin{eqnarray}
\tilde \pi(a \Box_{con} b)&=&\alpha \sum_{k=0}^{wt\,a} {wt\,a \choose k} \delta_{t,1} {wta\!\!+\!\!wtb\!\!-\!\!k\!\!-\!\!s\!\!-\!\!1 \choose r\!\!-\!\!1}  * \nonumber \\
&&*\, \tilde \pi(\bar b_{-wt\,a-wt\,b+k+s}\bar b_{-s} \bar b_{-1}\vert0 \!\!>) \nonumber\\
&=& 0 \qquad \qquad \qquad \mbox{as} \quad r < wt\,a \quad (\mbox{since} \, N\!+\!M>0) \ . \label{3nd case7}
\end{eqnarray}
\\
Finally we take 
\begin{eqnarray*}
a&=& b_{-n_1} \dots b_{-n_N} \bar b_{-\bar m_1} \dots \bar b_{-\bar m_M}  \vert 0 \!\!> \\
b&=& b_{-m_1} \dots b_{-m_M} \bar b_{-\bar n_1} \dots \bar b_{-\bar n_N} \bar b_{-r} \bar b_{-s}  \bar b_{-t} \vert 0 \!\!>  \qquad (N \equiv M \, mod \ 3) \ .
\end{eqnarray*}
This time the only possible contribution is vanishing, as $k$ would have to be outside its range.
This completes the proof that $\pi$ vanishes on $O(V)$.


\begin{thebibliography}{99}
\bibitem{bpz} A. Belavin , A. Polyakov , A. Zamolodchikov, 
Infinite conformal symmetry in two-dimensional quantum field theory,
Nucl. Phys. {\bf B241}, 333-380 (1984)
\bibitem{b} R. Borcherds, Vertex algebras, Kac-Moody algebras, and the Monster, Proc. Nat. Acad. Sci. USA {\bf 83} (1986), 3068
\bibitem{itzykson} A. Capelli, C. Itzykson, J. Zuber, The A-D-E Classification of Minimal and $A_1^{(1)}$ Conformal Invariant Theories, Commun. Math. Phys. {\bf 113} , 1-26 (1987)
\bibitem{dgm}  L. Dolan, P. Goddard, P. Montague, Conformal Field
Theory of Twisted Vertex Operators, Nucl. Phys., {\bf B338}, 529
(1990)
\bibitem{dgm95}  L. Dolan, P. Goddard, P. Montague, Conformal Field
Theories, Representations and Lattice Constructions, preprint
hep-th/9410029
\bibitem{fhl} I. Frenkel, Y.Huang, J. Lepowsky, On axiomatic approaches to vertex operator algebras and modules, Memoirs Amer. Math. Soc., {\bf 104}, 1 (1993) 
\bibitem{flm} I. Frenkel, J. Lepowsky, A. Meurman, Vertex Operator Algebras and the Monster, Pure Appl. Math. {\bf 134}, Academic Press, New York, 1988
\bibitem{fz} I. Frenkel, Y. Zhu, Vertex Operator Algebras Associated
to Representations of Affine and Virasoro Algebras, Duke Math. Journal
, {\bf 66}, 123 (1992)
\bibitem{g} P. Goddard, Meromorphic Conformal Field Theory, in {\it Infinite dimensional Lie algebras and Lie groups: Proceedings of the CIRM-Luminy Conference, 1988}, World Scientific, Singapore (1989)
\bibitem{montague} P. Montague, On the Uniqueness of the Twisted Representation in the $\Zop_2$ Orbifold Construction of a Conformal Field Theory from a Lattice, hep-th/9507085
\bibitem{z3montague} P. Montague, Third and Higher Order NFPA Twisted Constructions of Conformal Field Theories from Lattices, Nucl. Phys. B441 (1995) 337-382, hep-th/9502138  
\bibitem{ms} G. Moore , N. Seiberg, Classical and quantum conformal
field theory. Commun. Math. Phys.  {\bf 123}, 177-254 (1989)
\bibitem{prudnikov} A. Prudnikov, Y. Brychkov, O. Marichev, Integrals and Series {\bf Vol 1}, Gordon and Breach, New York (1986) (entries 4.2.5,36 and 4.2.5,47)
\bibitem{w} W. Wang, Rationality of Virasoro Vertex Operator Algebras, Int. Math. Research Notices No 7 (1993) 197 (in Duke Math. Journal)
\bibitem{zhu} Y. Zhu, Vertex Operator Algebras, Elliptic Functions and
Modular Forms, Caltech preprint (1990)
\end{thebibliography}
\end{document}